
\input harvmac.tex
\def\bar{\overline}
\def\tilde{\widetilde}
\def\np#1#2#3{Nucl. Phys. {\bf B#1} (#2) #3}
\def\pl#1#2#3{Phys. Lett. {\bf #1B} (#2) #3}

\def\physrev#1#2#3{Phys. Rev. {\bf D#1} (#2) #3}

\def\prep#1#2#3{Phys. Rep. {\bf #1} (#2) #3}

\def\cmp#1#2#3{Comm. Math. Phys. {\bf #1} (#2) #3}

\def\semidirect{\mathbin{\hbox{\hskip2pt\vrule height 4.1pt depth -.3pt
width .25pt \hskip-2pt$\times$}}}

\font\zfont = cmss10 
\font\litfont = cmr6

\def\bigone{\hbox{1\kern -.23em {\rm l}}}
\def\ZZ{\hbox{\zfont Z\kern-.4emZ}}
\def\half{{\litfont {1 \over 2}}}

\def\CM{{\cal M}}
\def\CP{{\cal P}}

\def\Im{{\rm Im ~}}
\def\Pf{{\rm Pf ~}}
\def\mod{{\rm mod ~}}

 \def\bar{\overline}
\Title{hep-th/9408099, RU-94-60, IASSNS-HEP-94/55}
{\vbox{\centerline{Monopoles, Duality and Chiral Symmetry Breaking}
\smallskip
\centerline{in N=2 Supersymmetric QCD}}}
\bigskip
\centerline{N. Seiberg}
\smallskip
\centerline{\it Department of Physics and Astronomy}
\centerline{\it Rutgers University, Piscataway, NJ 08855-0849, USA}
\smallskip
\centerline{and}
\smallskip
\centerline{\it Institute for Advanced Study}
\centerline{\it Princeton, NJ 08540, USA}
\medskip
\centerline{and}
\smallskip
\centerline{E. Witten}
\medskip
\centerline{\it Institute for Advanced Study}
\centerline{\it Princeton, NJ 08540, USA}
\bigskip
\baselineskip 18pt
\noindent
We study four dimensional $N=2$ supersymmetric gauge theories with
matter multiplets.  For all such models for which the gauge group is
$SU(2)$, we derive the exact metric on the moduli space of quantum vacua
and the exact spectrum of the stable massive states.  A number of new
physical phenomena occur, such as chiral symmetry breaking that is
driven by the condensation of magnetic monopoles that carry global
quantum numbers.  For those cases in which conformal invariance is
broken only by mass terms, the formalism automatically gives results
that are invariant under electric-magnetic duality.  In one instance,
this duality is mixed in an interesting way with $SO(8)$ triality.
\Date{8/94}

\newsec{Introduction}

\nref\russians{M.A. Shifman and A.I. Vainshtein, \np{277}{1986}{456};
\np{359}{1991}{571}.}%
\nref\cern{D. Amati, K. Konishi, Y. Meurice, G.C. Rossi and G.
Veneziano, \prep{162}{1988}{169} and references therein.}%
\nref\nonren{N. Seiberg, \pl{318}{1993}{469}.}%
\nref\moduli{N. Seiberg, Phys. Rev. {\bf D49} (1994) 6857,
hep-th/9402044.}%
\nref\exact{K. Intriligator, R. Leigh and N. Seiberg,
\physrev{50}{1994}{1092}, hep-th/9403198.}%
\nref\kl{V. Kaplunovsky and J. Louis, hep-th/9402005.}%
The holomorphic properties of supersymmetric field theories
\refs{\russians -\nonren}
can be a powerful tool in deriving exact results about them
\refs{\nonren - \kl}.  In $N=1$ theories in four dimensions,
the superpotential and the coefficients of the gauge kinetic terms are
holomorphic and are constrained by these considerations.  In
four-dimensional $N=2$ theories, the Kahler potential is also
constrained by holomorphy
\ref\specialg{S. J. Gates, Jr., \np{238}{1984}{349}; B. De Wit and
A. Van Proeyen, \np {245}{1984}{89}.}
and therefore it can be analyzed similarly.  This was used in the weak
coupling analysis of
\ref\natint{N. Seiberg, \pl{206}{1988}{75}.}
and in the exact treatment of
\ref\paperi{N. Seiberg and E. Witten, RU-94-52, IAS-94-43,
hep-th/9407087.}.
The ability to make exact statements in these four-dimensional strongly
coupled field theories makes them interesting laboratories where various
ideas about quantum field theory can be tested, as has been seen in
various $N=1$ \moduli\ and $N=2$ \paperi\ theories.

An important element in the analysis of these theories is the fact that
supersymmetric field theories often have a continuous degeneracy of
inequivalent ground states.  Classically, they correspond to flat
directions of the potential along which the squarks acquire expectation
values which break the gauge symmetry.  The singularities in the
moduli space of classical ground states
are the points where the gauge symmetry is
enhanced.  Quantum mechanically, the vacuum degeneracy can be lifted by
non-perturbative effects
\ref\ads{I. Affleck, M. Dine, and N. Seiberg, \np{241}{1984}{493};
\np{256}{1985}{557}.}.
Alternatively, the vacuum degeneracy can persist and the theory then has
a quantum moduli space of vacua.  One then would like to know whether
this space is singular and if so what is the physics of the
singularities.  This question has been studied in some $N=1$ theories in
\moduli.  The holomorphy of the superpotential enables one to determine
the light degrees of freedom, and the quantum moduli space.  In \paperi\
the pure gauge $N=2$ theory has been analyzed.  In this case, the masses
of the stable particles, the low energy effective interactions, and the
metric on the quantum moduli space can be determined.

The classical moduli space of the $N=2$ $SU(2)$ gauge theory is
parametrized by $u=\langle\Tr \,\phi^2\rangle $ where $\phi$ is a
complex scalar field in the adjoint representation of the gauge group.
For $u\not=0$ the gauge symmetry is broken to $U(1)$.  At $u=0$ the
space is singular and the gauge symmetry is unbroken.
Our main goal is to determine -- as quantitatively as possible -- how
this picture is modified quantum mechanically.

The quantum
moduli space is described by the global supersymmetry version of special
geometry.  The Kahler potential
\eqn\kamet{K= \Im a_D(u) \bar a(\bar u)}
determines the metric or equivalently the kinetic terms.  The pair
$(a_D, a)$ is a holomorphic section of an $SL(2,{\bf Z})$ bundle over
the  punctured
complex $u$ plane.  They are related by $N=2$ supersymmetry to a
$U(1)$ gauge multiplet.  $a$ is related by $N=2$ to the semiclassical
``photon'' while $a_D$ is related to its dual -- ``the magnetic
photon.''

For large $|u|$ the theory is semiclassical and
\eqn\semiclass{\eqalign{&a \cong \sqrt{2 u} \cr
&a_D \cong i{2\over \pi} a \log a .}   }
These expressions are modified by instanton corrections \natint.  The
exact expressions were determined in \paperi\ as the periods on a torus
\eqn\torpu{y^2=(x^2 - \Lambda^4)(x-u)} of the
meromorphic one-form
\eqn\oneform{\lambda={\sqrt 2
\over 2\pi}{dx \,\,(x-u)\over y}  .}
In \torpu, $\Lambda$ is the dynamically generated mass scale of the
theory.

The spectrum contains dyons labeled by various magnetic and electric
charges.  Stable states with magnetic and electric charges $(n_m,n_e)$
have masses given by the BPS
\nref\prasad{M. K. Prasad and C. M. Sommerfield, Phys. Rev. Lett.
{\bf 35} (1975) 760.}%
\nref\bog{E. B. Bogomol'nyi, Sov. J. Nucl. Phys. {\bf 24} (1976) 449.}%
\nref\wittenol{E. Witten and D. Olive, \pl {78}{1978}{97}.}%
formula \refs{\prasad - \wittenol}
\eqn\dyonmass{M^2=2|Z|^2= 2|n_e a(u) + n_m a_D(u)|^2 .}
There are two singular points on the quantum moduli space at $u=\pm
\Lambda^2$; they are points at which a magnetic monopole becomes
massless.  When an $N=2$ breaking but $N=1$ preserving mass term is
added to the theory, these monopoles condense, leading to confinement
\paperi.

Here, we extend our analysis to theories with additional $N=2$ matter
multiplets, known as hypermultiplets.  As in \paperi, we limit ourselves
to theories with gauge group $SU(2)$.  If matter multiplets are to be
included while keeping the beta function zero or negative, there are the
following possibilities.  One can consider a single hypermultiplet in
the adjoint representation.  If the bare mass is zero, this actually
gives a theory with $N=4$ supersymmetry; it is possible to add a bare
mass breaking the symmetry to $N=2$.  Or one can add $N_f$
hypermultiplets in the spin one-half representation of $SU(2)$, with an
arbitrary bare mass for each multiplet; this preserves asymptotic
freedom for $N_f\leq 3$, while the $\beta$ function vanishes for
$N_f=4$.  (It has been known for some time that the perturbative beta
function vanishes for $N_f=4$; our results make it clear that this is
true nonperturbatively.)  For all these theories, we will obtain the
same sort of exact results that we found in \paperi\ for the pure $N=2$
gauge theory.

One motivation for studying these systems is that just as the pure gauge
theory taught us something about confinement, the theory with matter may
teach us about chiral symmetry breaking.  In fact, we will find a new
mechanism for chiral symmetry breaking -- it arises in some of these
models (those with $N_f=2,3,4$) {}from the condensation of magnetic
monopoles which carry global quantum numbers.  These magnetic monopoles
can be continuously transformed into elementary quanta as parameters are
varied!  That bizarre-sounding statement, which is possible because of
the non-abelian monodromies, also means that one can interpolate
continuously {}from the confining phase (triggered by condensation of
monopoles) to the Higgs phase (triggered by condensation of elementary
quanta).  The fact that these two phases are in the same universality
class is, of course, expected for $N_f>0$ as elementary doublets are
present
\ref\higgscon{T. Banks, E. Rabinovici, \np{160}{1979}{349};
E. Fradkin and S. Shenker, \physrev{19}{1979}{3682}.}.

In studying confinement in \paperi\ and confinement and chiral symmetry
breaking in the present paper, we are mainly working in a region close
to a transition to a Higgs phase (in which the unbroken symmetry group
is abelian) and apparently far from the usual strongly coupled gauge
theories of gauge bosons and fermions only.  However, one can reduce to
a more usual situation by adding suitable $N=1$-invariant perturbations
of the superpotential.  For instance, in \paperi\ we exploited the
possibility of perturbing the superpotential by $m\Tr \Phi^2$, with $m$
a complex parameter and $\Phi$ an $N=1$ chiral multiplet related to the
gauge bosons by $N=2$ supersymmetry.  (The absolute value of $m$ is a
bare mass, and its phase determines the parity violation in certain
Yukawa couplings.)  Our analysis of confinement was valid for small $m$,
while a theory much more similar to ordinary QCD would emerge for large
$m$.

But all experience indicates that as long as supersymmetry is unbroken,
supersymmetric theories in four dimensions do not have phase transitions
in the usual sense as a function of the complex parameters such as $m$;
such transitions are more or less prevented by holomorphy. In the usual
study of phase transitions, one meets singularities -- phase boundaries
-- of real codimension one.  By contrast, one gets in the supersymmetric
case singularities of complex codimension $\geq 1$ in the space of
vacua, permitting continuous interpolations {}from one regime to
another.  (The only situation in which one meets a singularity in
interpolating from one regime to another is that in which the moduli
space of vacua has several branches, which intersect somewhere, and one
wishes to interpolate from one branch to another.)  Note that, as long
as supersymmetry is unbroken, the energy vanishes and so the usual
mechanism behind ordinary phase transitions -- minimizing the energy --
does not operate.  The structures that have emerged for small $m$ in
\paperi\ and the present paper are qualitatively similar to what one
would guess (by analogy with QCD) for large $m$, and we do believe that
the large $m$ and small $m$ theories are in the same universality class.

Another motivation for the present work is quite different: we will gain
new insights about electric-magnetic duality in strongly interacting
gauge theories.  We will see, for instance, that in those theories in
which conformal invariance is broken only by mass terms, the formalism
is inevitably invariant under electric-magnetic duality.  There are two
relevant examples.  One is the $N=4 $ theory, which is the original
arena of Olive-Montonen duality
\ref\om{C. Montonen and D. Olive, \pl {72}{1977}{117}; P. Goddard,
J. Nuyts, and D. Olive, Nucl. Phys. {\bf B125} (1977) 1.}.
\nref\cardy{J. Cardy and E. Rabinovici, Nucl. Phys. {\bf B205} (1982) 1;
J. Cardy, Nucl. Phys. {\bf B205} (1982) 17.}%
\nref\wilczek{A. Shapere and F. Wilczek, Nucl. Phys. {\bf B320}
(1989) 669.}%
\nref\sen{A. Sen, TIFR-TH-94-08, hep-th/9402002; G. Segal, to appear}%
This duality was originally formulated as a ${\bf Z}_2$ symmetry, in
terms of the coupling constant only, but when the $\theta$ angle is
included it can be extended to an action of $SL(2,{\bf Z})$ on $\tau$
\refs{\cardy-\sen}.  The other relevant case is the $N_f=4$ theory.
In both of these examples we will find a full $SL(2,{\bf Z})$ symmetry
exchanging electric and magnetic charges.  (However, because of a factor
of two in the conventions that will be explained below, $SL(2,{\bf Z})$
is defined differently in the two cases.)  In many ways, the richest
behavior that we find is that of the $N_f=4$ theory; it has an $SO(8)$
global symmetry, and it turns out that $SL(2,{\bf Z})$ duality is mixed
with $SO(8)$ triality.

We begin our discussion in section 2 with a warm-up example of an
abelian theory with $N=2$ supersymmetry.  It exhibits some of the new
elements which will be important later on.  Then we turn to $N=2$ QCD
and discuss its classical properties in section 3.  In section 4 we
begin the analysis of the quantum theory.  Section 5 is devoted to the
stable particles in the theory -- the BPS saturated states.  We
discuss their masses and quantum numbers.  Section 6 deals with
duality transformations.  Here we mention only the differences
compared with the situation in the pure gauge theory which we
discussed in \paperi.  In section 7 we motivate our suggestion for the
qualitative structure of the moduli space for $N_f=1,2,3$ -- the
number of singularities and their nature.  Section 8 describes the low
energy theory near the singularities and exhibits non-trivial
consistency checks of our suggestion.  In section 9 we break $N=2$ to
$N=1$ supersymmetry and recover the results in
\moduli.  In section 10 we suggest the qualitative structure on the
moduli space for the theory with $N_f=4$.  Section 11 is an introduction
to the more quantitative discussion of the metric in the remaining
sections.  In sections 12, 13 and 14 we find the metric on the moduli
space for $N_f=1$, 2 and 3 respectively.  Then, the masses of the
particles and some consistency checks are discussed in section 15.  In
section 16 we analyze two scale invariant theories -- the theory with
$N=4$ supersymmetry and the $N=2$ theory with four flavors.  In both
cases we can turn on $N=2$ preserving mass terms and solve for the
metric on the moduli space.  Our previous answers for the metric are
obtained by taking appropriate scaling limits as some masses go to
infinity.  Duality invariance of the results is manifest.  In section 17
we show a highly non-trivial consistency check of our answers.

\bigskip
\noindent{\it A Note On Conventions}

The $N_f>0$ theories have fields in the two-dimensional representation
of $SU(2)$, so they have particles of half-integral electric charge if
we use the same normalization as in \paperi.  Instead, we will multiply
$n_e$ by 2, to ensure that it is always integral, and compensate by
dividing $a$ by 2.  The asymptotic behavior is thus
\eqn\semiclassn{\eqalign{&a \cong \half \sqrt{2 u} \cr
&a_D \cong i{4\over \pi} a \log a}}
Because of this change of normalization, the effective coupling constant
$\tau = {\partial a_D \over \partial a}$ is also rescaled and now $\tau
= {\theta \over \pi} + {8\pi i \over g^2}$.  Correspondingly, the family
of curves \torpu\ is replaced by a different family that will be
described later.

\newsec{A warm-up example: QED with matter}

\subsec{QED}

For background, we first consider abelian gauge theories with $N=2$
supersymmetry and charged matter hypermultiplets - that is, the $N=2$
analog of ordinary QED.

The ``photon,'' $A_\mu$ is accompanied by its $N=2$ superpartners -- two
neutral Weyl spinors $\lambda$ and $\psi$ that are often called
``photinos,'' and a complex neutral scalar $a$.  They form an
irreducible $N=2$ representation that can be decomposed as a sum of two
$N=1$ representations\foot{We use the conventions and notation of
\ref\wb{J. Wess and J. Bagger, {\it Supersymmetry and Supergravity}
(Princeton University Press, 1982).}.}:
$a$ and $\psi$ are in a chiral representation, $A$, while $A_\mu$ and
$\lambda$ are in a vector representation, $W_\alpha$.

We take the charged fields, the ``electrons,'' to consist of $k$
hypermultiplets of electric charge one.  Each hypermultiplet, for
$i=1\dots k$, consists of two $N=1$ chiral multiplets $M^i$ and $\tilde
M_i $ with opposite electric charge; of course such an $N=1$ chiral
multiplet contains a Weyl fermion and a complex scalar.

The renormalizable $N=2$ invariant Lagrangian is described in an $N=1$
language by canonical kinetic terms and minimal gauge couplings for all
the fields as well as a superpotential
\eqn\qedsup{W=\sqrt{2} A M^i \tilde M_i + \sum_i m_i M^i
\tilde M_i  .}
The first term in related by $N=2$ supersymmetry to the gauge coupling
and the second one leads to  $N=2$ invariant mass terms.

Consider first the massless theory ($m_i=0$).  Its global symmetry is
$SU(k)\times SU(2)_R \times U(1)_\CR$.  $SU(k)$ acts on the $k$
hypermultiplets, while $SU(2)_R\times U(1)_\CR$ is the $R$ symmetry
group.  The two supercharges are in a doublet of $SU(2)_R$ and their
chiral components have charge one under $U(1)_\CR$. The $4k$ real
scalars in $M$ and $\tilde M$ transform like $({\bf k},{\bf 2},0)
\oplus ({\bf \bar k}, {\bf 2},0)$ and the scalar $a$ transforms as
$({\bf 1},{\bf 1},2)$.  Here and throughout the paper we denote
representations by their dimensions, except for representations of $U(1)$
which are labeled by their charge.

\subsec{The classical moduli space}

The classical moduli space of this theory has a branch with non-zero
$a$.  Along this flat direction the gauge symmetry is unbroken and all
the $M$ particles acquire a mass.  Since the photon is massless, we will
refer to this branch of the moduli space as the Coulomb branch.

For $k \ge 2$ and $m_i=0$ there are also flat directions where $M$ is
non-zero.  Along these directions the $U(1)$ gauge symmetry is broken
and therefore we will refer to this branch of the moduli space as the
Higgs branch.  Up to gauge and global symmetry transformations these
flat directions are
\eqn\mbranch{\eqalign{&M=(B, 0,...)\cr
&\tilde M=(0,B,0,...) .}}
(Vanishing of the $D$ terms requires $|M|=  |\tilde M|$.)
The global $U(1)_\CR$ is unbroken and we will ignore it.  The other
global symmetry is broken as
\eqn\globsymb{SU(k)\times SU(2)_R \rightarrow
\cases{SU(2)_{R'} & for $k=2$ \cr
U(1)\times SU(2)_{R'} & for $k=3$ \cr
SU(k-2) \times U(1) \times SU(2)_{R'} & for $k \ge 4$}}
where $SU(2)_{R'}$ is a diagonal subgroup of $SU(2)_R$ and an $SU(2)
\subset SU(k)$.  The light fields on the moduli space are in
hypermultiplets of $N=2$ and therefore, the moduli space is a
hyper-Kahler manifold.  They transform like
\eqn\lightuone{\matrix{
{\bf 3}\oplus {\bf 1}& \quad {\rm for ~} k =2 \cr
(3,{\bf 2}) \oplus (-3, {\bf 2}) \oplus (0,{\bf 3})+ (0,{\bf 1})
&\quad {\rm for ~} k =3 \cr
({\bf k-2}, 1, {\bf 2}) \oplus ({\bf \overline{k-2}} , -1, {\bf 2})
\oplus  ({\bf 1} ,0, {\bf 3} )\oplus ({\bf 1},0,{\bf 1} ) &\quad {\rm
for~} k \ge 4}}
The boson in the last representation labels inequivalent vacua and the
other bosons are the Goldstone bosons.

\subsec{The quantum moduli space}

Quantum mechanically, this theory is probably not well-defined because
it is not asymptotically free.  However, it is often the case that such
a theory is embedded in a larger theory which is asymptotically free.
In \paperi, we have encountered such a theory with $k=1$ as the low
energy limit of an asymptotically free $SU(2)$ gauge theory.  Below we
will see more examples.

The metric on the Higgs branch cannot be corrected quantum mechanically.
The reason for that is that the space is an almost homogeneous space
which has a unique hyper-Kahler metric invariant under the symmetries.
The metric on the Coulomb branch can be corrected.  Since the low energy
theory at the generic points on the Coulomb branch includes only the
gauge multiplet, the metric on this branch is of the special geometry
type, i.e. it is determined by the Kahler potential
\eqn\kabe{K= \Im a_D(a) \bar a  .}
The gauge kinetic energy is proportional to
\eqn\gaugecoue{\int d^2 \theta ~ {\partial a_D \over \partial a} ~
W_\alpha^2  .}
In this $N=2$ theory, the one loop approximation to $K$ is exact (there
are no higher order perturbative corrections and there are no $U(1)$
instantons on ${\bf R}^4$) leading to
\eqn\aadabe{a_D = - {ik \over 2\pi} a \log (a/\Lambda) .}
The lack of asymptotic freedom appears here as a breakdown of the theory
at $|a|= \Lambda/e $ where the metric on the moduli space $\Im {\partial
a_D \over \partial a}$ vanishes and the effective gauge coupling is
singular.  This is the famous Landau pole.

When the masses in \qedsup\ are not zero the moduli space changes.  The
singularities on the Coulomb branch can move.  Whenever $a=-{1 \over
\sqrt{2}} m_i$ one of the electrons becomes massless.  Therefore
\eqn\aadabem{a_D = - {i \over 2\pi} \sum_i( a+ m_i/\sqrt{2})
\log \left({ a+m_i/ \sqrt{2}\over \Lambda} \right)  .}

If some of the masses are equal, the corresponding singularities on the
Coulomb branch coincide and there are more massless particles there.  In
this case a Higgs branch with non-zero expectation values for these
electrons touches the Coulomb branch at the singularity.  When there is
only one massless electron hypermultiplet, the $|D|^2$ term in the
potential prevents a Higgs branch {}from developing.

\subsec{BPS-saturated states}

The $N=2$ algebra has ``large'' representations with 16 states and
``small'' ones with only four states.  As explained in \wittenol, the
masses of particles in small representations are determined by their
quantum numbers.  Indeed, in terms of the central extension $Z$ in the
$N=2$ supersymmetry algebra, the mass of a particle in a small
representation is $M=\sqrt{2}|Z|$.

The $N=2$ algebra requires that $Z$ is a linear combination of conserved
charges.  In \wittenol\ and also in \paperi, $Z$ was a linear
combination of the electric and magnetic quantum numbers $n_e$ and
$n_m$.  The classical expression can be written $Z=n_e a + n_m a_D$ and
this way of writing it is also valid quantum mechanically in the pure
gauge theory, as was explained in \paperi.  When there are additional
abelian conserved charges, they might conceivably appear in the formula
for $Z$.

Such a modification is present in the version of QED described above.
One can easily deduce this as follows.  First, of all, by counting
states one sees that the ``electrons'' are in ``small'' representations.
(In fact, any multiplet in which all states are of spin $\leq 1/2$ is
automatically a sum of ``small'' representations.)  On the other hand,
the masses of the electrons are not $\sqrt 2 |a|$, as would follow
{}from the ``old'' formula for $Z$, but rather the mass of the $i^{th}$
hypermultiplet is $|\sqrt 2 a+m_i|$.  To give such a result, the $U(1)$
charges $S_i$ of the hypermultiplets must appear in $Z$ as follows:
\eqn\zmodex{Z=n_e a + n_m a_D + \sum_i S_i m_i /\sqrt{2}  .}
This result can easily be verified by computing the Poisson brackets
of the supercharges.  The new term in $Z$ will pervasively affect
the analysis below.

\subsec{Breaking $N=2$ to $N=1$}

We will later need a deformation of this theory breaking $N=2$ to $N=1$
supersymmetry by adding a term $\mu A$ to the superpotential \qedsup.
This breaks the global $SU(2)_R$ symmetry.  Consider for simplicity the
case where all the electrons are massless ($m_i=0$).  Using the
equations of motion and the $U(1)$ D terms, it is easy to see that the
Coulomb branch collapses to a point $a=0$.  For $k=1$ there is a unique
ground state with $M=-\tilde M = \left( {\mu \over \sqrt{2}
}\right)^\half$ and the $U(1)$ gauge symmetry is spontaneously broken.
For $k \ge 2$ there are also $M$ flat directions.  Up to symmetry
transformations they have the form
\eqn\mflatwitm{\eqalign{&M=(C,0,...) \cr
&\tilde M=(-{\mu \over \sqrt{2}C} ,B  ,0,...) }}
with $|B|^2 + |{\mu \over \sqrt{2}C}|^2 = |C|^2$.  At the generic point
the global symmetry is broken as
\eqn\globsymbb{SU(k) \rightarrow
\cases{1 & for $k=2$ \cr
U(1) & for $k=3$ \cr
SU(k-2) \times U(1) & for $k \ge 4$}}
and at the special point $B= 0 $ it is broken as
\eqn\globsymbbs{SU(k)\rightarrow
\cases{U(1) & for $k=2$ \cr
SU(2) \times U(1) & for $k=3$ \cr
SU(k-1) \times U(1) & for $k \ge 4$ .}}

\newsec{Classical moduli space of QCD with matter}

We now turn to QCD with an $SU(2)$ gauge group.  The gluons are
accompanied by Dirac fermions and complex scalars $\phi$ in the adjoint
representation of the gauge group.  We also add $N_f$ hypermultiplets of
quarks in the fundamental representation.  (We will also consider the
case of a single hypermultiplet in the adjoint representation, this
being the $N=4$ theory.)  As in the previous section, each
hypermultiplet contains a Dirac fermion and four real scalars.  In terms
of $N=1$ superfields the hypermultiplets contain two chiral superfields
$Q^i{}_a$ and $\tilde Q_i{}_a$ ($i=1,...,N_f$ is the flavor index and
$a=1,2$ the color index) and the $N=2$ gauge multiplets include $N=1$
gauge multiplets and chiral multiplets $\Phi$.  The superpotential for
these chiral superfields is
\eqn\nonabtr{W=\sqrt {2} \tilde Q_i \Phi Q^i + \sum_i m_i\tilde Q_i Q^i}
with color indices suppressed.

When the quarks are massless the global symmetry of the classical theory
is a certain quotient of $O(2N_f)\times SU(2)_R \times U(1)_\CR $.  The
reason for the $O(2N_f)$ symmetry (rather than $SU(N_f) \times U(1)$) is
that for an $SU(2)$ gauge theory the quarks $Q$ and the antiquarks
$\tilde Q$ are in isomorphic representations of the gauge group.
Therefore, we will also often denote these $N=1$ chiral superfields by
$Q^r$ with $r=1,...,2N_f$ labeling the components of an $SO(2N_f)$
vector.  The squarks $(Q, \tilde Q)$ transform like $({\bf 2N_f}, {\bf
2}, 0)$ and the scalar in the gauge multiplet as $({\bf 1}, {\bf 1},
2)$.  It will be important that the symmetry of the hypermultiplets is
$O(2N_f)$ and not $SO(2N_f)$; for instance there is a ``parity''
symmetry, ${\bf Z}_2 \subset O(2N_f),$ acting as
\eqn\paritty{\rho:Q_1 \leftrightarrow \tilde Q_1}
with all other squarks invariant.

Globally, the symmetry group is not quite the product of $O(2N_F)\times
SU(2)_R \times U(1)_\CR$ with the Lorentz group.  A ${\bf Z}_2 \subset
U(1)_\CR$ is isomorphic to $(-1)^F$ which is in the Lorentz group.
Also, when combined with the center of the $SU(2)_R$, this ${\bf Z}_2$
acts the same as the ${\bf Z}_2$ in the center of $O(2N_f)$.

As in the abelian example, there is always a flat direction with non
zero $\phi$.  Along this direction the gauge symmetry is broken to
$U(1)$ and all the quarks are massive.  Only $U(1)_\CR $ is
spontaneously broken there.  We will refer to this branch of the moduli
space as the Coulomb branch.  For $N_f=0, 1$ there are no other flat
directions, but such directions appear (when $m_i=0$) for $N_f\ge 2$.
Since the gauge symmetry is completely broken along these flat
directions, we will refer to them as the Higgs branches.  The Higgs
branches can be analyzed as follows.

First, it follows {}from requiring that the superpotential be stationary
and the $D$ terms vanish that on the Higgs branches, $\Phi$ must be zero.
The flat directions in $Q$ space are found by setting to zero the $D$
terms, dividing by the gauge group $SU(2)$, and asking for the
superpotential to be stationary.  The combined operation of setting the
$D$ terms to zero and dividing by $SU(2)$ is equivalent to dividing by
$SL(2,{\bf C})$.  The quotient by $SL(2,{\bf C})$ of the space of
squarks can be parametrized by the $SL(2,{\bf C})$-invariant functions
$V^{rs}=Q^r{}_aQ^{sa}$ of the squarks; here $r,s=1\dots 2N_f $ and $a$
are the flavor
and color indices, and $V^{rs}=-V^{sr}$.  $V$ generates the ring of
$SL(2,{\bf C})$-invariant polynomials in the $Q$'s.  The $V$'s are not
independent but obey certain  quadratic equations  stating that
$V$ is of rank two.  For $N_f=2$, there is  a single such
equation
\eqn\snon{\epsilon_{rstu}V^{rs}V^{tu}=0.}

We still must impose the condition that the superpotential should be
stationary.  Since the superpotential is linear in $\Phi$, and $\Phi=0$,
the only non-trivial condition is $\partial W/\partial \Phi=0$, or
$X_{ab}=0$ with $X_{ab}=\sum_rQ^r{}_aQ^r{}_b.$ This is
equivalent to
\eqn\knon{0=Q^{ra}Q^{sb}X_{ab}=-V^{rt}V^{ts}}
for arbitrary $r,s$.

For example, the above equations can be analyzed as follows for $N_f=2$.
The symmetry group $O(4)$ is locally $SU(2)\times SU(2)$, and the
antisymmetric tensor $V^{rs}$ decomposes as $({\bf 3},{\bf 1})\oplus
({\bf 1},{\bf 3})$; we will call the two pieces $V_L$ and $V_R$.  The
symmetric tensor in \knon\ transforms as $({\bf 1},{\bf 1})\oplus ({\bf
3},{\bf3})$.  The $({\bf 3},{\bf 3})$ piece is bilinear in $V_L$ and
$V_R$ and so vanishes if and only if $V_L=0$ or $V_R=0$.  There are thus
two Higgs branches, with $V$ being self-dual or anti-self-dual.  The
$({\bf 1},{\bf 1})$ part of \knon\ gives $V_L{}^2=0$ (or $V_R{}^2=0$,
for the other branch), which actually duplicates the content of \snon.
By $V_L{}^2$ we mean of course $v_1{}^2+v_2{}^2+v_3{}^2$ where
$v_r=V^{r4}$.  The manifold given by the equation
$v_1{}^2+v_2{}^2+v_3{}^2=0$ for three complex variables $v_i$ is
equivalent to the quotient ${\bf C}^2/{\bf Z}_2$,\foot{Introduce new
variables $a,b$, defined up to an overall sign, by $v_1+iv_2 =a^2,
\,\,v_1-iv_2=b^2,\,\,v_3=iab$.} and so admits a flat hyperkahler metric
with a ${\bf Z}_2$ orbifold singularity at the origin.

Thus in particular, for $N_f=2$, there are two Higgs branches which meet
each other and the Coulomb branch at the origin.  The two branches are
exchanged by the ``parity'' symmetry generated by $\rho$ of \paritty.
For $N_f>3$, instead, there is a single irreducible Higgs branch (this
follows, for instance, {}from the symmetries) which meets the Coulomb
branch at the origin.

Along the Higgs branches the $U(1)_\CR$ symmetry is unbroken and the
other global symmetry is broken as
\eqn\glosymb{O(2N_f) \times SU(2)_R \rightarrow \cases{
SU(2) \times SU(2)_{R'} & for $N_f = 2$ \cr
O(2N_f-4)\times SU(2) \times SU(2)_{R'} & for $N_f \ge 3$ }}
where the first $SU(2)$ is a diagonal subgroup of an $SU(2) \subset
SO(4) \subset SO(2N_f)$ and the $SU(2)$ gauge symmetry. $SU(2)_{R'}$ is
a diagonal subgroup of the other $SU(2) \subset SO(4) \subset SO(2N_f)$
and the original $SU(2)_R$ symmetry.  The massless fields transform like
\eqn\masslesq{\cases{
({\bf 1},{\bf 3})\oplus ({\bf 1},{\bf 1}) & for $N_f =2$ \cr
({\bf 2N_f-4},{\bf 2},{\bf 2}) \oplus
 ({\bf 1},{\bf  1},{\bf  3}) \oplus
 ({\bf 1},{\bf 1},{\bf 1}) & for $N_f \ge 3$ }}
The boson in the  last representation labels inequivalent ground states.
The other bosons are the Goldstone bosons.

\newsec{A first look at the quantum theory}

\subsec{Symmetries of the quantum theory}

Now we consider the quantum modifications to the symmetry structure.
Since the one-loop beta function of the theory is proportional to
$4-N_f$ (higher order perturbative corrections to it vanish), we limit
ourselves to $N_f=0,1,2,3$ where the theory is asymptotically free and
to $N_f=4$ where the theory will turn out to be scale invariant.

The global $U(1)_\CR$ and the ``parity'' $Z_2 \subset O(2N_f)$ are
anomalous.  For $N_f>0$, a discrete ${\bf Z}_{4(4-N_f)}$ anomaly free
subgroup is generated by
\eqn\nonandis{\eqalign{
& W_\alpha  \rightarrow e^{i \pi \over 2(4-N_f)} W_\alpha (e^{- i \pi
\over 2(4-N_f) }\theta) \cr
& \Phi \rightarrow e^{i \pi \over 4-N_f} \Phi (e^{- i \pi \over
2(4-N_f)}\theta ) \cr
& Q^1 \rightarrow \tilde Q_1 (e^{- i \pi \over 2(4-N_f)}\theta ) \cr
& \tilde Q_1 \rightarrow Q^1 (e^{- i \pi \over 2(4-N_f)}\theta )}}
with all other squarks invariant.  (For $N_f=0$, the $Q$'s are absent
and cannot be used to cancel an anomaly; the anomaly-free global
symmetry is ${\bf Z}_8$ rather than ${\bf Z}_{16}$, and is generated by
the square of the above.)  A ${\bf Z}_2 \subset {\bf Z}_{4(4-N_f)}$ is
equal to $(-1)^F$.  We can combine this symmetry with an $SU(2)_R$
transformation to find a ${\bf Z}_{4(4-N_f)} $ symmetry which commutes
with $N=1$ supersymmetry
\eqn\nonandist{\eqalign{
& \Phi \rightarrow e^{i \pi \over 4-N_f} \Phi \cr
& Q^1 \rightarrow e^{- i \pi \over 2(4-N_f)}\tilde Q_1  \cr
& \tilde Q_1 \rightarrow e^{- i \pi \over 2(4-N_f)} Q^1 \cr
& Q^i \rightarrow e^{- i \pi \over 2(4-N_f)} Q^i  \cr
& \tilde Q_i \rightarrow e^{- i \pi \over 2(4-N_f)} \tilde Q_i
\cr }}
for $i\ge 2$.  In this form it is clear that a ${\bf Z}_2$ subgroup of
this group acts the same as a ${\bf Z}_2$ in the center of $SO(2N_f)$.

Since $u= \Tr\,\phi^2$ transforms as $u\to e^{2i\pi/(4-N_f)}$, the
global symmetry acting on the $u$ plane is ${\bf Z}_{4-N_f}$ for
$N_f>0$, or ${\bf Z}_2$ for $N_f=0$.

\subsec{A first look at the quantum moduli space}

We now begin the analysis of the quantum moduli space. The first basic
fact is that for large fields, the theory is weakly coupled and the
quantum moduli space is well approximated by the classical moduli space.

Consider first the Higgs branches.  The $SO(2N_f)\times SU(2)/
SO(2N_f-4) \times SU(2)\times SU(2)$ structure should persist quantum
mechanically.  This structure admits a unique hyper-Kahler metric up to
a constant multiple (and the multiple is fixed by the behavior for large
fields).  For instance, for $N_f=2$, the metric is the orbifold metric
on ${\bf R}^4/{\bf Z}_2$.  Therefore, there are no quantum corrections
to the metric, and the singularity cannot be removed.

If these manifolds continue to touch the Coulomb branch in the quantum
theory (as we will claim), one might be tempted to guess that the
$SU(2)$ gauge symmetry should be restored there.  The reason for that is
that on the Coulomb branch there is always a massless photon while on the
Higgs branches the three gauge bosons are degenerate.  However, this
assumes that the three massive gauge bosons always exist as stable
particles.  If this is not so, it could be that (as we will eventually
argue) the photon of the Coulomb branch is the only massless gauge boson
at the point where the branches meet.

We now turn to discuss the Coulomb branch.  We parametrize it by the
gauge invariant coordinate $u= <\Tr \phi^2>$.

For $N_f=0$ the anomaly free discrete symmetry \nonandis\ acts on $u$ as
${\bf Z}_2$, as explained at the end of the last section.  In analyzing
instanton corrections to the metric on the $u$ plane, we can treat the
$U(1)_\CR$ symmetry as unbroken by assigning charge 4 to $u$ and charge
8 to the single instanton factor $\Lambda_0^4$.

For $N_f =1, 2, 3$ the anomaly free discrete symmetry ${\bf
Z}_{4(4-N_f)}$ described in \nonandis\ acts on $u$ as ${\bf Z}_{4 -
N_f}$.  The expectation value of $u$ breaks the discrete symmetry to the
${\bf Z}_4$ that acts trivially on $u$.  We can treat the $U(1)_\CR
\times {\bf Z} _2$ (this $ {\bf Z} _2$ is the $\rho$ symmetry \paritty)
as unbroken by assigning charge 4 and even parity to $u$ and charge
$2(4-N_f)$ and odd parity to the instanton factor
$\Lambda_{N_f}^{4-N_f}$.

For $N_f=4$ the $U(1)_\CR$ symmetry is anomaly free and $u$ has charge
4. The ``parity'' ${\bf Z}_2 \subset O(8)$ is still anomalous.  We can
still treat it as unbroken by assigning odd parity to the instanton
factor $q^\half = e^{i\pi \tau}=e^{-{8\pi^2\over g^2} + i \theta}$.

As for $N_f=0$, the metric and the dyon masses are determined by a
holomorphic section of an $SL(2,{\bf Z})$ bundle:
\eqn\aadexp{\eqalign{
&a={1\over 2} \sqrt{2u} + \dots \cr
&a_D= i{4-N_f \over 2 \pi} a(u) \log {u\over\Lambda_{N_f}^2} + \dots}}
where the ellipses represent instanton corrections and $\Lambda_{N_f}$
is the dynamically generated scale of the theory with $N_f$ flavors (we
will later rescale it to a convenient value).  The metric is
$ds^2= \Im ( a_D' \bar a') du \,d\bar u$ and the dyon masses $M^2=2|Z^2|$
are expressed in terms of $Z=n_e a + n_m a_D$ where $(n_m,n_e)$ are the
electric and magnetic charges.  As we said above, we use a normalization
such that all electric charges are integers.

For $N_f \not=0$, the contributions to \aadexp\ {}from terms with an odd
number of instantons vanish:  this follows {}from the anomalous ${\bf
Z}_2$ in $O(2N_f)$.  The amplitudes with odd instanton number are odd
under this ${\bf Z}_2$, and so cannot generate contributions to the
metric for $u$, which is even.  However, there is no reason why the even
instanton contributions of the form $(\Lambda_{N_f}^2/u)^{n(4-N_f)}$
should vanish.  We therefore expect that
\eqn\aadexpf{\eqalign{
&a={1 \over 2} \sqrt{2u}\left( 1 + \sum_{n=1}^\infty a_n(N_f)
\left({\Lambda_{N_f}^2 \over u}\right)^{n(4-N_f)} \right) \cr
&a_D= i{(4-N_f)\over 2\pi} a(u) \log {u\over\Lambda_{N_f}^2} +
\sqrt{u}\sum_{n=0}^\infty {a_D}_n(N_f) \left({\Lambda_{N_f}^2 \over u}
\right)^{n(4-N_f)} }}
Under the ${\bf Z}_2$ or ${\bf Z}_{4-N_f}$ symmetry of the $u$ plane,
$a$ transforms linearly and $a_D$ picks up a multiple of $a$; the latter
fact means that the symmetry shifts the electric charges of magnetic
monopoles.

Use of a gauge invariant order parameter $u$ is convenient because one
can, for instance, determine the unbroken global symmetry without
worrying about the possibility that a broken global symmetry becomes
unbroken when combined with a gauge symmetry.  However, since the
$SU(2)$ gauge symmetry is not completely broken, there are massive
charged states in the spectrum and we should be careful in determining
the way the unbroken global symmetry acts on them.  In particular, the
unbroken ${\bf Z}_4$ is generated by the operator in \nonandist\ raised
to the $4- N_f$ power.  This transformation changes the sign of $\phi$
and so acts as charge conjugation on the charged fields.  For $N_f=1,3$
this transformation acts as parity in $O(2N_f)$ (because an odd power of
$\rho$ appears), so the parity transformation is realized on the
spectrum but reverses all electric and magnetic charges; the states of
given charge are in representations of $SO(2)$ or $SO(6)$ only.  For
$N_f=2,4$, the parity symmetry is altogether spontaneously broken (since
the unbroken symmetries all contain the parity raised to an even power),
so the states are only in $SO(4)$ or $SO(8)$ representations.

\newsec{BPS-saturated states}

As we explained in our discussion of QED, a special role is played by
the BPS-saturated states which are in ``small'' representations of the
$N=2$ algebra and are frequently the only stable states in the spectrum.

Since on the Higgs branches the gauge group is completely broken, there
are no electric and magnetic charges that can appear as central charges
in the $N=2$ algebra.  The only central charge could be a $U(1)$ charge
of a hypermultiplet.  A Higgs branch exists only when there are at least
two degenerate hypermultiplets (not necessarily with zero bare mass).
Then, if there is an abelian symmetry acting on these hypermultiplets,
the massless fields carry the corresponding charge, and it cannot appear
in the central extension.  Therefore, a contribution to the central
extension exists only when there is also another hypermultiplet, which
is not degenerate with the others (its bare mass might be zero).  We do
not know of striking physical phenomena associated with these states,
mainly because in the absence of electric and magnetic charge there do
not appear to be monodromies.

On the Coulomb branch, the simplest BPS-saturated states are the
elementary quarks whose mass (if the bare masses of the hypermultiplets
are set to zero, as we do until further notice) is $M=\sqrt{2}|a|$.
They are in the vector representation of $SO(2N_f)$.

\nref\ug{R. Jackiw, C. Rebbi, \physrev {13}{1976}{3398}.}%
\nref\cug{C. R. Nohl, \physrev {12}{1975}{1840}.}%
\nref\bug{M.F. Atiyah, N.J. Hitchin {\it The Geometry and Dynamics
of Magnetic Monopoles}, (Oxford U. 1988).}%
Since the $SU(2)$ gauge symmetry is spontaneously broken to $U(1)$ along
the Coulomb branch, there are also magnetic monopoles in the spectrum.
For $N_f\not=0$, the quark fields have fermion zero modes in the
background of the monopoles \refs{\ug - \bug}.  To be precise, each
$SU(2)$ doublet of fermions has a single zero mode.  With $N_f$
hypermultiplets and therefore $2N_f$ doublets, there are $2N_f$ zero
modes transforming in the vector representation of $SO(2N_f)$.  Rather
as in the quantization of the Ramond sector of superstrings, the
quantization of these fermions zero modes turns the monopoles into
spinors of $SO(2N_f)$.  The states of definite charge furnish a
representation of $SO(2N_f)$, not of $O(2N_f)$, for a reason noted at
the end of the last section.  Note that the occurrence of spinors (in
addition to the hypermultiplets, which are vectors) means that at the
quantum level the symmetry group is really the universal cover of
$SO(2N_f)$.

\nref\moncharge{E. Witten, Phys. Lett. {\bf 86B} (1979)  283.}
There is, however, an important subtlety here. Monopoles can carry
electric charge because the classical monopole solution is not invariant
under electric charge rotations.  There is a collective coordinate
associated with electric charge rotations, and quantizing it gives a
spectrum of states of various electric charge.  A $2\pi$ rotation by the
electric charge operator does not give the identity; in a state of
$n_m=1$, it gives a topologically non-trivial gauge transformation,
whose eigenvalue is $e^{i\theta}(-1)^H$; here $\theta$ is the usual
theta angle and $(-1)^H$ is the center of the $SU(2)$ gauge group (it
acts as $-1$ on the elementary hypermultiplets and $1$ on the vector
multiplet).  This is the effect described in \moncharge\ (where the pure
gauge theory was considered, so $(-1)^H$ was equivalent to 1). If the
electric charge operator, which we will temporarily call $Q_0$, is
normalized so that a $W$ boson has unit charge, the operator statement
is
\eqn\rino{e^{2\pi i Q_0}=e^{in_m\theta}(-1)^H.}
In the present context, we want to normalize
the charge operator so that the eigenvalues for the hypermultiplets
are $\pm 1$ (so $W$ bosons have charge $\pm 2$).  The normalized
charge operator is thus $Q=2Q_0$, and formula \rino\ becomes
\eqn\ino{e^{i\pi Q}=e^{in_m\theta}(-1)^H.}
It has the following significance: if we write the charge as
$Q=n_e+n_m\theta/\pi$ with $n_e\in {\bf Z}$, then the states
of even $n_e$ have $(-1)^H$ even, and the states of odd $n_e$ have
$(-1)^H$ odd.

$(-1)^H$ is the ``chirality'' operator in the spinor representation of
$SO(2N_f)$, so the above statement means that the monopoles of even
$n_e$ are in one spinor representation, and the monopoles of odd $n_e$
are in the other spinor representation.  This result ensures the
following: if $M$ is a monopole with $n_e=q$ and $M'$ is a monopole with
$n_e=q+1$, then the state of $(n_m=0,n_e=1)$ produced in $M'\bar M$
annihilation has $(-1)^H=-1$ and can be formed {}from the elementary
fields.  If there were no correlation between electric charge and
$SO(2N_f)$ chirality, then monopole-antimonopole annihilation would
produce states that do not in fact exist.

For $N_f=1,3$, the internal ``parity'' ensures that a dyon transforming
as a positive chirality spinor representation of $SO(2N_f)$ is
degenerate with a particle with opposite electric and magnetic charge
and opposite $SO(2N_f)$ chirality.  There is no such relation for
$N_f=2,4$ where the internal parity is simply spontaneously broken.

As in \sen, the spectrum may also include states with magnetic charge
$n_m\geq 2$.  Certain general restrictions on the quantum numbers of
these states (some of which can be deduced {}from \ino) ensure that they
can be interpreted as bound states of already known particles and
that particle-antiparticle annihilation gives consistent results.  For
$N_f\geq 2$, these restrictions are conveniently stated in terms of the
quantum numbers of the states under the center of the universal cover of
$SO(2N_f)$.  For $N_f=2$, the universal cover is $SU(2)\times SU(2)$ and
the center is ${\bf Z}_2\times {\bf Z}_2$.  We write a representation of
${\bf Z}_2\times {\bf Z}_2$ as $(\epsilon,\epsilon')$, where
$\epsilon=0$ for the trivial representation of ${\bf Z}_2$ and
$\epsilon=1$ for the non-trivial representation.  Then for $N_f=2$,
states of arbitrary $(n_m,n_e)$ transform under the center as
$((n_e+n_m)\,\mod 2, n_e\, \mod 2)$.  For $N_f=3$, the universal cover of
$SO(6)$ is $SU(4)$, and its center is ${\bf Z}_4$.  ${\bf Z}_4$ acts by
$\exp ( {i \pi \over 2} (n_m+ 2n_e) )$.  For $N_f=4$, the universal
cover is ${\rm Spin}(8)$, with center ${\bf Z}_2\times {\bf Z}_2$.  The
four representations of ${\bf Z}_2\times {\bf Z}_2$ are conveniently
labeled by representations of ${\rm Spin}(8)$ that realize them; we will
call them $o$ (associated with the trivial representation of ${\rm
Spin}(8)$), $v$ (associated with the vector), and $s$ and $c$ (the two
spinors).  ${\rm Spin}(8)$ has a ``triality'' group of outer
automorphisms which is isomorphic to the permutation group ${\bf S}_3$
of the three objects $v,s$, and $c$.  The quantum numbers of all
particles under the center of ${\rm Spin}(8)$ are determined by $(n_m\,
\mod 2, n_e \,\mod 2)$: $(0,1)$ corresponds to $v$, e.g. the elementary
quark; $(1,0)$ corresponds to $s$, e.g. the fundamental neutral
monopole; $(1,1)$ corresponds to $c$, e.g. the first excited dyon with
magnetic charge 1; $(0,0)$ corresponds to $o$, e.g. the elementary gauge
fields.

We now turn on an $N=2$ invariant mass term, $m_{N_f}$, for one of the
quarks.  The global $SO(2N_f)$ symmetry is explicitly broken to
$SO(2N_f-2)\times SO(2)$.  Since the global symmetry includes an abelian
continuous symmetry, it can contribute to the central extension in the
algebra.  As in QED (see section 2), the mass of BPS-saturated
states is given by $M=\sqrt{2} |Z|$ with
\eqn\massform{Z=n_e a + n_m a_D + S{ m_{N_f}\over \sqrt{2}}}
where $S$ is the $SO(2)$ charge. Just as in QED, the appearance of the
extra term can be deduced {}from the fact that the hypermultiplets are in
``small'' representations.  It follows {}from \massform\ that for $a=\pm
m_{N_f}/\sqrt{2}$ one of the elementary quarks is massless.  This fact
can be easily verified using the classical Lagrangian.

\newsec{Duality}

As in the pure gauge theory \paperi, we can perform $SL(2,{\bf Z})$
duality transformation on the low energy fields.  Although they are
non-local on the photon field $A_\mu$, they act simply on $(a_D,a)$.
Several new issues appear when matter fields are present.

First, consider the situation of one massive quark with mass $m_{N_f}$
and examine what happens when $a$ approaches $m_{N_f}/\sqrt 2$ where
one of the elementary quarks becomes massless.  As in the discussion in
section 2, loop diagrams in which this quark propagates make a
logarithmic contribution to $a_D$.  The behavior near $a=m_{N_f}/\sqrt
2$ is thus
\eqn\hidoc{\eqalign{
a & \approx a_0 \cr
a_D& \approx c -{i\over 2\pi}(a-a_0)\ln (a-a_0) \cr}}
with $a_0=m_{N_f}/\sqrt 2$ and $c$ a constant.
The monodromy around $a=a_0$ is thus
\eqn\nidoc{\eqalign{
a & \to a\cr
a_D & \to a_D+a-a_0=a_D+a-{m_{N_f}\over \sqrt 2}\cr}}
Thus, under monodromy, the pair $(a_D,a)$ are not simply transformed by
$SL(2,{\bf Z})$; they also pick up additive constants.  It was explained
in section 3.1 of \paperi\ that the duality symmetry of the low energy
theory permits such constants to appear; but it was also shown in
\paperi, section 4, that this possibility is not realized for the pure
$N=2$ gauge theory.  The above simple consideration of a massless
quark shows that this possibility does enter for $N_f>0$.

If one arranges $a_D, $ $a$, and the bare mass $m$ as a three
dimensional column vector $(m/\sqrt 2 , a_D, a)$, then the monodromy in
\nidoc\ can be written in the general form
\eqn\mondfo{\CM=\pmatrix{
1 & 0 & 0 \cr
r& k& l \cr
q &n & p \cr}}
with $\det\CM=kp-nl=1$.  This is the most general form permitted by the
low energy analysis of \paperi.  The specific form of the first row in
\mondfo\ means that $m$ is monodromy-invariant; intuitively this
reflects the fact that $m$ is a ``constant,'' not a ``field.''

Since the central charge in \massform\ must be monodromy-invariant, one
can deduce at once how the charges transform.  If one arranges the
charges as a row vector $W= (S,n_m, n_e)$, then $W$ transforms by $W\to
W\CM^{-1}$.  Explicitly,
\eqn\expli{\CM^{-1}=\pmatrix{
1 & 0  &  0 \cr
lq-pr & p & -l \cr
nr-kq &-n & k  \cr}.}
Thus, the electric and magnetic charges $n_e$ and $n_m$ mix among
themselves but do not get contributions proportional to the global
symmetry charge $S$.  On the other hand, the $S$ charge can get
contributions proportional to gauge charges $n_e$ or $n_m$.
Equivalently, the global symmetry can be transformed to a linear
combination of itself and a gauge symmetry but not the other way around.
Notice that the monodromy matrix mixing the charges in this way survives
even if the bare mass $m$ vanishes.

Here we get an elementary example of the situation suggested in section
4 of \paperi\ -- the spectrum of BPS-saturated one-particle states is
not transformed in the expected way under monodromy.  As is clear {}from
\massform, one of the elementary quarks is massless at
$a=m_{N_f}/\sqrt{2}$.  For large mass this happens at large $|u|$ where
semiclassical techniques are reliable.  The monodromy around that point
shifts the $S$ value of a magnetic monopole by an amount proportional to
the magnetic charge; this follows upon using the monodromies in \nidoc\
to determine the matrices $\CM$ and hence $\CM^{-1}$.  However, in this
regime the spectrum of magnetic monopoles can be worked out explicitly
using semiclassical methods and in particular the values of $S$ are
bounded.  (In semiclassical quantization of the monopole, the only zero
modes carrying $S$ are the fermion zero modes, and there are only
finitely many of them.  The boson zero modes, which could carry an
arbitrary charge, are $S$-invariant.)  This means that a phenomenon
first considered in two dimensions by Cecotti et. al.
\ref\cecotti{S. Cecotti, P. Fendley, K. Intriligator, and C. Vafa,
\np{386}{1992}{405}; S. Cecotti and C. Vafa, \cmp{158}{1993}{569}.}
must be operative: as we circle around the singularity, the monopole
crosses a point of neutral stability where it can decay to two other
states.  Then monodromy indeed changes the quantum numbers in the
expected fashion but what starts as a one particle state comes back as a
multiparticle state; meanwhile, the spectrum of BPS-saturated
one-particle states jumps.  In this example, since everything is
happening in the semiclassical regime, it must be possible to exhibit
the jumping very explicitly.

\bigskip
\noindent{\it Self-Duality For $N_f=4$?}

$N=4$ super Yang-Mills theory has a spectrum of BPS-saturated states
that seems to be invariant under $SL(2,{\bf Z})$ acting on $(n_m,n_e)$.
This is one of the main pieces of evidence for Olive-Montonen duality in
that theory.

For the pure $N=2$ theory, the spectrum is not $SL(2,{\bf
Z})$-invariant, though it was seen in \paperi\ that an $SL(2,{\bf
Z})$-invariant formalism arises naturally in determining many of its
properties.  The same situation will prevail, as we will see, for the
$N_f=1,2,3$ theories.

But as we will now explain, the $N_f=4$ theory is a candidate as another
theory that may possess manifest $SL(2,{\bf Z})$ symmetry in the
spectrum, though with the details rather different {}from the case of
$N=4$.

The states with $(n_m,n_e)=(0,1)$ are the elementary hypermultiplets,
which transform in the vector representation $v$ of ${\rm Spin}(8)$.
The states with $(n_m,n_e)=(1,0)$ transform as one spinor representation
$s$, and the states with $(n_m,n_e)=(1,1)$ transform as the other spinor
representation $c$.  All of these representations are eight dimensional,
and they are permuted by the ${\bf S}_3$ group of outer automorphisms of
${\rm Spin}(8)$. One could think of these states as being permuted by
certain $SL(2,{\bf Z})$ transformations together with triality.  If one
is willing to optimistically assume that suitable multi-monopole bound
states exist, generalizing the one found in \sen, for every relatively
prime pair of integers $(p,q)$, then an $SL(2,{\bf Z})$-invariant
spectrum is possible.  One wants for each such $(p,q)$ to have eight
states of $(n_m,n_e)=(p,q)$, transforming according to an eight
dimensional representation of ${\rm Spin}(8)$ that depends on the
reduction of $(n_m,n_e)$ modulo 2.  In this way, the theory could have
an $SL(2,{\bf Z})$ symmetry, mixed with ${\rm Spin}(8)$ triality.

In fact, in the latter part of this paper, when we solve quantitatively
for the low energy structure of the $N_f=4$ theory with arbitrary bare
masses, we will get a triality and $SL(2,{\bf Z})$-invariant answer,
strongly indicating that this possibility is realized.

Clearly, since $SL(2,{\bf Z})$ permutes the various ${\rm Spin}(8)$
representations, it does not commute with ${\rm Spin}(8)$.  In fact, the
four classes of ${\rm Spin}(8)$ representations are permuted under
$SL(2,{\bf Z}) $ like the four spin structures on the torus -- $o$ is
like the odd spin structure while the other three are like the even
ones.  The full group is a semidirect product ${\rm Spin}(8) \semidirect
SL(2,{\bf Z})$.

An explicit description of this semidirect product is as follows.  The
outer automorphism group (triality) of ${\rm Spin}(8)$ is the group
${\bf S}_3$ of permutations of three objects.  It permutes the three
eight dimensional representations.  This group can be regarded as the
group of $2\times 2$ matrices of determinant one with entries that are
integers mod 2.  Therefore (by reduction mod 2) there is a homomorphism
{}from $SL(2,{\bf Z}) \rightarrow {\bf S}_3$.  The kernel consists of
matrices congruent to 1 mod 2.  $SL(2,{\bf Z})$ acts on ${\rm Spin}(8)$
by mapping to ${\bf S}_3$ which then acts on ${\rm Spin}(8)$; using this
action of $SL(2,{\bf Z})$ on ${\rm Spin}(8)$, one constructs the
semidirect product ${\rm Spin}(8)\semidirect SL(2,{\bf Z})$.

It should be noted that for $N_f\not=0$ the elementary massive gluons
are only neutrally stable against decay to the massive quarks.
Apparently, they are analogous to bound states at threshold in
nonrelativistic quantum mechanics (which can exist as discrete states).
If this is the right interpretation and the theory is indeed dual, there
should also be massive ${\rm Spin}(8)$-invariant BPS-saturated states of
spins $\leq 1$ and charges $(n_m,n_e)= (2k,2l)$ with arbitrary
relatively prime $k$ and $l$.

\newsec{The singularities for $N_f=1,2,3$}

In this section, we will begin our study of the singularities of the
quantum moduli spaces, using a method that is natural for the
asymptotically free theories of $N_f\leq 3$.  We first consider the case
of very large bare masses compared to the dynamical mass scale
$\Lambda$, where the theory reduces to the $N_f=0$ theory and the vacuum
structure is known {}from \paperi.  Then we extrapolate to small masses.
The conformally invariant $N_f=4$ theory involves somewhat different
issues and its structure will be determined later.

We start with the $N_f=3$ theory with three equal mass quarks $m_r=m \gg
\Lambda$.  The mass terms break the global ${\rm Spin}(6)=SU(4)$ flavor
symmetry to $SU(3) \times U(1)$.  Classically there is a singularity at
$a= m/\sqrt{2}$ where some of the elementary quarks are massless.  The
massless fields there are electrically charged and they form a triplet
of $SU(3)$.  Since we consider the case $m \gg \Lambda$, this
singularity is in the semiclassical region $u \approx 2a^2 =m^2 \gg
\Lambda^2$ and it persists quantum mechanically.  For $u \ll m^2 $ the
three quarks are massive and can be integrated out semiclassically.  The
low energy theory is the pure gauge ($N_f=0$) theory.  The scale
$\Lambda_0$ of the low energy theory can be determined at the one loop
approximation in terms of the masses and the scale of the high energy
theory $\Lambda_3$ to be $\Lambda_0^4=m^3\Lambda_3$.  Therefore, the
moduli space at small $u$ is given approximately by that of the pure
gauge theory with scale $\Lambda_0$.  It has two singular points where,
respectively, monopoles of $(n_m,n_e)= (1,0)$ and $(n_m,n_e)= (1,1)$ are
massless.  These two monopoles are $SU(3)$ invariant.

As the mass $m$ of the quarks is reduced, the singular point at large
$u$ moves toward the origin and the location of the two other singular
points can change.  As we discussed in the previous section, the values
of $(n_m,n_e)$ and the charges under the abelian symmetries of the
massless particles at the singularities can change.  However, their
non-abelian global charges cannot change.  The states massless at the
various singularities transform, respectively, as ${\bf 3}$, ${\bf 1}$,
and ${\bf 1}$ of the global $SU(3)$ symmetry.  For $m=0$ the global
symmetry is enlarged {}from $SU(3) \times U(1)$ to $SU(4)$, Therefore,
the massless particles at the different singularities must be in
representations of $SU(4)$.  The only way for this to happen is that two
of the singularities with a massless ${\bf 3}$ and ${\bf 1}$ of $SU(3)$
must combine into a singularity with a massless ${\bf 4}$ of $SU(4)$
while the third singularity goes elsewhere\foot{The three singularities
cannot all combine together, since the fundamental group of the
once-punctured $u$ plane is abelian, and an abelian representation of
the fundamental group, when combined with the known behavior at
infinity, will lead (as we saw in section 5.2 of \paperi) to an
indefinite metric on the quantum moduli space.  Nor can any of the
singularities go to infinity without changing the coefficient of the
logarithm $a_D\sim a\ln a$; this coefficient cannot be changed as it is
determined by the one loop beta function.}.  Therefore, the $N_f=3$,
$m=0$ theory has precisely two singularities in the $u$ plane, with
massless particles that are respectively a ${\bf 4}$ and ${\bf 1}$ of
$SU(4)$.

The $SU(4)$ quantum numbers of the particles at the singularities can be
used to constrain their electric and magnetic charges. As explained in
section 5, the smallest choice of $(n_m,n_e)$ for a state in the ${\bf
4}$ of $SU(4)$ is $(1,0)$ and for an $SU(4)$ singlet it is $(2,1)$.
As we will discuss at the end of section 14, it is possible to show
that if our picture is correct, then
 the states that become massless at the singularities
are continuously connected to BPS-saturated states with the same global
quantum numbers that exist in the semiclassical region of large $u$.
The $(1,0)$ in the ${\bf 4}$ of $SU(4)$ certainly exists
semiclassically.  It is not obvious whether an $SU(4)$ singlet bound
state of two monopoles exists semiclassically (a somewhat similar state
was found in \sen\ for $N=4$), but we conjecture that such a state must
exist.

The same procedure can be used to determine the singularities of the
massless $N_f=1,2$ theories.  One starts with large equal masses for the
quarks and analyzes the singularities at large $u$.  As the masses get
smaller the singularities move toward the origin.  Finally, the global
symmetries determine the full structure.

An alternate procedure, which we will use here, is to follow the
singularities of the massless $N_f=3$ theory as some quarks become
heavy.  First we give a mass $m_3$ to only one of the hypermultiplets.
It breaks $SO(6) \rightarrow SO(2) \times SO(4)$ (or equivalently
$SU(4)\to SU(2)\times SU(2)\times U(1)$).  Of the six fermion
zero modes, two for each hypermultiplet, two are lifted, say $\eta_1$
and $\eta_2$, giving a perturbation $im_3\eta_1\eta_2/2$ to the monopole
Hamiltonian.  As the $\eta_i$ act as gamma matrices upon quantization,
the perturbation has eigenvalues $\pm m_3/2$, with equal multiplicities;
it can be diagonalized to give
\eqn\massmaf{{1\over 2}\pmatrix{
m_3 & & & \cr
  & m_3 & & \cr
  & &-m_3& \cr
  & & &-m_3. }}
This breaks $SU(4) \rightarrow SU(2) \times SU(2) \times U(1)$ which is
of course the right pattern.  The four monopoles are split to two pairs.
One pair transforms as $({\bf 2}, {\bf 1},\half) $ and the other
as $({\bf 1},{\bf 2}, -\half)$ under the unbroken $SU(2)\times
SU(2)\times U(1)$.

Given the $U(1)$ charges, the mass formula \massform\ for BPS-saturated
states shows that one pair becomes massless at $a_D+m_3/2\sqrt 2=0$, and
one at $a_D-m_3/2\sqrt 2=0$.  Since, in the $m_3=0$ limit, $a_D$ is a
good local coordinate near its zero, there is one nearby point obeying
the first of these equations and one nearby point obeying the second.

Therefore, for small non-zero $m_3$ there are three singularities.  At
two of them there are two massless particles transforming as one or the
other spinor of $SO(4)$, while at the third singularity there is a
massless $SO(4)$ singlet state.

Now we increase $m_3$. For large $m_3$, we expect one singularity at
large $u$ where the elementary quark becomes massless.  This state is an
$SO(4)$ singlet, so we identify the  singularity it generates
with the continuation
to large $m_3$ of the singularity of the small $m_3$ theory that
is generated by a massless singlet.

We can now integrate out the heavy quark, eliminating the singularity
just described (since it goes to infinity for $m_3\to \infty$) and
leaving the other singularities.  The low energy theory is the massless
$N_f=2$ theory.  Its scale $\Lambda_2$ is determined by a one loop
calculation to be $\Lambda_2^2=m_3 \Lambda_3$.  As we take $m_3$ to
infinity holding $\Lambda_2$ fixed we are left with the two other
singularities of the ($m_1=m_2=0,m_3\not= 0$) $N_f=3$ theory.  Each of
those singularities has two massless states in one or the other spinor
representation of $SO(4)$.  Given these $SO(4)$ quantum numbers and
assuming that these states are continuously connected to states that
exist semiclassically for large $u$, the minimal choices for their
electric and magnetic charges are $(n_m,n_e)=(1,0)$ and
$(n_m,n_e)=(1,1)$.

As we explained above, the ${\bf Z}_2$ symmetry which changes the sign
of $u$ does not commute with $SO(4)$.  It exchanges the two spinors.
Therefore, the two singularities at finite $u$ are related by the ${\bf
Z}_2$ symmetry.  Their different electric charges can also be determined
by rotating $u$ continuously by $\pi$.  This amounts to changing
$\theta$ by $\pi$, an operation which adds one unit of electric charge
to the monopole; that is why they have equal $n_m$ and have $n_e$
differing by 1.

We now repeat this analysis for the flow {}from $N_f=2$ to $N_f=1$.
Consider one of the pairs of massless monopoles (in $({\bf 2},{\bf 1})$
of $SO(4) \cong SU(2) \times SU(2)$).  A mass term for the second quark
$m_2$ breaks $SO(4) \rightarrow SO(2) \times SO(2)$.  For small $m_2$
the mass term in the monopole theory is
\eqn\massmatt{{1\over 2}\pmatrix{
m_2 & \cr & -m_2}} and it splits the pair.  Similarly, the other pair in
$({\bf 1},{\bf 2})$ of $SO(4)$ is also split and the massive theory has
four singularities.  As $m_2$ becomes large, precisely one of these
singularities moves to large $u$ (since the quark of large $m_2$ has a
component that is massless at one value $u\approx m_2{}^2$) where a
semiclassical analysis is reliable.  Three singularities are left
behind.  Now we integrate out the massive quark.  The low energy theory
is the massless $N_f=1$ theory with scale $\Lambda_1^3= m_2
\Lambda_2^2$. This theory has the three singularities that do not go to
infinity as $m_2\to \infty$.  They are related by the discrete ${\bf Z}
_3$ global symmetry of the $N_f=1$ theory.  As we move {}from one
singularity to the other, the monopole acquires one unit of electric
charge.  Therefore, the values of $(n_m,n_e)$ for the massless states at
the singularities are $(1,0)$, $(1,1)$ and $(1,2)$.

Similarly, we can follow the monopoles {}from $N_f=1$ to $N_f=0$.  One
of the monopole points moves to infinity and the other two remain,
giving the structure proposed in \paperi.

It is curious to note that for small quark mass the massless fields at
the singularity are magnetic monopoles whereas for large mass some of
them are electric charges -- the elementary quarks.  This continuous
transformation {}from an elementary particle to a magnetic monopole is
possible because of the nonabelian monodromies.  The values of
$(n_m,n_e)$ of a massless particle at a singularity are determined by
the monodromy around the singularity.  This monodromy depends on a
choice of base point and a path around the singularity.  When these
choices are changed, the monodromy is conjugated.  As the mass changes
the singularities move on the moduli space and correspondingly, the
natural choice for the path along which the monodromy is computed is
changed.  Therefore, although the conjugacy class of the monodromy
cannot change, the natural labeling by quantum numbers $(n_m,n_e)$ can
change.  We are used to this phenomenon in the case of a magnetic
monopole acquiring electric charge \moncharge\ by changing $\theta
\rightarrow \theta + 2 \pi$.  Here, because of the other singularities,
also the magnetic charge can change.

To summarize this discussion, for the theories with zero bare masses one
has the following:

\smallskip
\noindent
For $N_f=0$ the global symmetry acting on the $u$ plane is ${\bf Z}_2$.
There are two singularities related by this symmetry with massless
states $(n_m,n_e)=(1,0)$ and $(n_m,n_e)=(1,1)$.

\smallskip
\noindent
For $N_f=1$ the global symmetry of the $u$ plane is ${\bf Z}_3$.  There
are three singularities related by this symmetry with massless states
$(n_m,n_e)=(1,0)$, $(n_m,n_e)=(1,1)$ and $(n_m,n_e)=(1,2)$.

\smallskip
\noindent
For $N_f=2$ the global symmetry of the $u$ plane is  ${\bf Z}_2$.
There are two singularities related by this symmetry.  The massless
states at one singularity have $(n_m,n_e)=(1,0)$ and are in one spinor
of $SO(4)$ while the massless states at the other singularity have
$(n_m,n_e)=(1,1)$ and are in the other spinor of $SO(4)$.

\smallskip
\noindent
For $N_f=3$ the $u$ plane has no global symmetry.  There are two
singularities.  In one of them there are four massless states with
$(n_m,n_e)=(1,0)$ in a spinor representation of $SO(6)$ with definite
chirality; in the other there is a single state with $(n_m,n_e)=(2,1)$.

Although the arguments we suggested for the singularity structure are
quite plausible, they are certainly not a rigorous proof.  In the next
sections we will study the consequences of this picture and supply
what we regard as convincing evidence.

\newsec{Low energy theory near the singularities}

In this section we study the low energy effective field theory near the
singularities.  For simplicity, we will focus on the massless theories.
It is straightforward to extend our considerations to the massive ones.

The light states near all the singularities are a photon multiplet and
some charged fields.  Using the duality transformation, the low energy
theory near any of the singularities is an abelian gauge theory with
some light hypermultiplets. This theory was studied in section 2, but
two important points should be noted.  First, the $N=2$ photon multiplet
we encounter in our low energy effective theories is not the
semiclassical photon.  It is related to the semiclassical photon by an
appropriate duality transformation.  Similarly, the scalar $a$ in
section 2 should be identified with a linear combination of $a$ and
$a_D$ of the non-abelian theory.  The second important difference is
that the low energy theory also contains higher dimension operators.
These include terms of the form $\left( {A/\Lambda} \right)^n
W_\alpha^2$ in the gauge coupling.  These terms break the $U(1)_\CR$
symmetry of the abelian theory.

The flavor symmetry of our original massless theory is $SO(2N_f)$ and
the light states at the singularities are in a spinor of $SO(2N_f)$
(except for one of the singularities of the $N_f=3$ theory at which the
massless multiplet is a flavor singlet).  At first look, this might seem
in contradiction with the $SU(k)$ symmetry of the effective low energy
theory under which the $k$ light fields transform as the fundamental
representation of $SU(k)$.  Fortunately, special properties of the
relevant $SO$ groups make this consistent.  For $N_f=2$ the two light
states transform as $({\bf 2},{\bf 1})$ or $({\bf 1},{\bf 2})$ under
$SO(4) \cong SU(2) \times SU(2)$ and, either way, only a single $SU(2)$
acts on the light fields and the two hypermultiplets transform as a
doublet. Similarly, for $N_f=3$ the four states at the singularity are
in a spinor of $SO(6) \cong SU(4)$ and transform as the ${\bf 4}$ of
$SU(4)$.

Having established that the symmetries of the low energy theory act
correctly, we will now study the flat directions of this theory.  They
are the quantum moduli space of the original theory and should be
connected smoothly to the semiclassical picture.

The low energy theories in all the singularities have flat directions of
the $a$ field along which all the hypermultiplets acquire a mass.  As
such they are smoothly connected with the Coulomb branch of the original
non-abelian theory.

For $N_f=1$ there is a single light hypermultiplet at every singularity
and therefore there are no other flat directions.  This is consistent
with the absence of Higgs branches in the moduli space of the original
theory.

For $N_f=2$ the low energy theory at the singularity is QED with $k=2$
hypermultiplets.  As discussed in section 2, this theory has $M$ flat
directions along which $SU(2) \times SU(2)_R$ is broken to $SU(2)_{R'}$
and the light fields transform as ${\bf 3} \oplus {\bf 1}$ of the
unbroken symmetry (see equations \globsymb\ and \lightuone).  Adding to
this symmetry the other $SU(2)$ global symmetry which does not act on
the light fields, we conclude that along these $M$ flat directions the
symmetry breaking pattern is $SO(4) \times SU(2)_R \rightarrow SU(2)
\times SU(2)_{R'}$ and the light fields transform like $({\bf 1},{\bf
3}) \oplus ({\bf 1},{\bf 1})$ of the unbroken symmetry.  This is exactly
the pattern of symmetry breaking and light spectrum observed
on the Higgs branch of the original theory in \glosymb\ and \masslesq.

Classically, the $N_f=2$ theory had two Higgs branches touching the
Coulomb branch at the origin.  Quantum mechanically, the two branches
touch the Coulomb branch at different points but the metric on them and
the pattern of symmetry breaking are the same as they are classically.

For $N_f=3$ there are two singularities.  In one of them $k=1$; there
are no $M$ flat directions emanating {}from that point.  In the other one
there are $k= 4$ hypermultiplets in a spinor of $SO(6)$.  Along the $M$
flat directions $SU(4) \times SU(2)_R$ is broken to $SU(2) \times U(1)
\times SU(2)_{R'}$ and the light fields transform as $({\bf 2}, 1, {\bf
2})\oplus ({\bf 2},-1,{\bf  2})\oplus ({\bf 1}, 0, {\bf 3})\oplus ({\bf
1}, 0,{\bf 1})$ (see equations \globsymb\ and \lightuone).  Again, this
is precisely the pattern of symmetry breaking and light spectrum
observed in \glosymb\ and \masslesq\ on the Higgs branch of the original
theory.

Note how the Higgs branch of the quantum moduli space has two weakly
coupled limits.  In one of them the weakly coupled particles are
magnetically charged in a spinor of $SO(2N_f)$ and in the other limit
they are quarks which are doublets of the $SU(2)$ gauge group and are
components of a vector of $SO(2N_f)$ (which is broken to a subgroup).
The original gauge symmetry looks like it is confined at one end because
magnetic monopoles condense there.  At the other end it looks like it is
completely broken by the Higgs mechanism.  Since our theory includes
matter fields in the fundamental representation, there is no strict
gauge invariant distinction between confinement and complete gauge
symmetry breaking\higgscon\ and therefore there is no contradiction
here.  The gauge invariant order parameters - $\tilde M M$ at one end
and $V^{rs}=Q^rQ^s$ at the other end -- transform the same way under the
global symmetry ($({\bf 3} , {\bf 1})$ under $SO(4)\cong SU(2) \times
SU(2)$ for $N_f=2$ and ${\bf 15} $ of $SO(6)\cong SU(4)$ for $N_f=3$)
and hence lead to the same pattern of symmetry breaking and to the same
massless spectrum.

\newsec{Breaking $N=2$ to $N=1$}

In this section we break $N=2$ supersymmetry to $N=1$ by adding a mass
term $m\,\Tr \Phi^2$ to the tree level superpotential \nonabtr.  When $m
\gg \Lambda$, the $N=1$ chiral multiplet in the adjoint representation
$\Phi$ is heavy and can be integrated out.  The resulting theory is
$N=1$ SUSY with gauge group $SU(2)$ and $2N_f$ chiral doublets $Q^r$
with $r=1,...,2N_f$.  An interesting term at tree level is a quartic
term ${1 \over m} (Q^rQ^s)^2$ in the superpotential which breaks the
global $SU(2N_f)$ symmetry of the $N=1$ theory to $SO(2N_f)$. At one
loop, the scale $\tilde \Lambda_{N_f}$ of the $N=1$ theory is given by
$\tilde \Lambda_{N_f}^{6-N_f}= m^2 \Lambda_{N_f}^{4-N_f}$.  As
$m\rightarrow \infty$ with $\tilde \Lambda_{N_f}$ held fixed the quartic
term in the superpotential is negligible and we should recover the known
results of the $N=1$ theory \moduli.

For small $m$ we can use the low energy effective theory.  The mass term
is represented as a term $m U $ in the superpotential.  Since it has no
critical points as a function of $U$, the only reason that there are any
supersymmetric ground states at all is that new degrees of freedom
become light and have to be included near the singularities.  Near the
singularities, one can use an effective Lagrangian like that of
subsection 2.5 and approximate $U\approx u_0\Lambda_{N_f}^2 + u_1
\Lambda_{N_f} A + \CO(A^2) $ where $u_0$ and $u_1$ are dimensionless
constants and $A$ is the chiral superpartner of the light photon.  This
is exactly the Lagrangian we studied in subsection 2.5 with $\mu= m u_1
\Lambda_{N_f}$.  As we saw there, the value of $a$ is fixed at zero and
therefore $u=u_0\Lambda_{N_f}^2$.  The matter fields $M$ and $\tilde M$
acquire expectation values breaking the $U(1)$ gauge symmetry.  Since
these are magnetic monopoles, this means confinement of the original
charges.

We see that the continuum of vacua on the Coulomb branch has disappeared
and the surviving ground states are at the singularities.  Every
singularity leads to a vacuum.

Next we should identify what happens to the Higgs branches.  We continue
to use the effective Lagrangian of subsection 2.5. For $N_f=2$ we have
two regions described by QED with $k=2$ where generically $SO(4)$ is
broken to $SU(2)$ (see equation \globsymbb) and at a special point it is
broken to $SU(2) \times U(1)$ (see equation \globsymbbs).  For $N_f=3$
there is an isolated ground state (related to the condensation of the
$(n_m,n_e)=(2,1)$ monopoles) as well as a continuum.  At the generic
point in the continuum $SO(6)$ is broken to $SU(2) \times U(1)$ (see
equation \globsymbb) and at a special point it is broken to $SU(3)
\times U(1)$ (see equation \globsymbbs).

This low energy effective Lagrangian is a good description of the
physics for small $m$.  However, it might not be appropriate in the
limit $m \rightarrow \infty$.  There are two reasons for that.  First,
in that limit new states which are massive for any finite $m$ can become
massless and should be included in the Lagrangian.  Second, as $m
\rightarrow \infty$ we have to take $\Lambda_{N_f}$ to zero in order to
keep the low energy scale $\tilde \Lambda_{N_f}$ fixed.  This means that
the different ground states on the Coulomb branch approach each other.
The appropriate effective Lagrangian should describe all of them.

The degrees of freedom that we expect are those of the $N=1$ theory that
is obtained by integrating out $\phi$.  This theory can be usefully
described \moduli\ by an effective theory for the gauge invariant
composite field $V^{[rs]}=Q^rQ^s$.  It would be equivalent to use the
composite  monopole fields  $Y_a^b=\tilde M_a M^b$ with
$a,b=1,...,k$.  For $N_f=3$ the fields $V$ are in the {\bf 15} of
$SO(6)$ and so are the fields $Y$ if $\Tr \,Y$ is removed.  For $N_f=2$
the fields $V$ are in $({\bf 3},{\bf 1}) \oplus ({\bf 1},{\bf 3})$ of
$SO(4)$.  In terms of the monopole fields, these representations are
obtained by considering the monopole bilinears $Y$ in the two branches
and removing their traces.

The effective superpotential for $V$ can be constrained along the lines
of \nonren.  We can require it to respect all the symmetries of the
theory - including those explicitly broken by $m$ or the anomaly - if we
assign appropriate transformation laws under such symmetries to $m$ and
$\Lambda$.  We also demand that it be locally holomorphic in $V$, $m$
and $\Lambda_{N_f}$, and that it has a finite limit as $m
\rightarrow \infty$ with $\tilde \Lambda_{N_f}$ held fixed which
coincides with that of the $N=1$ theory \moduli.

For $N_f=2$ these considerations determine the superpotential
\eqn\nftvl{W=X(\Pf V - m^2\Lambda_2^2) + {1 \over m} V^2}
where $X$ is a Lagrange multiplier.  For finite $m$ it leads to two
branches: $X=\pm {1 \over m} $ with $V^{rs}= \pm \half \epsilon^{rstu}
V^{tu}$ and $\Pf V = m^2\Lambda^2$.  These are the two Higgs branches we
found before.  As $m \rightarrow \infty$ more fields become massless, and
we recover the full moduli space of the $N=1$ theory which is constrained
by $\Pf V = m^2\Lambda_2^2= \tilde \Lambda_2^4$ \moduli.

For $N_f=3$ the superpotential is
\eqn\supervo{-{1 \over m^2 \Lambda_3} \Pf V  + {1 \over m} V^2 .}
For finite $m$ the equation of motion of $V$ has two types of solution.
There is a continuum of states which we associate with the Higgs branch.
There is also an isolated state with unbroken $SU(4)$ at $V=0$; we
interpret this as the vacuum with the condensation of the
$(n_m,n_e)=(2,1)$ monopole.  As $m \rightarrow \infty$, the isolated
state merges into the continuum to form the moduli space of the $N=1$
theory.

Note the following crucial point.  For $N_f=2$ the ground states of the
massive theory are on the quantum moduli space of the corresponding
$N=1$ theory $\Pf V= m^2 \Lambda_2^2$.  This is not the case for
$N_f=3$.  The states can be described by the order parameter $V$ of the
$N=1$ theory but they occur for values of $V$ that do not obey the
equations of motion of the $N=1$ theory (namely $\epsilon_{r_1...r_{6}}
V^{r_1r_2} V^{r_3r_4}=0$).  The same phenomenon happens when other
perturbations of the massless $N=1$, $N_f=3$ theory (like adding mass
terms for the quarks \moduli\ or gauging a subgroup of the global
symmetry \exact) are considered.  It arises because all the components
of $V$ are massless in the $N=1$ theory at $V=0$.  Therefore, all of
them should be kept in the effective Lagrangian.

In sum, we have found two different low energy effective Lagrangians for
the theory broken to $N=1$.  One of them includes a photon and some
monopole fields.  The other includes only the fields $V$.  For finite
non-zero $m$ they lead to the same physics for the massless modes and
differ in the way they describe the massive fields. A low energy
effective Lagrangian with a finite number of terms cannot be expected to
describe massive fields correctly.  At best it can give an approximate
description of the light fields.  The monopole Lagrangian has a smooth
$m \rightarrow 0$ limit because it includes the fields which become
massless in this limit.  On the other hand, the Lagrangian with $V$ has
a smooth $m \rightarrow \infty$ limit because it includes the fields
which become massless in that limit.

This picture also explains the phenomenon observed in \moduli\ in the
$N=1$, $N_f=3$ theory where at the origin of field space confinement (to
the extent that it is well defined in a theory with matter fields in the
fundamental representation) occurred without chiral symmetry breaking.
This is due to the condensation of monopoles which do not carry global
quantum numbers, notably the $(n_m,n_e)=(2,1)$ monopole.

We see here a new phenomenon in quantum field theory.  Magnetic
monopoles acquire global charges because of the existence of fermion
zero modes.  When these monopoles condense, they lead to chiral symmetry
breaking.  This leads us to suggest that to the extent that condensation
of monopoles can be used to describe confinement in QCD, it can also be
used to describe chiral symmetry breaking.

\newsec{The singularities for $N_f=4$}

When the number of flavors is four the one loop beta function vanishes.
Because of the properties of $N=2$, the beta function is also zero to
all orders in perturbation theory.  Does the exact beta function
vanish?  The non-perturbative contributions
to the beta function can be studied by examining the low energy
effective coupling $\tau$.  As in \natint\ we can examine ${\partial^2
\tau(a) \over \partial a^2}$ by computing a matrix element of four
fermions.  However, as we explained above, because of the parity
symmetry \paritty\ the one instanton contribution to this matrix element
vanishes when the number of flavors is non-zero.  For $N_f=0,1,2,3$ the
two instanton contribution to $\tau$ is non-zero.  If it is also non-zero
for $N_f=4$, it is logarithmic in $a$, so that $ {\partial^2 \tau(a)
\over \partial a^2} \sim e^{-{16 \pi^2 \over g^2} + 2i\theta}/a^2=
e^{2\pi i \tau_{cl}} /a^2$.  Including also multiple instanton
contributions, $\tau(a) $ can have the form $c(e^{2\pi i \tau_{cl}})
\ln (a /\Lambda)$ leading to a term $\int d^2 \theta \, c(e^{2\pi i
\tau_{cl}}) \ln (A /\Lambda) W_\alpha{}^2$ in the low energy effective
Lagrangian.  The appearance of the scale $\Lambda$ in this theory
signals a non-perturbative anomaly both in conformal invariance and in
$U(1)_\CR$ (the anomaly in $U(1)_\CR$ can be seen by performing a
$U(1)_\CR$ transformation on the low energy effective action).  When
the theory is put on manifolds with non-zero second Betti number,
instantons in the low energy abelian theory lead to explicit
exponentially small breaking of $U(1)_\CR$ as a result of which all
amplitudes in certain topological sectors would vanish.  This seems
bizarre.  Another consequence of the logarithm in $\tau$ is that the
metric on the moduli space, namely $\Im \tau \, da \, d \bar a$, is not
positive definite.  Although it could perhaps be modified by other
non-perturbative effects, we find this unlikely.  So we will assume
that $c=0$ and the exact quantum theory is scale invariant.  In any
event, the results that we will obtain add considerably to the
plausibility of our assumption of exact scale invariance.

Such a scale invariant theory is characterized by the classical
dimensionless coupling constant $\tau= {\theta \over \pi} + {8\pi i
\over g^2}$ because there is no dimensional transmutation.  Its quantum
moduli space is by scale invariance the same as the classical space, and
the absence of corrections to $\tau$ implies that there are no
corrections to
\eqn\clascuf{\eqalign{
&a= \half \sqrt{2u}\cr
&a_D=\tau a.\cr}}
Clearly, the only monodromy is the one around the origin, which is
$P=-1$.

The situation is more interesting when some masses $m_i$ are not zero.
Then, both scale invariance and $U(1)_\CR$ are explicitly broken by the
masses.  As some of the masses go to infinity (with a suitable limit of
$\tau$) we should recover the quantum moduli space of the asymptotically
free theories with $N_f=0,1,2$ or 3.

More explicitly, if some masses $m_i$ for $i=n+1,...,4$ are taken to
infinity, the right scaling limit is obtained by taking $\tau
\rightarrow i \infty$ holding fixed
\eqn\contlimi{\Lambda_n^{4-n} \sim e^{\pi i \tau}\prod_i m_i= q^\half
\prod_i m_i }
and $u$.  Then, the low energy theory has $N_f=n$ flavors and
scale parameter $\Lambda_n$.  Various definitions of this scale (e.g.\
with other subtraction schemes) differ
by a multiplicative constant of order one.

For example, if only one mass $m_4$ is not zero, the scaling limit
should be that of the $N_f=3$ theory.  For weak coupling, its moduli
space has two singularities at $u \ll m_4$ (where the theory flows to
strong coupling) with four massless particles in one of them and a
single massless particle in the other.  Since for large $u$ we recover
the fourth flavor, there is another singularity where the fourth quark
is massless.

More generally, if we weight each singularity by the number of massless
hypermultiplets at that point, then the total weighted number of
singularities in the complex $u$ plane is always six.\foot{This will
ultimately follow {}from the fact that the low energy physics is
described by a curve $y^2=F(x,u)$ where the discriminant of $F(x,u)$
with respect to $x$ is a sixth order polynomial in $u$; its zeroes are
the singularities.} If the masses are generic, there are six
singularities each of weight one.  If some masses are degenerate, some
of the singularities can be combined to a smaller number of
singularities of higher weight. Denote by $k_l$ the weight of the
$l^{th}$ singularity, so $\sum k_l=6$.  The value of the $k_l$ at the
singularities are constrained by the symmetries of the massive theory.
We mention a few examples:

\noindent
(1) $m_i=(m,0,0,0)$ with global symmetry $SU(4)\times U(1)$.  This is
the case discussed above with three singularities and $k_l=(4,1,1)$.
The four massless particles in the first singularity transform according
to the fundamental representation of $SU(4)$.

\noindent
(2) $m_i=(m,m,m,m)$ with global symmetry $SU(4)\times U(1)$.  There are
three singularities with $k_l=(1,1,4)$ where again the massless
particles in the last singularity transform according to the fundamental
representation of $SU(4)$.  Note the similarity between this example and
the previous one.

\noindent
(3) $m_i=(m,m,0,0)$ with global symmetry $SU(2)\times SU(2) \times
SU(2)\times U(1)$.  There are three singularities with $k_l=(2,2,2)$
where the massless particles in every singularity are in a doublet of
one of the $SU(2)$ factors.

\noindent
(4) $m_i=(m+\mu,m-\mu,0,0)$ with non-zero $\mu \not=m$.  The global
symmetry is $SU(2)\times SU(2)\times U(1)\times U(1)$.  For non-zero
$\mu$ one of the singularities in case (3) splits and there are four
singularities with $k_l=(1,1,2,2)$ where the massless particles in the
singularities with $k=2$ are in a doublet of one of the $SU(2)$ factors.
As $\mu$ varies between $0$ and $m$, this example interpolates between
examples (3) and (1).

\noindent
(5) $m_i=(m,m,\mu,\mu)$ with non-zero $\mu \not=m$.  The global symmetry
is $SU(2)\times SU(2)\times U(1)\times U(1)$.  For non-zero $\mu$ one of
the singularities in case (3) splits and there are four singularities
with $k_l=(1,1,2,2)$ where the massless particles in the singularities
with $k=2$ are in a doublet of one of the $SU(2)$ factors.  As $\mu$
varies between $0$ and $m$, this example interpolates between examples
(3) and (2).  Note the similarity between this example and the previous
one.

\noindent
(6) $m_i=(m,m,m,0)$ with global symmetry $SU(3)\times U(1)\times U(1)$.
There are four singularities with $k_l=(1,1,1,3)$ where the massless
particles in the singularity with $k=3$ transform according to the
fundamental representation of $SU(3)$.

As $m \rightarrow \infty$ with an appropriate shift of $u$ to remove
the last of these singularities, we should recover the quantum moduli
space of theories with fewer flavors.

The similarities between examples (4) and (5) (and between their special
cases (1) and (2)) can be used as evidence for the triality of the
theory.  The ${\bf S}_3$ automorphism of ${\rm Spin}(8)$ is generated by
two transformations.  They act on the masses (which are in the adjoint
representation of ${\rm Spin}(8)$) as
\eqn\triagen{\eqalign{
m_1 &\rightarrow m_1 \cr
m_2 &\rightarrow m_2 \cr
m_3 &\rightarrow m_3 \cr
m_4 &\rightarrow -m_4 \cr}}
which exchanges the two spinors keeping the vector fixed and
\eqn\triagent{\eqalign{
m_1 &\rightarrow \half(m_1+m_2+m_3+m_4) \cr
m_2 &\rightarrow \half(m_1+m_2-m_3-m_4) \cr
m_3 &\rightarrow \half(m_1-m_2+m_3-m_4) \cr
m_4 &\rightarrow \half(m_1-m_2-m_3+m_4) \cr}}
which exchanges the vector with one of the spinors keeping the second
spinor fixed.  The transformation in \triagent\ is particularly
interesting.  Since it exchanges electrons and monopoles, to map the
theory to itself it must also act on $\tau$.  However, the action on
$\tau$ should not affect the qualitative structure of the moduli space:
the number of singularities and the values of the $k_l$ up to
permutation are independent of $\tau$ at least for generic $\tau$.
Returning to our examples, it is easy to check that the transformation
\triagent\ exchanges $m_i=(m,m,\mu,\mu) \leftrightarrow
m_i=(m+\mu,m-\mu,0,0)$ and therefore triality would relate their moduli
spaces and  explain the similarity between them.

\newsec{Preliminaries for determining the metric}

In the remainder of this paper we will find the exact solution for the
low energy effective action, metric on moduli space, and particle masses
for the various theories.  The basic idea, as in \paperi, involves
introducing a suitable family of elliptic curves, and interpreting
$(a_D,a)$ as periods of an appropriate family of meromorphic one-forms.

\subsec{$N_f=0$}

It is appropriate to start first by reconsidering the $N_f=0$ theory.
In section 6 of \paperi, we described this theory by the family of curves
\eqn\famofcurves{y^2=(x-\Lambda^2)(x+\Lambda^2)(x-u).}
(The renormalization scale $\Lambda$ was set to 1 in that reference.)
This is the modular curve for the group $\Gamma(2)$ consisting
of integer-valued matrices
\eqn\mval{M=\pmatrix{ m & n \cr p & q \cr}}
with $\det M=1$ and with $n$ and $p$ even.

However, if we wish to compare the $N_f=0$ theory to theories with
$N_f>0$, it is natural to make a change of conventions that was
mentioned at the end of the introduction.  The magnetic and electric
charges $(n_m,n_e)$ were normalized in \paperi\ so that they were
arbitrary integers in the pure $N=2$ (or $N=4$) theory.  However, for
$N_f>0$, there are particles with half-integral electric charge in those
units.  Therefore, for $N_f>0$, it is convenient to multiply $n_e$ by 2
-- to restore its integrality -- and divide $a$ by 2 -- to preserve the
structure $Z=a_Dn_m+an_e$.  In fact, it is all but necessary to make
this change of conventions if one wishes to exhibit all the $SL(2,{\bf
Z})$ symmetry for $N_f=4$, while the original conventions are natural
for exhibiting the duality symmetry for $N=4$.  The reason is that the
$N=4$ theory has an $SL(2,{\bf Z})$ duality symmetry acting on
$(a_D,a)$, while the $N_f=4$ theory turns out to have an $SL(2,{\bf Z})$
symmetry acting on $(a_D,a')$, where $a'=a/2$.

Dividing $a$ by 2 has the effect of conjugating the monodromy matrix
$M$ of \mval\ by
\eqn\deft{W=\pmatrix{ 1 & 0 \cr 0 & 2},}
replacing $M$ by
\eqn\neft{W^{-1}MW=\pmatrix{m & 2n\cr {p/2} & q\cr}.}
The matrices of this form make up the group $\Gamma_0(4)$ consisting
of unimodular integer-valued matrices with the upper right entry
divisible by four.

With the new conventions, the monodromy at infinity is
\eqn\lefto{ \pmatrix{ -1 & 4 \cr 0 & -1}}
(that is, under monodromy, a magnetic monopole picks up electric
charge 4 instead of 2 since we are using a smaller unit of charge),
and the monodromy due to a massless monopole is
\eqn\tefto{\pmatrix{ 1 & 0 \cr -1  & 1}.}
These matrices generate $\Gamma_0(4)$, and in the new conventions,
the $N_f=0$ theory is described by the modular curve of $\Gamma_0(4)$.
This is the family of curves
\eqn\ramily{y^2=x^3-ux^2+{1 \over 4}\Lambda^4x,}
as we will explain presently.  Like \famofcurves, \ramily\ has a
${\bf Z}_4$ symmetry generated by $y\to iy$, $x\to-x$, $u\to -u$; only a
${\bf Z}_2$ quotient acts on the $u$ plane.  The two families of curves
are related by what is called an ``isogeny''; for fixed $u$ the curve
described by \ramily\ is a double cover of the curve described by
\famofcurves, and vice-versa.

Is it more natural to describe the pure $N=2$ theory using the
conventions that lead to \famofcurves\ or the conventions that lead to
\ramily?  That depends on the context.  If one is viewing this theory as
the low energy limit of the $N=4$ theory perturbed by an $N=2$ invariant
mass term, then \famofcurves\ is more natural; if one is viewing the
same theory as the low energy limit of a theory with $N_f=4$ (and bare
masses for the hypermultiplets) then \ramily\ is more natural.

\subsec{Some convenient facts}

To prove that \ramily\ is the modular curve of $\Gamma_0(4)$ is a fairly
simple exercise using the addition law on a cubic curve.\foot{
$\Gamma_0(4)$ parametrizes elliptic curves with a cyclic subgroup of
order four.  Such a subgroup is generated by the points $P_\pm $ with
coordinates $x=\Lambda^2/2$, $y=\pm {1\over 2}\Lambda^2
\sqrt{\Lambda^2-u}$; this follows {}from the fact that the tangent line
to $P_+$ or $P_-$ passes through the point of order two with coordinates
$x=y=0$.  (The fact that there is no natural way to pick a sign of
$\sqrt{\Lambda^2-u}$ means that only a subgroup of order 4, not a point
of order 4, is naturally determined; this is appropriate for
$\Gamma_0(4)$.)  Conversely, given an elliptic curve with a cyclic
subgroup $T$ of order 4, one can pick coordinates so that the generators
of $T$ have $x=\Lambda^2/2$ and the point of order two has $x=y=0$; then
the curve takes the form of \ramily\ with a uniquely determined $u$.}
For our purposes, it will be more helpful to simply work out the
singularities and the monodromies of the family.

A general cubic curve of the form
\eqn\onn{y^2=F(x)=x^3+\alpha x^2+\beta x +\gamma =(x-e_1)(x-e_2)(x-e_3)}
describes a double cover of the $x$ plane branched over $e_1,e_2,e_3$,
and $\infty$.  This curve becomes singular when two of the branch points
coincide -- thus for $e_i=e_j$, or $e_i\to\infty$.  For instance, for
the family \ramily\ the branch points are $0, $ $\half (u\pm
\sqrt{u^2-\Lambda^4}) $, and $\infty$, and singularities occur precisely
for $u=\pm\Lambda^2$ and $u\to\infty$.

The singularities we will meet on the finite $u$ plane will always be
singularities at which precisely two branch points coincide.  For
$u\to\infty$, we will often have a more complicated configuration.
Mathematically, a singularity at which precisely two branch points
coincide is said to be ``stable''; any singularity can be put in this
form by reparametrizing $x$ and $y$ by $u$-dependent factors, as we will
frequently do to understand the behavior for $u\to\infty$.  A family of
curves in which precisely two branch points are coinciding at, say,
$u=0$, always looks locally near the singularity like the family
\eqn\tung{y^2=(x-1)(x^2-u^n) }
for some integer $n$.  The monodromy is then conjugate to $T^n$
where as usual
\eqn\mung{T=\pmatrix{ 1 & 1 \cr 0 & 1 \cr}. }
This can be proved, for instance, by
computing the periods of the holomorphic differential $\omega=dx/y$,
that is the functions
\eqn\pung{\eqalign{\omega_1&=\int_{-u^{n/2}}^{u^{n/2}}{dx\over y}\cr
                  \omega_2& =\int_{u^{n/2}}^1{dx\over y},\cr}}
near $u=0$.

An important role in our analysis will be played by the discriminant,
$\Delta$, of a polynomial $F$.  It is defined as
\eqn\discr{\Delta=\prod_{i<j}(e_i-e_j)^2}
 with $e_i$ the roots of the polynomial. Since $\Delta$ is symmetric
under permutations of the roots, it can be expressed as a polynomial in
the coefficients of $F$.  For instance, for the cubic polynomial
\eqn\cubpol{F=x^3+Bx^2+Cx+D}
the discriminant is
\eqn\discub{\Delta=-27D^2+18BCD+B^2C^2-4B^3D-4C^3.}

Obviously, the branch points of $y^2=F(x,u)$ coincide precisely when
$\Delta=0$, so singularities (apart {}from $u=\infty$) are at zeroes of
$\Delta$.  We can be somewhat more precise.  Notice that the curve in
\tung\ has branch points at $1,\infty$, and $\pm u^{n/2}$.  In
particular, its discriminant behaves as $\Delta\sim u^n$ for $u$ near 0.
Therefore, in the stable case the exponent of the monodromy is the order
of vanishing of the discriminant.

For example, for the family of curves \ramily, the branch points are at
$0,\infty,$ and $\half(u\pm \sqrt{u^2-\Lambda^4})$, as we noted above.
If $u=\Lambda^2+\epsilon$, the branch points for $\epsilon$ small are
located at approximately $0$, $\half(\Lambda^2 \pm \Lambda \sqrt{2
\epsilon})$ and $\infty$.  Thus, the discriminant is proportional to
$\epsilon$, so $n=1$ and the monodromy near $u=\Lambda^2$ is conjugate to
$T$.  In view of the ${\bf Z}_2$ symmetry of the $u$ plane, the same
is true of the monodromy near $u=-\Lambda^2$.

For $u$ near $\infty$, the structure is slightly more complicated.  The
branch points for large $u$ are at approximately $0, \Lambda^4/4u, u$
and $\infty$.  The singularity at infinity is not stable since two pairs
of branch points ($0$ and $\Lambda^4/4u$, and also $u$ and $\infty$)
coincide for $u\to\infty$.  By a change of variables
\eqn\wungo{\eqalign{x & = x'u\cr y & =y'u^{3/2} \cr}}
we get a stable situation with branch points approximately $0$,
$\Lambda^4/4u^2$, $1$, and $\infty$ and only one pair that coincides for
$u\to\infty$.  This gives $\Delta\sim u^{-4}$, so the monodromy at
infinity for the family of curves in the $x'-y'$ plane is conjugate to
$T^4$ if one circles counter-clockwise around the origin in the $u^{-1}$
variable; a counter-clockwise circuit in $u$ gives $T^{-4}$.  Going back
to the $x-y$ plane, we must note that the differential form $dx/y$ picks
up a factor of $u^{1/2}$ {}from the change of variables in \wungo; hence
its periods $da_D/du$ and $da/du$ pick up a factor of $u^{1/2}$.  Since
$u^{1/2}$ is odd under the monodromy at infinity, this gives an extra
factor of $P$ (the operator that acts as $-1$ on $(a_D,a)$) in the
monodromy at infinity, so that the monodromy for the original family of
curves is $PT^{-4}$.

It can be shown straightforwardly that there is only one representation
of the fundamental group of the $u$ plane punctured at $\pm\Lambda^2$
and $\infty$ with the monodromies conjugate to $T$, $T$, and $PT^{-4}$
and a ${\bf Z}_2$ symmetry exchanging the first two singular points;
this gives another way to prove that \ramily\ is the right family.  In
fact, one can work out the required monodromies for all $N_f\leq 3$
using the general properties unearthed in sections 7, 8.  The
monodromies at infinity are determined by the one loop beta functions in
the microscopic theory to be
\eqn\moninfty{\CM_\infty=PT^{N_f-4},}
just as we discussed for $N_f=0$ in \paperi.  According to our proposal
in the previous sections, with one exception
for $N_f=3$, the singularities at finite points in the
$u$ plane correspond to massless magnetic monopoles with one unit of
magnetic charge.  The monodromy around such a point is determined by the
one loop beta function in QED with $k$ hypermultiplets.  {}From the
infrared behavior of QED as in \paperi, section 5.4, it is conjugate to
$T^k$.  We pick a base point at infinity and then the quantum numbers
$(n_m,n_e)$ of the massless monopoles are well defined.  The monodromy
around a point with $k$ magnetic monopoles with $( n_m=1,n_e)$ is
$(T^{n_e}S)T^k(T^{n_e}S)^{-1} $ as $T^{n_e}S$ conjugates a
hypermultiplet with $(0,1)$ to $(1,n_e)$.  The monodromy
for the $(2,1)$ state in $N_f=3$ can be determined similarly.
Altogether the monodromies are
\eqn\monmon{\matrix{
STS^{-1}, \quad (T^2S)T(T^2S)^{-1} ; &\quad\quad {\rm for}~ N_f=0 \cr
STS^{-1} , \quad (TS)T(TS)^{-1}, \quad (T^2S)T (T^2S)^{-1}; \quad  &
\quad\quad {\rm for}~  N_f=1 \cr
ST^2S^{-1}, \quad (TS)T^2(TS)^{-1} ; &\quad\quad {\rm for}~  N_f=2 \cr
(ST^2S)T (ST^2S)^{-1}, \quad  ST^4S^{-1} ; \quad  & \quad\quad {\rm
for}~  N_f=3 \cr}}
It is straightforward to check that the product of the monodromies is
$\CM_\infty$ of \moninfty.  This is another
check of our assertions about the singularities.

\subsec{General structure of the curve}

In trying to generalize \ramily\ to $N_f>0$, it is useful to first
extract some properties of the $N_f=0$ curve that we will try to
generalize:
\smallskip
\noindent (1) The family is of the form $y^2=F(x,u,\Lambda)$, where
$F$ is a polynomial in $x,u,$ and $\Lambda$ that is at most cubic
in $x$ and $u$.
\smallskip
\noindent (2) The part of $F(x,u,\Lambda)$ that is cubic
in $x$ and $u$ is $F_0=x^2(x-u)$.
\smallskip
\noindent (3) If one assigns the $U(1)_\CR$ charges $4$ to $u$ and $x$,
$2$ to $\Lambda$, and 6 to $y$, then the family is $U(1)_\CR$ invariant;
and in particular $F$ has charge 12.
\smallskip
\noindent (4)
$F$ can be written $F=F_0+\Lambda^4 F_1$ where $F_1={1\over 4} x$.
\smallskip

For the time being we will simply assume that property (1) remains valid
in the presence of matter hypermultiplets.  (This property can be at
least partly deduced {}from our later consideration of the $N=4$ and
$N_f=4$ theories.)

As for property (2), it is easy to see that it must be valid for
$N_f\leq 3$.  The cubic part of $F$ (after absorbing a possible
multiplicative constant in $y$) would in general take the form
$(x-e_1u)(x-e_2u)(x-e_3u)$ for some $e_i$.  For large $u$ the branch
points are at $e_1u,e_2u$, $e_3u$, and $\infty$.  After dividing $x$ by
$u$, the branch points for $u\to\infty$ are at the $e_i$ and $\infty$,
so they are distinct if the $e_i$ are distinct.  Thus, if the $e_i$ are
distinct, the family of curves has no singularity for $u=\infty$.  This
is actually the correct behavior for the conformally invariant theories,
but for $N_f\leq 3$ it would prevent one {}from getting for $u\to\infty$
the logarithm associated with asymptotic freedom as in \aadexp.  On the
other hand, one does not want all three $e_i$ equal, for then $u$ could
be eliminated {}from the cubic term by changing variables {}from $x$ to
$x'=x-e_1u$; it would then again be impossible to get a logarithm.  So
we can assume for asymptotically free theories that $e_1=e_2\not= e_3$.
By a redefinition of $x$ by $x\to ax+bu$ and a rescaling of $y$, one can
assume that $e_1=0$, $e_3=1$, and hence that the cubic part of $F$ is
$x^2(x-u)$.

As far as the third property is concerned, one expects that $U(1)_\CR$
can be treated as a symmetry if $\Lambda$ is assigned the correct
charge.  4 and 2 are indeed the correct $U(1)_\CR $ charges of $u$ and
$\Lambda$.  For $u$, this is just a statement about the classical
theory.  To understand the charge of $\Lambda$, one notes that the one
loop beta function and $U(1)_\CR$ anomaly are such that a one-instanton
amplitude is proportional to $\Lambda^4$ and violates $U(1)_\CR$ by
eight units.  The physical meaning of $x$ and $y$ is somewhat
mysterious, so we will have to accept their $U(1)_\CR$ charges of 4 and
6 as an empirical fact.

It is now easy to interpret property (4).  Given the anomalous
 $U(1)_\CR $ conservation and the fact that $F$ is polynomial, it is
clear that in the weak coupling limit of the theory, obtained by taking
$\Lambda\to 0$, $F$ reduces to the cubic term $F_0$.  Moreover, since it
is only instantons that violate the $U(1)_\CR$ charge if we do {\it not}
assign a $U(1)_\CR$ charge to $\Lambda$, and a one-instanton amplitude
has the quantum numbers of $\Lambda^4$, the expansion of $F$ in powers
of $\Lambda^4$ can be interpreted as an expansion in the instanton
number.  The fact that $F=F_0+\Lambda^4F_1$ means that for $N_f=0$, $F$
has only the classical contribution $F_0$ and the one instanton
contribution $\Lambda^4F_1$.

\newsec{The curve for $N_f=1$}

\subsec{Massless $N_f=1$}

Our next goal is to adapt the principles just described to determine the
families of curves that control the low energy behavior of the $N=2$
theories with matter.  In doing so, we let $\Lambda_{N_f}$ be the
renormalization scale parameter of the theory with $N_f$ flavors.  We
consider first the case of $N_f=1$ with zero bare mass.

For $N_f=1$, the one-instanton amplitude is proportional to
$\Lambda_1{}^3$, and an amplitude with $r$ instantons is proportional to
$\Lambda_1{}^{3r}$.  However, for $N_f>0$ and one or more massless
hypermultiplets, the equation defining the curve can only receive
contributions {}from terms with an even number of instantons.  This is
because an amplitude with an odd number of instantons violates the
internal ``parity'' symmetry of equation \paritty, while the curve is
invariant under this symmetry.  For $N_f=1$, given that $\Lambda_1$ has
$U(1)_\CR$ charge 2 and $F$ has charge 12, the only possible term in $F$
that depends on $\Lambda_1$ is a two-instanton term, which must be a
constant times $\Lambda_1{}^6$.  Therefore, the curve of the massless
$N_f=1$ theory must be
\eqn\nfonecurve{y^2=x^2(x-u) + t \Lambda_1{}^6.}
The constant $t$ can be absorbed in a redefinition of $\Lambda_1$
\eqn\lambtil{\tilde \Lambda_1{}^6= t\Lambda_1{}^6.}
Below we will determine $t$ for a particular definition of $\Lambda_1$.

Let us now see that this curve has the right properties.  First of all,
we have the expected ${\bf Z}_3$ symmetry of the $u$ plane, as
\nfonecurve\ is invariant under multiplying $x$ and $u$ by a cube root
of unity, with $y$ invariant.  ({}From the formulas below for $a$ and
$a_D$, it will be clear that $a$ and $a_D$ transform correctly under the
symmetry.)  To find the singularities, we compute the discriminant of
the polynomial on the right hand side of \nfonecurve, with the result
\eqn\deltaq{\Delta = \tilde \Lambda_1^6(4u^3-27 \tilde \Lambda_1^6).}
Consequently, on the finite $u$ plane there are three singular points at
$u=e^{2\pi i n \over 3} 3\tilde \Lambda_1^2/ 4^{1\over 3} $,
($n=1,2,3$), permuted by the ${\bf Z}_3$ symmetry, and with monodromy
conjugate to $T$.  To find the behavior for large $u$, we note that for
large $u$ the branch points are approximately at $x=u, \infty$, and $\pm
\tilde\Lambda_1{}^3/\sqrt {u}$.  Upon absorbing a factor of $u$ in $x$
and a factor of $u^{3\over 2}$ in $y$, we get a stable situation with
branch points at $1, \infty$ and $\pm \tilde \Lambda_1{}^3/ u^{3/2}$.
The discriminant is hence of order $u^{-3}$ for $u\to \infty$, so the
monodromy is conjugate to $PT^{-3}$ (the factor of $P$ arises {}from the
factor of $u^{3\over 2}$ in $y$).

\subsec{Massive $N_f=1$}

Now we consider the $N_f=1$ theory with a non-zero bare mass $m$ for the
hypermultiplet.  $m$-dependent terms must vanish for $\Lambda_1\to 0$,
since classically the mass $m$ of the hypermultiplet does not affect the
low energy couplings of the vector multiplet.  $m$ is odd under the
internal ``parity,'' so with $m\not=0$, one can have contributions of
odd instanton number to the equation defining the curve; they simply
have to be odd in $m$.  Since $m$ has $U(1)_\CR$ charge 2, the most
general possibility is
\eqn\uncco{y^2=x^2(x-u)+t \Lambda_1{}^6+m\Lambda_1{}^3(ax+bu)
+cm^3\Lambda_1{}^3}
with constants $a$, $b$, and $c$ that must be determined.

To do so we note that when $m$ is large the quark can be integrated out.
The low energy theory is the pure gauge $N=2$ theory; its scale
$\Lambda_0$ is determined by the one loop beta function to be
\eqn\onetoz{\Lambda_0^4=m\Lambda_1^3.}
$\Lambda_0$ for the pure gauge theory was defined such that the
singularities are at $u=\pm \Lambda_0^2$.  Then we can use equation
\onetoz\ as a definition of $\Lambda_1$.  Other definitions
corresponding to different subtraction schemes differ by a
multiplicative constant in \onetoz.

An obvious constraint on \uncco\ is that in the limit of large $m$ we
recover the curve of the $N_f=0$ theory.  More precisely, we should
consider the scaling limit $m \rightarrow \infty$, $\Lambda_1
\rightarrow 0$ with $\Lambda_0^4=m\Lambda_1^3$ held fixed.  Comparison
of \uncco\ and \ramily\ shows that $a={1 \over 4}$ and $b=c=0$.
Therefore, the curve is
\eqn\unccot{y^2=x^2(x-u)+{1 \over 4} m\Lambda_1{}^3x+ t \Lambda_1{}^6.}
By computing the discriminant, one can verify that in the double
scaling limit, one of the singularities moves to infinity;
in fact the singularity in question  is
at
\eqn\furtag{u\approx -{m^2 \over 64t},~~~~ x\approx -{8t\Lambda_1{}^3
\over m}.}
(It is not necessary here to actually consider the discriminant of
the cubic polynomial on the right of \unccot; the term cubic in $x$ is
unimportant near the singularity, which is controlled by the discriminant
and zeroes of the quadratic polynomial $-ux^2+{1\over 4}m\Lambda_1{}^3x
+t\Lambda_1{}^6$.)

The constant $t$ cannot be determined by the singularities and
monodromies of the curve, since they are invariant under scaling of $m$
or $\Lambda_1$.  To fix $t$, we note that for large $m$, there should be
a singularity at $u\approx m^2$ where one of the elementary quarks
becomes massless.  Comparing to \furtag, we see that $t=-1/64$, so the
curve is
\eqn\unccof{y^2=x^2(x-u)+{1 \over 4} m\Lambda_1{}^3x- {1 \over 64}
\Lambda_1{}^6.}

Of course, we see here the merits of the change of normalization that
led us {}from \famofcurves\ to \ramily.  If one wishes to study the
$N_f=0$ theory as a low energy limit of a massive theory with $N_f>0$,
which has fields of one-half the $W$ boson charge, it will appear
naturally in the form of \ramily.  Similarly (as we will see later),
renormalization group flow {}from the mass-deformed $N=4$ theory, which
does not have isodoublets, will lead naturally to \famofcurves.

Note that given the $N_f=1$ curve, the $N_f=0$ curve can be immediately
determined, but not the other way around.  Of course, that is a
manifestation of the usual irreversibility of the renormalization group.

\newsec{The curve for $N_f=2$}

\subsec{Massless $N_f=2$}

First we consider the $N_f=2$ theory without masses.  The instanton
amplitude is now $\Lambda_2{}^2$, and the internal parity implies that
without bare masses, only amplitudes with an even number of instantons
contribute.  Given that $\Lambda_2$ has $U(1)_\CR$ charge 2 and the
equation for the curve has charge 12, the curve can receive
contributions only {}from instanton number 2 and is of the form
\eqn\impo{y^2=x^2(x-u)+ax\Lambda_2{}^4+bu\Lambda_2{}^4.}
It remains to determine $a$ and $b$.

Let us denote the polynomial on the right hand side of \impo\ as $F$.
The discriminant of $F$ (with respect to $x$) is a polynomial in $u$
that is quartic (or lower order, for some special values of $a$ and
$b$).  Generically, this polynomial has four simple zeroes, so that the
family of curves \impo\ has four singularities in the $u$ plane, with
monodromies conjugate to $T$.  We want to find the values of $a$ and $b$
for which there are instead two singularities with monodromies conjugate
to $T^2$.  This can be done by studying the discriminant or more
directly as follows.

As we have noted before, near a point $u=u_0$ at which $F$ has a double
root at $x=x_0$, $F$ looks like $F={\rm const}\cdot ((x-x_0)^2-
(u-u_0)^n)$, for some $n$; the monodromy is then conjugate to $T^n$.
The condition that $n>1$ is that
\eqn\uncon{F={\partial F\over \partial x}={\partial F\over \partial
u}=0}
at $x=x_0, $ $u=u_0$.  (This is equivalent to saying that the two
dimensional variety defined by the equation $y^2=F(x,u)$ in three
variables $x,u,$ and $y$ has a singularity at $x=x_0$, $u=u_0$, $y=0$.)
One immediately sees that, with $F$ as above, the equations \uncon\
require $a=-b$.  As in our discussion of
$N_f=1$, these considerations cannot determine
the normalization of $\Lambda_2$ and we conclude that the
family of curves is
\eqn\famco{y^2=(x^2-\tilde \Lambda_2{}^4)(x-u)  }
for some $\tilde \Lambda_2$.  Below we will express it in terms of a
convenient definition of $\Lambda_2$.  This, indeed, is a family that we
have seen before -- it is  essentially our old friend
\famofcurves\ of the $N_f=0$ theory in the ``old'' normalization.
Note in particular that the expected ${\bf Z}_2$ symmetry of the $u$
plane has appeared.

\subsec{Massive $N_f=2$}

Now we allow arbitrary bare masses $m_1$, $m_2$ for the quarks.  The
masses transform like $({\bf 3}, {\bf 1}) \oplus ({\bf 1}, {\bf 3})$ of
the global $SO(4) \cong SU(2) \times SU(2)$.  The polynomial defining
the curve must be $SO(4)$ invariant.  Also, terms that are odd (or even)
under the internal parity which changes the sign of one of the $m$'s
must come {}from an odd (or even) number of instantons.  The most
general possible structure of the curve is then
\eqn\yspo{y^2=(x^2- t\Lambda_2{}^4)(x-u) +m_1m_2\Lambda_2{}^2(ax+bu)
+c(m_1{}^2+ m_2{}^2)\Lambda_2{}^4}
with constants $a$, $b$, and $c$ that must be determined.  We define
$\Lambda_2$ such that when $m_2$ is large, the scale of the low
energy $N_f=1$ theory is $\Lambda_1^3=m_2\Lambda_2^2$.  Then, we should
also determine the constant $t$ relating $\Lambda_2$ to $\tilde
\Lambda_2$.

We proceed as in our discussion of the massive $N_f=1$ theory.  The
scaling limit $m_2 \rightarrow \infty$, $\Lambda_2 \rightarrow 0$ with
$\Lambda_1^3=m_2\Lambda_2^2$ held fixed determines $a={ 1 \over 4}$,
$b=0$ and $c=-1/64$.  The curve is then
\eqn\yspoo{y^2=(x^2- t\Lambda_2{}^4)(x-u) + {1\over 4}
m_1m_2\Lambda_2{}^2x - {1 \over 64} (m_1{}^2+ m_2{}^2)\Lambda_2{}^4.}
As for $N_f=1$, the constant $t$ is determined by imposing that when
any mass, say $m_1$, is large there is a singularity at $u=m_1^2$.
Substituting $m_2=0$ and $u=m_1^2$ in the discriminant of \yspoo, the
leading term at large $m_1$ is $4m_1{}^8\Lambda_2{}^4(t-1/64)$; hence
$t=1/64$ and the curve is
\eqn\yspoof{y^2=(x^2- {1 \over 64} \Lambda_2{}^4)(x-u)
+{1\over 4} m_1m_2\Lambda_2{}^2x - {1 \over 64} (m_1{}^2+
m_2{}^2)\Lambda_2{}^4.}

As a check, we can compute the discriminant for $m_1=m_2=m$ and find
\eqn\testnft{\Delta= {\Lambda_2{}^4 \over 16} (u+m\Lambda_2+{1 \over 8}
\Lambda_2^2)(u-m\Lambda_2+ {1 \over 8} \Lambda_2^2) (u- m^2-{1 \over
8}\Lambda_2^2)^2.}
This shows that when the two quarks are degenerate there is a
singularity at $u=m^2+\Lambda_2^2/8$ with two massless particles (double
zero of the discriminant) which for large $m$ can be identified with
that of the massless quarks.  The reason for the two-fold degeneracy
there is that the massless particles at the singularity are in a doublet
of the global $SU(2)$ symmetry of the massive theory.  The other two
singularities - which coincide for $m=0$ - come {}from simple zeroes of
the discriminant corresponding to vacua with one massless multiplet
each.

\newsec{The curve for $N_f=3$}

\subsec{Massless $N_f=3$}

Now we will determine the curve for the $N_f=3$ theory with zero bare
masses.  We do this using the expected singularity structure: there
should be a singularity in the $u$ plane with monodromy conjugate to
$T^4$ and one with monodromy conjugate to $T$.

Since the $N_f=3$ theory has no symmetry in the $u$ plane, we may as
well add a constant to $u$ and assume that the singularity with
monodromy $T^4$ is at $u=0$.  The curve is then given by $y^2=F(x,u)$,
where $F$ is cubic in $x$ and $u$, and at $u=0$, $F(x,0)=0$ has a double
root; by adding a constant to $x$, we may as well assume that the double
root is at $x=0$.  In fact, $F$ must have the form
\eqn\jundo{F=a\Lambda_3{}^2x^2+bu^2x+cux^2 +x^3.}
This polynomial has two roots within $O(u^2)$ of the origin, so the
discriminant is proportional to $u^4$  for small $u$
and the monodromy  around $u=0$ is conjugate to $T^4$.  The
structure in \jundo\ was determined as follows.  Terms $u^n$ with $n\leq
3$ or $xu^m$ with $m<2$ would cause the zero of the discriminant at
$u=0$ to be of order lower than 4, so those terms have been suppressed.
The coefficient of $x^3$ was set to one to agree with the classical
limit.

To determine the parameters, we require that $b\not= 0$, since otherwise
the curve is singular for all $u$, not the situation we want.  We also
ask that \jundo\ has the right classical limit; this is equivalent to
saying that the cubic part of $F$, which is $x^3+cux^2+bu^2x$, is equal
to $(x-e_1u)^2(x-e_2u)$ with $e_1\not= e_2$.  Since $b\not= 0$, the only
way to achieve this is to have $x^2+cux+bu^2$ be a perfect square, say
$(x+\alpha u)^2$.  After rescaling $x$ and $y$ so that $\alpha=1$, and
renaming $x+u$ as $x$, we finally determine our curve:
\eqn\ysq{y^2=x^2(x-u)+t\Lambda_3{}^2(x-u)^2. }
 $t$ is a constant which can be absorbed in a redefinition of
$\Lambda_3$.

Notice that for $N_f=3$, a one-instanton amplitude is proportional to
$\Lambda_3$, so the term in \ysq\ proportional to $\Lambda_3{}^2$ can be
interpreted as a two-instanton effect.  Thus, the contributions {}from
an odd number of instantons vanish, as expected {}from the internal
``parity.''

The other singularity of this family, with monodromy conjugate to $T$,
is at $u=-t\Lambda_3{}^2/4$.  This singularity is presumably associated
with vanishing mass of a state of $(n_m,n_e)=(2,1)$.

\subsec{Massive $N_f=3$}

We now determine the curve for the massive $N_f=3$ theory.  We repeat
the steps we used for $N_f=1,2$.  The masses $m_1$, $m_2$, $m_3$ are in
the ${\bf 15}$ of the global $SO(6)$ symmetry.  The equation for the
curve must be compatible with $SO(6)$ invariance and the anomalous
$U(1)_\CR$, and must reduce to the previous result when the masses
vanish.  Also, it should be invariant under the parity transformation
which changes the sign of one of the masses and the sign of the
instanton factor $\Lambda_3$.  And it should flow to the massive $N_f=2$
curve in the scaling limit of $m_3 \rightarrow \infty$ with
$\Lambda_2^2=m_3\Lambda_3$ fixed.  The most general polynomial equation
with these properties is
\eqn\nfths{\eqalign{y^2=&x^2(x-u)+t\Lambda_3{}^2(x-u)^2
-{1 \over 64} (m_1{}^2+ m_2{}^2+m_3{}^2)\Lambda_3{}^2 (x-u)  \cr
&+ {1 \over 4} m_1m_2m_3\Lambda_3x
-{1 \over 64} (m_1{}^2m_2{}^2+ m_2{}^2m_3{}^2+m_1{}^2m_3{}^2)
\Lambda_3{}^2\cr
&+a(m_1{}^2+ m_2{}^2+m_3{}^2)\Lambda_3{}^4 + bm_1m_2m_3\Lambda_3{}^3.}}
Here we encounter a new element that did not arise for $N_f=1,2$.  The
two constants $a$ and $b$ are not determined by the previous
considerations nor are they a choice of scale (like $t$).  To determine
them we examine the small $m$ theory.

For $m_1=m_2=m_3=0$ there are two singularities.  One of them is at $u=0$
with four massless particles and the other at $u=-t\Lambda_3{}^2/4$
where there is a single massless particle.  Correspondingly, the
discriminant of the massless curve has a single zero at
$u=-t\Lambda_3{}^2/4$ and a fourth order zero at $u=0$. When some of the
masses are turned on the zero of the discriminant at
$u=-t\Lambda_3{}^2/4$ moves and the multiple zero at $u=0$ splits. We
will now examine this splitting and determine the constants $a$ and $b$.
It is enough to consider the case with $m_1=m_2=m_3=m$.  {}From our
analysis of the monopole structure, we know that the zero at $u=0$
should split to a single zero and a triple zero.  The discriminant is of
course holomorphic in all variables, and since the single and triple
zero near $u=0$ are unique, their positions also vary holomorphically.
(By contrast, if a fourth order zero splits to two double zeroes, the
formula for the locations of the two double roots can contain a square
root branch cut.)  Therefore, for small $m$ the discriminant of the
curve with three equal masses should have a zero at $u \sim m$.
Expanding the discriminant of \nfths\ in $m$ with $u=\lambda m$ we find
that actually
\eqn\discnfths{\Delta= -4m^2t^2\Lambda_3{}^8 \left[3\Lambda_3{}^2at +
m(t \Lambda b+18\lambda a) +\CO(m^2) \right].}
To get a zero of $\Delta$ for $m\to 0$ with $\lambda$ of order $1$, the
coefficient of $m^2$ has to vanish.  Since $t\not= 0$, this forces
$a=0$. Then, vanishing of the $\CO(m^3)$ term forces $b=0$.

To determine $t$ we impose that for large $m$ there is a singularity at
$u=m^2$. The leading  contribution to the discriminant at large $m$  with
$u=m^2$ is
${1 \over 16} m^{10} \Lambda_3{}^2(1+64t)$ and therefore $t=-1/64$.

We conclude that the curve is
\eqn\nfthsf{\eqalign{
y^2=&x^2(x-u)- {1 \over 64} \Lambda_3{}^2(x-u)^2 -{1 \over 64} (m_1{}^2+
m_2{}^2+m_3{}^2) \Lambda_3{}^2 (x-u)  \cr
&+ {1 \over 4} m_1m_2m_3\Lambda_3x -{1 \over 64} (m_1{}^2m_2{}^2+
m_2{}^2m_3{}^2 +m_1{}^2m_3{}^2)\Lambda_3{}^2.}}

As a test we evaluate the discriminant for three equal masses $m$ and
find that it has a triple zero at $u=m^2+m\Lambda_3/8$.  The three
particles at the singularity are in the fundamental representation of
the global $SU(3)$ symmetry. As $m$ varies {}from zero to infinity they
move {}from the origin where they are magnetic monopoles to infinity
where they are interpreted as the elementary quarks.

\newsec{Masses, periods and residues}

As in \paperi, the particle masses and the low energy metric and
couplings are now determined by equating $a$ and $a_D$ with periods of a
certain meromorphic one-form $\lambda$ on the curve $E$.  $\lambda$ has
two characteristics: (i) $\lambda$ may have poles but (as long as the
monodromies are in $SL(2,{\bf Z})$) its residues vanish; (ii) to achieve
positivity of the metric on the quantum moduli space, its derivative
with respect to $u$ is proportional to $dx\over y$.

Condition (i) means that the definition of $a$ and $a_D$ by contour
integrals
\eqn\hucxx{\eqalign{ &a = \int_{\gamma_1}\lambda \cr
                     & a_D = \int_{\gamma_2}\lambda,\cr}}
-- with $\gamma_1$ and $\gamma_2$ some contours on $E$ --
is invariant under deformation of the $\gamma_i$, even across poles
of $\lambda$.  This ensures that only the homology classes of the
$\gamma_i$ matter and reduces the monodromies to a group $SL(2,{\bf Z})$
that acts on $H_1(E,{\bf Z})$.  In the presence of bare masses, this is
too strong a condition since (as we saw originally in the discussion
of $N=2$ QED in section 2.4) when the bare masses are non-zero the
monodromies are not quite in $SL(2,{\bf Z})$.

As for condition (ii), the differential form $dx\over y$ has no poles
and represents a cohomology class on $E$ of type $(1,0)$.  Having
$d\lambda/du=f(u)\,dx/y$ leads to positivity of the metric as we
explained in \paperi.  The function $f(u)$ is determined by requiring
the right behavior at the singularities , for instance $a\approx {1\over
2}\sqrt{2u}$ for large $u$.  $f$ is a constant for the same reasons as
in \paperi.  The proper relation is in fact
\eqn\onso{{d\lambda\over du}={\sqrt 2\over 8\pi}{dx\over y}.}
Up to an inessential sign, this is $1/2$ the value in the ``old''
conventions.  By integration with respect to $u$, \onso\ determines
$\lambda$ (once the curve is known) for all values of $N_f$.
\onso\ is only supposed to hold up to a total differential in $x$;
$\lambda$ is supposed to be meromorphic in $x$.

But when one obtains $\lambda$ by integration of \onso, does one get a
result that obeys condition (i)?  As an example,\foot{We consider this
example because the theories with $N_f\geq 3$ require considerably more
powerful methods - developed in the last section of this paper - to
analyze condition (i).  On the other hand, once the conditions are
imposed for $N_f=2$, there is no need to consider $N_f<2$ separately
since the correct behavior for those cases follows {}from $N_f=2$ by the
renormalization group flow.} consider the
massive $N_f=2$ theory
with $y^2=(x^2- {1 \over 64} \Lambda_2{}^4)(x-u) + {1 \over 4}
m_1m_2\Lambda_2{}^2x - {1 \over 64}  (m_1{}^2+ m_2{}^2)\Lambda_2{}^4$.
In this case, a meromorphic $\lambda$ obeying \onso\ can be found by
inspection:
\eqn\lambdo{\lambda= {-\sqrt{2} \over 4\pi}
{dx\,\,y\over x^2-  { 1 \over 64} \Lambda_2{}^4} .}  Note that $\lambda$
has poles at $x=\pm { 1 \over 8} \Lambda_2{}^2$ whose residues
are\foot{Note that the residues are independent of $u$.  The reason for
this is that $\omega=d\lambda/du$ has zero residues.  The operation of
taking the residue in $x$ commutes with differentiation with respect to
$u$, so the fact that $\omega$ has zero residues means that the residues
of $\lambda$ are annihilated by $d/du$.}
\eqn\residunft{{\pm m_1 \pm m_2 \over (2\pi i) 2\sqrt{2}}.}
Thus, condition (i), as we have stated it so far, holds only when the
bare masses are zero.

More generally, when the bare masses are not zero, the residues of
$\lambda$ mean that the definition of $a$ and $a_D$ in \hucxx\ is {\it
not} invariant under deforming the contours $\gamma_i$ across the poles
of $\lambda$.  Under such a deformation, $a$ and $a_D$ would change by a
constant -- the residue of the pole.

It follows that the monodromies are no longer simply in $SL(2,{\bf Z})$.
In defining $a$ and $a_D$, one tries to take a smoothly varying family
of contours that keeps away {}from the poles of $\lambda$.  However, in
looping around a closed path in the $u$ plane, the contour might come
back on the wrong side of a pole.  When this happens, in addition to the
$SL(2,{\bf Z})$ action, $a$ and/or $a_D$ will jump by a constant.

But {}from our analysis in sections 2 and 6, we know that precisely such
constants are needed when the bare masses $m$ are not zero.  In fact,
according to equation \nidoc, the constant jump should be of the form
$\sum_i m_i S_i /\sqrt{2}$.  The $S_i$ are abelian conserved charges
that appear in the central extension of the $N=2$ algebra along with the
electric and magnetic charges.

In the $N_f=2$ theory, the $S_i$ are integers for the fundamental
particles and are non-zero half integers for monopoles.  Therefore, the
allowed jumps are linear combinations of $ (\pm m_1 \pm m_2 )/
2\sqrt{2}$ -- exactly as we see in \residunft.  (Note that since $a$ is
a contour integral of $\lambda$, the jump in $a$ is $2\pi i$ times the
residue of $\lambda$.)

In sum, in the massive theory, we do not want the residues of $\lambda$
to be zero.  We want them to be linear in the masses and in particular
independent of $\Lambda_k$.  It is a non-trivial test of our answers for
the curves that the residues of $\lambda$ have the right form.  In fact,
the detailed structure of the equations for the curves could all have
been determined {}from this one condition alone; that will be our
strategy in section 17.

\bigskip
\noindent
{\it Singularities From Semi-classical States}

In \paperi\ and the present paper, we have assumed that singularities of
the curve come not from massless non-abelian gauge fields but {}from
massless hypermultiplets of spins $\leq 1/2$.  $N=2$ multiplets of such
low spin are necessarily BPS-saturated, and -- as BPS-saturated states
have strong stability properties -- it is natural to expect that such
states would be the continuation to the strong coupling regime of states
that can be seen semiclassically; for brevity we will refer to such
states as semiclassical states.

Moreover, {\it ex post facto}, {}from the curves that we obtain, it is
possible to show that (if the curves are correct) the singularities must
be due to semiclassical states.  The approach to showing this was
explained in the last paragraph of \paperi: one shows that one can
interpolate {}from the semiclassical region of large $u$ to the
singularities along a path on which $a/a_D$ is never real.  The
importance of the requirement that $a/a_D$ is not real is that (see the
end of section 4 of \paperi) as long as $a/a_D$ never becomes real, the
spectrum of BPS-saturated states cannot jump and therefore the
BPS-saturated states are precisely the semiclassical ones.

\nref\manton{N. S. Manton and B. J. Schroers, Ann. Phys. {\bf 225}
(1993) 290.}%
The argument given at the end of \paperi\ can be carried over to the
models of the present paper, but for brevity we will do this only in the
case which is perhaps most interesting: the $(n_m,n_e)=(2,1)$ state for
$N_f=3$.  What makes this the most interesting example is that, in fact,
it is not known whether this state exists semiclassically; in view of
what we are about to say, its semiclassical existence is a prediction of
our analysis.  This prediction should be testable using the methods of
\refs{\sen,\manton}.  (Of course, our proposal of $SL(2,{\bf Z})$
symmetry for $N_f=4$ makes many similar predictions.)

In units with $\Lambda_3{}^2/64=1$, the massless $N_f=3$ curve is
\eqn\ununu{y^2=x^2(x-u)-(x-u)^2.}
The polynomial on the right hand side has zeroes at $x_0=u$ and
at
\eqn\xpm{x_\pm={1\over 2}\left(1\pm\sqrt{1-4u}\right).}
In particular, at $u=1/4$, $x_+=x_-$, giving the singularity that we
have attributed to a massless state of $(n_m,n_e)=(2,1)$.
To show that this
state is semiclassical, we will interpolate on the positive $u$ axis
{}from the semiclassical regime of $u\to\infty$ to the singularity at
$u=1/4$.  For $u>1/4$, $x_+$ and $x_-$ are complex conjugates.

We have
\eqn\inno{\eqalign{ {da\over du} &=\int_{\gamma_1 }\omega \cr
                     {da_D\over du} & =\int_{\gamma_2}\omega, \cr}  }
where $\omega=(\sqrt 2/8\pi)dx/y$, $\gamma_1$ is a circle in the $x$
plane that loops around $x_+$ and $x_-$ but not $x_0$, and $\gamma_2$ is
a contour that loops round $x_0$ and $x_+$ but not $x_-$.  Complex
conjugation leaves $x_0$ alone and exchanges $x_+$ with $x_-$; hence
$\gamma_1$ is invariant under complex conjugation, but complex
conjugation turns $\gamma_2$ into a contour $\gamma_3$ that loops around
$x_0$ and $x_-$ while avoiding $x_+$.  So $a$ is real but the complex
conjugate of $a_D$ is given by
\eqn\ginno{{d\overline a_D\over du}=\int_{\gamma_3}\omega.}
$\gamma_3$, however, is homotopic to the sum of $-\gamma_1$ and
$-\gamma_2$. (The minus sign comes {}from keeping track of the
orientations of the contours.) Hence, \ginno\ gives
\eqn\pinno{\overline a_D=-a-a_D.}
In other words,\foot{This equation has the following interpretation.
The curve \ununu\ is real for real $u$, that is, the coefficients in
the equation are real.  There are two types of real elliptic curve:
$\tau$ can have real part zero or $1/2$.  (Thus $\tau$ is either
invariant or transformed by the $SL(2,{\bf Z})$ transformation
$\tau\to \tau-1$ under the complex conjugation operation $\tau\to
-\bar\tau$.) The two possibilities correspond in a suitable basis to
$a$ real, $a_D$ imaginary, or $a$ real, $a_D=-a/2+{\rm imaginary}$.
For $u>1/4$ we have the second possibility.}
\eqn\rinno{a_D=-{a\over 2}+{\rm imaginary}.}

Now, jumping of BPS saturated can only occur when $a_D/a$ is real; in
other words, for $u$ positive and greater than $1/4$, it can only occur
when the imaginary part of $a_D$ vanishes and so $2a_D+a=0$.  When that
happens, a BPS-saturated state of $(n_m,n_e)=(2,1)$ becomes massless,
if there is such a state.

If our curve is correct, there must be such a state at $u=1/4$ since we
need it to generate the singularity.  Let $u'$ be the smallest value of
$u>1/4$ at which the imaginary part of $a_D$ vanishes.  Then, one can
interpolate {}from $u=1/4$ to $u=u'$ without any jumping of BPS-saturated
states; hence, the $(2,1)$ state would again be massless at $u'$.  This
should produce an extra singularity that our curve does not have.  So if
the curve is correct, $u'$ does not exist, ${\rm Im}\,a_D$ never
vanishes on the positive $u$ axis for $u>1/4$, one can interpolate {}from
$u=1/4$ to $u=\infty$ without jumping, and the $(2,1)$ state that gives
the singularity at $u=1/4$ must be semiclassical.

Of course, it would be desirable to sharpen this argument and prove
directly {}from \inno\ that $a_D $ is never real for $u>1/4$.

\newsec{Structure of the scale invariant theories}

In this section, we will analyze the $N_f=4$ and $N=4$ theories.  In the
absence of bare masses, those theories are conformally invariant.  In
the presence of $N=2$-invariant bare masses, their properties (or at
least the properties that we can analyze) are much richer.

The challenge that these theories present is that there is a
dimensionless coupling constant,
\eqn\taunf{\tau =\cases{
{\theta\over \pi}+{8\pi i\over g^2} & for $N_f=4$\cr
{\theta\over 2 \pi}+{4\pi i\over g^2} & for $N=4$.\cr}}
Therefore, in the curve $y^2=F(x,u, m_i,\tau)$ that controls the low
energy behavior, the coefficients are functions of $\tau$ that must be
determined.  This contrasts with $N_f<4$ where, instead of $\tau$, one
has the renormalization scale $\Lambda$; dimensional analysis ensures
that (if $F$ is holomorphic and free of singularities\foot{Holomorphy
of $F$ is needed in order for the coefficient $\tau (u,m_i, \Lambda)$ in
front of $W_\alpha{}^2$ in the low energy theory to be holomorphic in
the field $u$ and the parameters $m_i$ and $\Lambda$, as follows {}from
\nonren.  Absence of singularities in $F$ is less obvious.}) the
dependence on $\Lambda$ is polynomial, so that there are only finitely
many parameters to determine.

Of course, if one is willing to assume $SL(2,{\bf Z})$ duality, and if
one can guess the modular weights of $x,y,u$, and the $m_i$, then the
coefficients are modular forms, which are determined by $SL(2,{\bf Z})$
in terms of finitely many coefficients.  The assumption of $SL(2,{\bf
Z})$ would thus put us back in a situation similar to that which we have
already encountered for $N_f<4$.  We will not follow that road because
instead of assuming $SL(2,{\bf Z})$ invariance, we want to deduce it.
Therefore, we face a more difficult task.  We will in fact provide two
different routes to the goal; in this section we analyze the $N_f=4$ and
$N=4$ models by more careful application of the methods that we have
used for $N_f\leq 3$, while in the next section we use a method
suggested at the end of section 15.

\subsec{The massless case}

The first step is to find the right family of curves for the conformally
invariant case, that is when the bare masses are zero.  In this case,
the classical formulas
\eqn\elfo{
\eqalign{& a = \cases{
\half \sqrt{2u} ,& for $N_f=4$\cr
\sqrt{2u}& for $N=4$,\cr} \cr
& a_D = \tau a \cr}}
are exact.  So we wish to find a curve $y^2=F(x,u,\tau)$ such
that the differential form
\eqn\difform
{\omega=\cases{
{\sqrt 2 \over 8\pi} {dx \over y}  & for $N_f=4$\cr
{\sqrt 2 \over 4\pi} {dx \over y} & for $N=4$,\cr}}
has the periods $({\partial a_D \over \partial u},{\partial a \over
\partial u})$, with $(a_D,a)$ given in \elfo.

Now, a genus one curve $E$ and a differential form with periods a
multiple of $(\tau,1)$ can be found as follows.  Let $E$ be the quotient
of the complex $z$ plane by the lattice generated by $\pi $ and
$\pi\tau$.  Let $\omega_0=dz$.  Obviously, the periods of $\omega_0$
(integrated on contours that run {}from 0 to $\pi$ and {}from 0 to
$\pi\tau$, respectively) are $\pi $ and $\pi\tau$.

\nref\koblitz{N. Koblitz, {\it Introduction To Elliptic Curves And
Modular Forms} (Springer-Verlag, 1984).}
\nref\chandra{K. Chandrasekharan, {\it Elliptic Functions}
(Springer-Verlag, 1985).}
To find an algebraic description of $E$ (see section I.6 of \koblitz\
or section III.3 of \chandra),
one introduces the Weierstrass ${\CP}$ function, which obeys
\eqn\unno{\CP(z)= \CP(z+1)=\CP(z+\tau)=\CP(-z)}
and has for its only singularity on $E$ a double pole at the origin.
If one sets $x_0=\CP(z)$, $y_0=\CP'(z)$, then one finds
\eqn\unjo{y_0{}^2=4x_0{}^3 -g_2(\tau)x_0-g_3(\tau).}
Here $g_2=60\pi^{-4}
G_4(\tau)$, $g_3=140\pi^{-6}G_6(\tau)$, and $G_4, \,G_6$ are the usual
Eisenstein series
\eqn\punno{\eqalign{
G_4(\tau) & = \sum_{m,n\in {\bf Z}_{\not= 0}}{1\over (m\tau+n)^4}\cr
G_6(\tau) & = \sum_{m,n\in {\bf Z}_{\not= 0}}{1\over (m\tau+n)^6}\cr}}
which define modular forms for $SL(2,{\bf Z})$ of weight 4 and 6,
respectively.
Since the definition of $x_0$ and $y_0$ was such that $y_0=dx_0/dz$, one
also can rewrite $\omega_0=dz$ as
\eqn\yunno{\omega_0={dx_0\over y_0}.}

Now, set $x=x_0u$, $y=\half y_0u^{3/2}$, and
\eqn\pinno{\omega=\cases{
{\sqrt {2/u} \over 4\pi} \omega_0 ={\sqrt 2\over 8\pi}{dx\over y} & for
$N_f=4$ \cr
{\sqrt {2/u} \over 2\pi} \omega_0={\sqrt 2\over 4\pi}{dx\over y} & for
$N=4$.\cr}}
The equation for the curve becomes
\eqn\punjo{y{}^2=x^3 -{1 \over 4} g_2(\tau)xu^2-{1 \over 4}
g_3(\tau)u^3.}
This change of variables and in particular the normalization of $u$
is motivated by the following requirement.  For weak coupling $(\tau
\rightarrow i \infty)$ we should recover our curve $y^2=F_0(x,u)=
x^2(x-u)$.  It is easy to check {}from the asymptotic behavior $g_2={4
\over 3} + \CO(q)$, $g_3={8 \over 27} + \CO(q)$ that after replacing $x$
in \punjo\ by $x-u/3$ we find $F_0$.

The periods of $\omega$ are now ${\sqrt {2/u} \over 4 } (1,\tau)$ for
$N_f=4$ and ${\sqrt {2 /u} \over 2 } (1,\tau)$ for $N=4$ and (since in
general the periods of $\omega$ are $da/du$ and $da_D/du$), one has
\eqn\iggo{\eqalign{& a =\cases{
\half \sqrt{2u} & for $N_f=4$\cr
\sqrt{2u}       & for $N=4$\cr} \cr
                   &a_D  = \tau a,\cr}}
as desired.

Thus, we have determined the appropriate curve for the massless theory.
The coefficients are modular forms.  This is not really a new test of
$S$-duality; it is equivalent to the fact that the metric of the
classical theory is $S$-dual, which is one of the traditional pieces of
evidence for $S$-duality.

\bigskip
\noindent{\it Spin Structures}

The equation \punjo\ for the curve can be factored
\eqn\gunjo{y^2= \left(x-e_1(\tau)u\right)\left(x-e_2(\tau) u\right)
\left(x-e_3(\tau)u\right),}
with the $e_i$ being the roots of the cubic polynomial $4x^3-g_2x-g_3$;
they obey $e_1+e_2+e_3=0$.
The classical formulas for the $e_i$ (\chandra, p. 69; note we set
$\omega_1=\pi$) are
\eqn\milmo{\eqalign{e_1-e_2& = \theta_3{}^4(0,\tau) \cr
                   e_3-e_2& = \theta_1{}^4(0,\tau) \cr
                   e_1-e_3& = \theta_2{}^4(0,\tau), \cr}}
where $\theta_i$ are the $\theta$ functions
\eqn\upilmo{\eqalign{\theta_1(0,\tau)&=\sum_{n\in {\bf
Z}}q^{\half (n+1/2)^2}\cr
\theta_2(0,\tau)&=\sum_{n\in {\bf Z}}(-1)^nq^{\half n^2}\cr
\theta_3(0,\tau)&=\sum_{n\in {\bf Z}} q^{\half n^2}\cr}}
and hence
\eqn\eexpan{\eqalign{
e_1&={2 \over 3} + 16q+16q^2+\CO(q^3) \cr
e_2&=-{1 \over 3} - 8 q^\half -8q-32q^{3\over 2} -8q^2+\CO(q^{5\over
2}) \cr
e_3&=-{1 \over 3} + 8 q^\half -8q +32q^{3\over 2} -8q^2+\CO(q^{5\over
2})  \cr}}
(note that unlike \chandra, we use $q=e^{2\pi i \tau}$).  As $g_2$ and
$g_3$ are modular forms of weight 4 and 6, the $e_i$ are of weight 2
(indeed, the theta functions are of weight $1/2$).  However, the $e_i$
are not modular forms for $SL(2,{\bf Z})$ because there is no
modular-invariant way to select a particular root of the cubic.  Rather,
the $e_i$ are modular forms of three different (conjugate) subgroups of
$SL(2,{\bf Z})$, each of index three.

Actually, there is a natural one-to-one association of the $e_i$ with
the even spin structures on $E$.  This can be seen as follows, beginning
with the description of $E$ as the quotient of the $z$-plane by a
lattice.  The even spin structures of $E$ are in natural correspondence
with the non-zero half-lattice points $z=1/2,\tau/2$, and $(\tau+1)/2$.
Since $\CP(z)=\CP(-z)$, the derivative $y=\CP'(z)$ vanishes at these
points (indeed, up to a lattice translation those points are invariant
under $z\leftrightarrow -z$).  So the half-lattice points are zeroes of
$y$.  Looking at the equation \gunjo, we see therefore that the
half-lattice points have $x=e_iu$, for $i=1,2$, or 3.  So the even spin
structures (or half-lattice points) correspond to the $e_i$.  (And each
$e_i$ is a modular form for the subgroup of $SL(2,{\bf Z})$ that fixes
that spin structure -- these subgroups are conjugate to $\Gamma_0(2)$,
the subgroup of $SL(2,{\bf Z})$ obtained by requiring that the upper
right entry be even.)

For the $N=4$ theory, the relation of the $e_i$ to the spin structures
will have no particular importance.  For $N_f=4$, it is very important
since, as we know already, $SL(2,{\bf Z})$ permutes the three
eight-dimensional representations of ${\rm Spin}(8)$ in the same way
that it acts on the spin structures.  So the three $e_i$ are permuted
under ${\rm Spin}(8)$ triality.

\subsec{The curve for $N=4$}

We recall that $N=4$ supersymmetric Yang-Mills theory can be regarded as
$N=2$ super Yang-Mills theory with an additional matter field that is a
hypermultiplet in the adjoint representation of the gauge group.  One
can give a bare mass $m$ to that hypermultiplet, explicitly breaking
$N=4$ to $N=2$.  In this subsection we analyze the resulting theory, for
gauge group $SU(2)$.

We first consider the theory for weak coupling, that is for $|q|\ll
1$, with  $m\not= 0$.  There is one singularity at
$u\approx {1 \over 4} m^2$ where a component $H$ of the elementary
hypermultiplet is massless.  This gives a monodromy conjugate to
$T^2$.\foot{The elementary hypermultiplet for $N=4$ has twice the
electric charge of the hyperdoublets that we have considered earlier.
As a massless hyperdoublet gives monodromy $T$, and the one-loop beta
function which determines the monodromy is proportional to the square of
the charge, the massless $H$ particle would give monodromy $T^4$ in the
conventions of the $N_f=4$ theory.  With the $N=4$ conventions, the
monodromy is $T^2$.} In addition, at an energy of order $\Lambda_0 \sim
q^{1/4}m$, the theory evolves to a strongly coupled pure $N=2$ gauge
theory.  As was explained in \paperi, this theory has two singularities,
associated with massless monopoles, with monodromies conjugate to $T^2$.
So altogether, we get three singularities, each conjugate to
$T^2$.\foot{These three singularities are permuted under monodromies in
$q$ and $m$.  This is the reason the $N=4$ conventions in which they are
all conjugate to $T^2$ are preferable to the $N_f\not= 0$ conventions in
which one is conjugate to $T^4$ and the others to $T$.}

Of course, the above analysis was valid for very weak coupling.  Could
it be, for instance, that what we described above as one singularity
conjugate to $T^2$ is really a pair of singularities conjugate to $T$,
separated by an amount that vanishes for weak coupling?  $SL(2,{\bf Z})$
group theory alone would permit this, but it is impossible because each
of the singularities arises when a {\it single} hypermultiplet becomes
massless.

So we are looking for a family of curves
\eqn\polo{y^2=F(x,u)}
(with cubic $F$) that -- as $u$ varies -- has precisely three
singularities each conjugate to $T^2$.  The restriction that this places
on $F$ was explained in our discussion of the  $N_f=2$ theory in
section 13.1.  There is a singularity at $u_0$ with monodromy $T^n$ for
$n>1$ (generically $n$ will be 2) if and only if for some $x_0$,
\eqn\nolo{F={\partial F\over \partial x}={\partial F\over \partial u}=0}
at $x=x_0$, $u=u_0$.  Conditions \nolo\ mean that the curve
$F(x,u)=0$ has a singularity at $(x,u)=(x_0,u_0)$.

Therefore, we are looking for a plane cubic curve $F(x,u)=0$ with three
distinct singularities.  The possible singularities of a plane cubic
curve can be completely classified.  If $F$ is an irreducible
polynomial, there is at most one singularity (a node or cusp).  If
$F=F_1F_2$, with $F_1$ linear in $x$ and $u$ and $F_2$ quadratic and
irreducible, there are precisely two singularities (perhaps at
infinity), namely the points where $F_1=F_2=0$.  The only way to get
three singularities is to have $F=F_1F_2F_3$, where the three factors
are linear; the three singularities are the points $F_i=F_j=0$ for any
two distinct $i$ and $j$.

To reproduce the known $m=0$ limit of $F$, the $F_i$ must be (up to a
scalar multiple and a permutation of $i$) $F_i=x-e_iu-f_i$ where $f_i$
are functions of $m$ and $\tau$ only and vanish at $m=0$.  Moreover, by
adding constants (that is, functions of $m$ and $\tau$ only) to $x$ and
$u$, one can eliminate two of the three $f_i$.  Since we did not assign
any physical meaning to $x$ we can take the freedom to shift it.
However, we want to preserve $u= \langle \Tr \phi^2 \rangle$.
Therefore, we will denote the shifted $u$ by $\tilde u$ and will later
determine the relation between them.  To keep the symmetry under
permuting the $e_i$, we shift $x$ and $u$ such that $f_i=e_i{}^2f$.
Then the equation of the mass-deformed $N=4$ theory becomes
\eqn\eqofour{y^2=(x-e_1 \tilde u-e_1{}^2f)(x-e_2 \tilde u-e_2{}^2f)
(x-e_3 \tilde u-e_3{}^2f).}
We still need to find $f$ and to determine $\tilde u$ in terms of $u$.

The relation between $u$ and $\tilde u$ is determined by examining the
theory at weak coupling; i.e.\ in the limit $\tau \rightarrow i \infty$.
In this limit we should reproduce our weak coupling curve $y^2=
F_0=x^2(x-u)$.  This motivates us to change variables to
\eqn\shiftxu{\eqalign{
&\tilde u =  u - \half e_1f \cr
&x \rightarrow x- \half e_1 u + \half e_1^2f}}
in \eqofour.  The family of curves becomes
\eqn\eqofourt{y^2= \left( x- c_1  u\right)
\left( x+c_2  u- c_2(c_1+c_2)f \right)
\left( x-c_2 u+ c_2(c_1-c_2) f \right)}
with $c_1={3 \over 2} e_1$ and $c_2 =\half(e_3-e_2)$.  In this form it
is easy to study the weak coupling limit.  For a smooth limit,
$f_0=f(\tau=i\infty)$ should be finite.  Using $c_1(\tau = i\infty)= 1$
and $c_2(\tau = i\infty)= 0$, the exact curve \eqofourt\ becomes $y^2=
F_0=x^2(x-u)$ as required.  Therefore, in the form \eqofourt\ the family
of curves is expressed in terms of $u= \langle \Tr \phi^2 \rangle$.

We can now relate $f_0=f(\tau=i\infty)$ to the mass $m$.  We do that
by examining the singularities of \eqofourt. The roots of the equation
are at $x_1 =c_1u$ and $x_{2,3}= \pm c_2(-u+ (c_1\pm c_2)f)$.  A
singularity occurs when $x_i=x_j$ (for any two distinct $i$ and $j$).
This occurs for
\eqn\singtilu{\eqalign{
&u_1={3 \over 2} e_1f =c_1f \cr
&u_{2,3}=\pm \half (e_3-e_2) f= \pm c_2f .\cr}}
In the weak coupling limit $c_1 \approx 1$, $c_2 \approx 8q^\half$ and
hence $u_1 \approx f_0$ and $u_{2,3} \approx \pm 8q^\half f_0$.  We
interpret the singularity at $u_1$ as associated with a massless
elementary field.  It should be at $a=m/\sqrt 2$.  For weak coupling
and in the $N=4$ normalization this is at $u \approx\half a^2  = m^2/4$.
Therefore,
\eqn\scalconnf{f_0=f(\tau = i\infty)=m^2/4 . }

The other two singularities at $u_{2,3} \approx \pm 8q^\half f_0$ are
interpreted as the two singularities of the low energy pure gauge $N=2$
theory.  More precisely, we can now take the scaling limit $q
\rightarrow 0$, $f_0=m^2/4 \rightarrow \infty$ holding
$\Lambda_0^2=2q^\half m^2$ fixed.  In this limit \eqofourt\ becomes
\eqn\eqofourtst{y^2=(x- u)( x^2 - \Lambda_0^4)}
which is exactly the curve of the expected low energy pure gauge $N=2$
theory in the $N=4$ conventions with scale $\Lambda_0$.

We still need to determine $f(\tau)$.  To do that we will consider the
residues of the differential form $\lambda$.  As explained in section 15
(see the footnote preceding equation \residunft), they are independent
of $u$ and hence, on dimensional grounds, are proportional to $m$.  The
proportionality factor must be independent of $\tau$, since the residues
are related to the central extension in the $N=2$ algebra, which is
independent of $\tau$.  We will explain the general theory of these
residues in section 17, but for the moment, we give instead the
following indirect argument which shows that $f$ is independent of
$\tau$.

Let us assume first that $f$ is independent of $\tau$.  Then, equation
\eqofour\ has simple modular properties.  $y$, $x$, $\tilde u$, $e_i$
and $f$ have modular weights 6, 4, 2, 2 and 0 respectively.  Therefore
the differential form $\lambda$ determined by $d\lambda/du \sim dx /y$
transforms like a differential form of weight zero under $SL(2,{\bf Z})$
and the same is true of its residues.  We also know that the residues do
not diverge for $\tau\to i\infty$ (since in \scalconnf\ we showed that
at any rate the assumption that $f$ is constant is valid for $\tau\to
i\infty$).  A modular function of weight zero that is bounded at
infinity is a constant, so if $f$ is constant, the residues are
constants.

It is now trivial to determine the residues for arbitrary $f$.  Indeed,
on dimensional grounds, the residues are proportional to $m$, but $m$
only enters through the function $f(\tau,m)=m^2f_1(\tau)$.  Hence, the
residues are a constant times $m\sqrt{f_1(\tau)}$.  The residues,
therefore, are independent of $\tau$ if and only if $f_1$ is a constant.
The constant is known since we have determined the behavior for $\tau\to
i \infty$:
\eqn\concf{f={1 \over 4 } m^2. }

We conclude that the curve governing the low energy behavior of the
mass-deformed $N=4$ theory is
\eqn\eqofourf{y^2=(x-e_1 \tilde u-{1 \over 4 }e_1{}^2 m^2)(x-e_2 \tilde
u-{1 \over 4} e_2{}^2 m^2)(x-e_3 \tilde u-{1 \over 4 }e_3{}^2m^2)}
with
\eqn\defutu{u=\langle \Tr \phi^2 \rangle = \tilde u + {1 \over 8}
e_1m^2.}

This formula is completely $SL(2,{\bf Z})$ invariant; the coefficients
are modular forms.  Since the formula is not limited to weak coupling,
this $SL(2,{\bf Z})$ invariance is a genuine, new, strong coupling test
of $SL(2,{\bf Z})$ invariance of the $N=4$ theory.  (Also, we learn that
the $N=2$-invariant bare mass preserves $SL(2,{\bf Z})$.)

$SL(2,{\bf Z})$ invariance may, however, be lost in various weak coupling
limits.  For instance, for the weak coupling limit $\tau \rightarrow i
\infty$ the natural variable is $u=\langle \Tr \phi^2 \rangle $ which
differs {}from $\tilde u$ by an additive renormalization \defutu.
Unlike $\tilde u$, $u$ does not transform like a modular form.
Furthermore, in the $\tau \rightarrow i \infty$ limit we defined a
scaling theory by taking also $m \rightarrow \infty$ holding
$\Lambda_0^4 =4m^4e^{2\pi i \tau}$ fixed.  By $SL(2,{\bf Z})$ we can
find other weakly coupled theories and scaling limits in which
$\tau\to p/q$ with $p/q$ an arbitrary rational number.  The theory in
these limits is strongly coupled in the original variables -- the
elementary gauge fields -- but weakly coupled in dual variables,
$\phi_d$, which were interpreted as monopoles in the original theory.
The natural parameter is then $u_d=\langle \Tr \phi_d{}^2 \rangle $.
It is related to $u$ or $\tilde u$ by a shift and a modular
transformation. In a suitable limit with $\tau\to p/q$, $m \rightarrow
\infty$ one gets a pure gauge $N=2$ theory.  Thus, the pure $N=2$
theory that arises in the scaling limits does not have $SL(2,{\bf Z})$
symmetry; $SL(2,{\bf Z})$ merely permutes the possible equivalent
scaling limits.

\subsec{The curve for $N_f=4$}

In this subsection we determine the curve for the $N_f=4$ theory with
arbitrary masses $(m_1,m_2,m_3,m_4)$.  Among other things we will
establish triality and $SL(2,{\bf Z})$ invariance.

We start by considering the situation for $m_i=(m,m,0,0)$.  As we
discussed in section 10, in this case we expect to find three
singularities with two massless particles in each.  According to our
general discussion in section 11, this means that the monodromy around
any of them is conjugate to $T^2$.  This is exactly the situation we
encountered in our discussion of the curve for $N=4$, so with these
masses the curve is
\eqn\mmzz{y^2=\prod_i(x-e_i\tilde u-e_i^2f).}
Here $\tilde u$ is related to $u=\langle \Tr \phi^2 \rangle$ by a
constant shift as in \shiftxu\ and $f$ is proportional to $m^2$ and {\it
a priori} may depend on $\tau$.

As for $N=4$, since the residues of $\lambda$ should be $\tau$
independent, $f$ must be a constant.  In order to determine the constant
we again consider this curve in the weak coupling limit $\tau
\rightarrow i \infty$.  Again, we shift $x$ and $\tilde u$ as in
\shiftxu\ and take the limit $q \rightarrow 0$, $f \rightarrow \infty$
with $qf^2$ held fixed.  The low energy theory is that of the massless
$N_f=2$ theory.  Indeed, we find the curve \famco\ with $\tilde
\Lambda_2{}^2 = \Lambda_2{}^2/8=8q^\half f$.  We also find a singularity
at $u \approx f$ which we interpret as associated with a massless
elementary quark and hence it should be at $a=m/\sqrt{2}$.  In the
$N_f=4$ normalization this means that it is at $u \approx 2a^2= m^2$.
Hence, $f=m^2$.

We now turn to the theory with arbitrary $m_i$.  We first impose the
global $SO(8)$ symmetry and construct its low dimension invariants.
There is a unique quadratic invariant
\eqn\quadinv{R={1\over 2} \sum_i  m_i^2 .}
There are three linearly independent quartic invariants.  We take
them to be $R^2$ and
\eqn\soeinv{\eqalign{
T_1&={1 \over 12}\sum _{i>j}m_i^2m_j^2 - {1\over 24}\sum_i m_i^4 \cr
T_2&=- {1 \over 2}\prod_i m_i - {1 \over 24}\sum _{i>j}m_i^2m_j^2 +
{1\over 48}\sum_i m_i^4 \cr
T_3&={1 \over 2}\prod_i m_i - {1 \over 24}\sum _{i>j}m_i^2m_j^2 +
{1\over 48}\sum_i m_i^4 \cr}}
with $\sum T_i=0$.  The reason for writing them like that is that
the $T_i$ are permuted under the triality automorphism of $SO(8)$
(which acts on the masses as in equations \triagen\ and  \triagent);
we anticipate (but do not assume) that triality is a symmetry of the
theory.  There are four six order $SO(8)$ invariants.  We take
them to be $R^3$, $RT_i$ and
\eqn\sixtho{N={3 \over 16}\sum_{i>j>k}m_i^2m_j^2m_k^2 -
{1 \over 96}\sum_{i\not= j} m_i^2m_j^4 + {1 \over 96}\sum_i m_i^6.}
$R$ and $N$ are invariant under triality.

In trying to generalize the curve for $m_i=(m,m,0,0)$ to arbitrary
$m$ we impose that:

\noindent
1) The limit of the curve as any masses go to zero is smooth and
hence it is polynomial in $m_i$.

\noindent
2) $U(1)_\CR$ (or equivalently dimensional analysis) constrains
the powers of $m_i$.  This is achieved by assigning charges $4,4,6,2$ to
$\tilde u,x,y,m_i$.

\noindent
3) For $m_i=(m,m,0,0)$ we recover the curve \mmzz.

Since for $m_i=(m,m,0,0)$ we have $R\not= 0$,  $T_i=N=0$, the most
general form of the curve consistent with these conditions is
\eqn\mostgeneral{y^2=W_1W_2W_3 + x\sum_iT_if_i +\tilde u \sum_iT_ig_i
+ R\sum_iT_ih_i + pN}
with
\eqn\widef{W_i=x-e_i \tilde u - e_i^2 R}
and $f_i$, $g_i$, $h_i$ and $p$ are functions of $\tau $ to be
determined.

For $m_i=(m,m,0,0)$ the curve has three singularities at $ \tilde u_i
=e_im^2$ with two massless particles in each.  Therefore, the
discriminant of the curve $\Delta$ has three double zeros at these
values of $ \tilde u$.  We now consider the situation with
$m_i=(m+\mu,m-\mu,0,0)$.  According to the discussion in section 10, for
non-zero $\mu$ one of the three singularities should split and the other
two can move but remain double zeroes.  Our weak coupling limit above
identified the singularity at $ \tilde u_1$ with the one at which
elementary quarks are massless.  Therefore, this one should split.  We
will now determine some of the coefficients by demanding that the zero
at $ \tilde u_2$ moves but remains a double zero.

Since the curve is holomorphic in $\mu^2$, so is the discriminant
$\Delta$.  For $|\mu| \ll |m|$ its double zero at $ \tilde u_2$ starts
moving at order $\mu^2$ to $ \tilde u_2=e_2m^2+\mu^2\lambda$ .  For it
to remain a double zero, the order $\mu^2$ term in $\Delta( \tilde
u=e_2m^2+\mu^2\lambda)$ should vanish.  It is straightforward to
calculate this term.  It is proportional to
\eqn\consc{-e_1e_3(f_2+f_3-2f_1)+e_2(g_2+g_3-2g_1) +
(h_2+h_3-2h_1) +p.}
We get one constraint by setting \consc\ to zero.  Repeating this at
$\tilde u_3$ we get a similar equation with the subscripts 2 and 3
interchanged.  Four more equations (related by other permutations of the
subscripts) are obtained by studying the cases $m_i=(m,m,\mu,\mu)$, and
$m_i=(m,m,\mu,-\mu)$.  To organize the equations, we break the symmetry
between $ \tilde u_2$ and $ \tilde u_3$ by deciding that in the first of
these cases the zero at $ \tilde u_2$ splits and in the other the zero
at $ \tilde u_3$ splits.  The opposite choice leads to similar results
with $e_2$ and $e_3$ interchanged.  Using these six equations we
determine the ten unknowns $f_i,g_i,h_i,p$ in terms of four unknowns
$F,G,H,A$:
\eqn\solveeq{\eqalign{
f_1&=A(e_2-e_3)+F\cr f_2&=A(e_3-e_1)+F\cr f_3&=A(e_1-e_2)+F\cr
g_i&=-f_ie_i+G\cr h_i&=-f_ie_i^2+H\cr
p&=-(e_1-e_2)(e_2-e_3)(e_3-e_1)A.\cr}}
Since $\sum T_i=0$, the values of $F,G,H$ do not affect the curve and
hence can be set to zero.

In order to determine the unknown function $A(\tau)$ we consider the
$\mu$ dependence more fully.  One approach would be to study higher
order terms in the expansion around $\mu=0$, and require that the double
zeroes of $\Delta$ remain double zeroes.  Instead, we examine the curve
for $\mu=m$.  For this value there are three massless quarks, and we
expect the two unsplit double zeros to merge, giving $\Delta$ a fourth
order zero (which we have previously encountered in the $N_f=3$ theory).
$\Delta$ has this fourth order zero if and only if
\eqn\solveforf{A=(e_1-e_2)(e_2-e_3)(e_3-e_1)}
leading to the curve
\eqn\mostgeneralf{y^2=W_1W_2W_3
+ A\left(W_1T_1(e_2-e_3)+W_2T_2(e_3-e_1)+W_3T_3(e_1-e_2) \right)
- A^2 N}
again with
\eqn\widefa{\eqalign{
&W_i=x-e_i \tilde u - e_i^2 R \cr
&A=(e_1-e_2)(e_2-e_3)(e_3-e_1)\cr
&R={1\over 2} \sum_i  m_i^2 \cr
&T_1={1 \over 12}\sum _{i>j}m_i^2m_j^2 - {1\over 24}\sum_i m_i^4 \cr
&T_2=-{1 \over 2}\prod_i m_i - {1 \over 24}\sum _{i>j}m_i^2m_j^2 +
{1\over 48}\sum_i m_i^4 \cr
&T_3={1 \over 2}\prod_i m_i - {1 \over 24}\sum _{i>j}m_i^2m_j^2 +
{1\over 48}\sum_i m_i^4 \cr
&N={3 \over 16}\sum_{i>j>k}m_i^2m_j^2m_k^2 -
{1 \over 96}\sum_{i \not= j} m_i^2m_j^4 + {1 \over 96}\sum_i m_i^6 .
\cr}}

We see that our final answer is modular invariant. To be precise, full
$SL(2,{\bf Z})$ invariance, which permutes the $e_i$, is a symmetry if
combined with ${\rm Spin}(8)$ triality, which permutes the $T_i$.  This
is a strong indication that the full theory is dual in the way described
in sections 6 and 10.

\bigskip
\noindent
{\it Weak coupling and scaling limits}

As for $N=4$, the theory has infinitely many weakly coupled limits
related by $SL(2,{\bf Z})$.  The obvious one is $\tau \rightarrow i
\infty$; others are at $\tau = p/q$ with $p/q$ rational.  The weakly
coupled variables are different in the various limits (the monopoles in
one limit are the quarks in another  limit).  Correspondingly,
$u=\langle \Tr \phi^2 \rangle$ is different in the different scaling
theories.  Let us focus on $\tau \rightarrow i \infty$.  We define $u$
and shift $x$ in a way similar to \shiftxu
\eqn\xushift{\eqalign{
u&= \tilde u + \half e_1 R\cr
x&\rightarrow x-\half e_1 u+ \half e_1^2 R.\cr}}
Substituting \xushift\ in \mostgeneralf\ we find
\eqn\nonsyma{\eqalign{
y^2&=(x^2- c_2^2 u^2) (x- c_1 u) -
c_2^2(x-c_1 u)^2 \sum_i m_i^2 - c_2^2 (c_1^2 -c_2^2) (x- c_1 u )
\sum_{i>j} m_i^2 m_j^2 \cr
& +2 c_2( c_1^2 - c_2^2)(c_1 x - c_2^2 u) m_1m_2m_3m_4 -
c_2^2 (c_1^2 -c_2^2 )^2 \sum_{i>j>k} m_i^2 m_j^2 m_k^2 \cr}}
where, as before, $c_1 = {3 \over 2 }e_1$ and $c_2=\half( e_3-e_2)$.

We can now analyze the renormalization group flow {}from $N_f=4$ to
$N_f<4$.  In sections 12-14 we have already verified the flows {}from
$N_f=3$ to $N_f=0,1,2$, so it is sufficient here to consider the flow
{}from $N_f=4$ to the massive $N_f=3$ theory.  To do this, we take the
limit $\tau\to i\infty$, $m_4\to\infty$, keeping fixed $m_1,m_2,m_3$ and
\eqn\murtag{\Lambda_3=64 q^{1/2}m_4.}
(Recall that $q^{1/2}$ is the one instanton factor.  The reason for the
factor of 64 is that in section 14 we took $\Lambda_2{}^2=\Lambda_3m_3$
but the above discussion of \mmzz\ implies that the flow {}from $N_f=4$
to massless $N_f=2$ gives $\Lambda_2{}^2=64 q^{1/2}m_3m_4$. Thus the 64
in \murtag\ is needed to agree with our previous definition of
$\Lambda_3$.)  Taking this limit using $c_1 \approx 1$ and $c_2 \approx
8q^\half$ leads to
\eqn\nfthsff{\eqalign{
y^2=&x^2 (x-u)- {1 \over 64} \Lambda_3{}^2(x-u)^2 -{1 \over 64}
(m_1{}^2+ m_2{}^2+m_3{}^2) \Lambda_3{}^2 (x-u)  \cr
&+ {1 \over 4} m_1m_2m_3\Lambda_3 x -{1 \over 64} (m_1{}^2m_2{}^2+
m_2{}^2m_3{}^2 +m_1{}^2m_3{}^2)\Lambda_3{}^2.}}
which is the same as the massive $N_f=3$ curve \nfthsf.

As for the $N=4$ theory, scaling limits around other weakly coupled
points lead to equivalent theories in terms of dual degrees of freedom.

\newsec{The Theory Of The Residues}

One important difference between the mathematical structure of the $N=2$
theories with matter considered in this paper and the ``pure gauge
theory'' studied in \paperi\ is that in the presence of matter the
monodromies do not simply transform $(a_D,a)$ linearly, by $SL(2,{\bf
Z})$ transformations; $a$ and $a_D$ also pick up additive constants
under monodromy.  As we explained in section 15, these constants can be
detected as the residues of the differential form $\lambda$.  Since the
jumps in $a$ or $a_D$ are integral linear combinations of $m_i/\sqrt 2$
(with $m_i$ the bare masses) and are $2\pi i$ times the residues of
$\lambda$, the residues of $\lambda$ should be of the form
\eqn\reslam{{\rm Res}\,\lambda=\sum_i{n_im_i\over 2\pi
i\sqrt 2},~~{\rm with}~~ n_i\in {\bf Z}.}

We have not so far verified or exploited this condition in full.  It was
verified in section 15 for the massive $N_f=2$ theory and exploited only
in a very limited way in section 16 for $N=4$ and $N_f=4$.  The reason
that we have not yet used the full force of the residue condition is
that in fact, except in special cases in which $\lambda$ can be found by
inspection, implementing this condition requires a fairly elaborate
machinery.  This machinery will be developed in the present section and
used to give a new derivation of the curves for $N_f=4$ and $N=4$ (the
others can be obtained, of course, by renormalization group flow).
Since the new derivation does in fact give results that agree with what
we obtained previously, this will also show that the previously obtained
curves do have residues of the right form.

In general, in this paper up to the present point we have followed a
scenic route, starting with simple cases (the $N_f=0$ theory), gradually
adding and understanding new ingredients, climbing upstream to larger
$N_f$ (against the renormalization group current that flows to smaller
$N_f$), and finally understanding what {}from this point of view are the
most challenging cases of $N_f=4$ and $N=4$.  The analysis of the
residues presented in the present section has the opposite flavor; after
building up the necessary apparatus, the machinery is easily applied
directly to $N_f=4$ and $N=4$, and gives the answer after a very short
calculation.  This approach certainly provides additional insight into
some questions like why triality holds for $N_f=4$ and may also be
useful in generalizing to gauge groups other than $SU(2)$.

\subsec{The Meaning Of The Residues}

The $N_f=4$ theory is controlled by a curve $y^2=F(x,u,m_i,\tau)$
and a differential form $\lambda$ obeying
\eqn\pnson{{d\lambda\over du}=\omega+{\rm exact~form~in }~x}
with
\eqn\immop{
\omega={\sqrt 2\over 8\pi}{dx\over y}.}
For $N=4$ the structure is the same, except that $8\pi$ is replaced
by $4\pi$.  $F$ should be such that the residues of $\lambda$
are linear in the quark bare masses.  This is a severe restriction
on $F$; we will see that it determines $F$ uniquely (up to the usual
changes of variables) independently of most of the arguments that we have
used up to this point.

As a preliminary, let us write \pnson\ in a more symmetrical form.
If $\lambda=dx \,\,a(x,u)$, then \pnson\ means
\eqn\qnson{{\sqrt 2\over 8\pi}{dx\over y}=dx\,\,\, {\partial
a\over\partial u}
+dx\,\,{\partial\over\partial x}f(x,u);}
the arbitrary total $x$-derivative $dx\,\,\partial f/\partial x$ is
allowed because it does not contribute to the periods.  \qnson\ can be
understood much better if written symmetrically in $x$ and $u$.
Henceforth, instead of using a one-form $\omega = (\sqrt 2/8\pi)\cdot
dx/y$, we will use a two-form
\eqn\defomega{\omega={\sqrt 2\over 8\pi}{dx\,\,\,du\over y}.}
Similarly, we combine the functions $a,f$ appearing in \qnson\ into a
one-form $\lambda=-a(x,u)dx+f(x,u)du$.  The change in notation for
$\omega$ and $\lambda$ should cause no confusion.  Then equation \qnson\
can be more elegantly written
\eqn\bnson{\omega = d\lambda.}

The meaning of the problem of finding $\lambda$ can now be stated.  Let
$X$ be the (noncompact) complex surface defined by the equation
$y^2=F(x,u)$ (we suppress the parameters $m_i$ and $\tau$).  Being
closed, $\omega$ defines an element $[\omega]\in H^2(X,{\bf C})$.  A
smooth differential $\lambda$ obeying \bnson\ exists if and only if
$[\omega]=0$.  Moreover, by standard theorems, in the absence of
restrictions on the growth of $\lambda$ at infinity, if $\lambda$ exists
it can be chosen to be holomorphic and of type $(1,0)$.

If on the other hand $[\omega]\not= 0$, then \bnson\ has no smooth, much
less holomorphic, solution.  However, $X$ has the property that if one
throws away a sufficient number of complex curves $C_a$, then
$X'=X-\cup_aC_a$ has $H^2(X',{\bf C})=0$. (The necessary $C_a$ are
explicitly described later.)  So if we restrict to $X'$, the cohomology
class of $\omega$ vanishes and $\lambda$ exists.  $\lambda$ may however
have poles on the $C_a$, perhaps with residues, which we call ${\rm
Res}_{C_a}(\lambda)$.\foot{The residues of $\lambda$ along $C_a$ are
constants, since $d{\rm Res}_{C_a}(\lambda) ={\rm Res}_{C_a}d\lambda
={\rm Res}_{C_a}\omega=0$.} If $\lambda$ does have residues, then
$d\lambda$ contains delta functions, and if one works on $X$ instead of
$X'$, one really has not \bnson\ but
\eqn\pinno{\omega=d\lambda-2\pi i\sum_a{\rm Res}_{C_a}(\lambda)
\,\cdot\,[C_a]}
where $[C_a]$ (which represents the cohomology class
known as the Poincar\'e dual of $C_a$)
is a delta function supported on $C_a$.

In cohomology, \pinno\ simply means
\eqn\ginno{[\omega]=-2\pi i \sum_a
{\rm Res}_{C_a}(\lambda)\cdot [C_a].}
Thus, if we pick the $C_a$ so that the $[C_a]$ are  a basis
of $H^2(X,{\bf C})$, then the residues ${\rm Res}_{C_a}(\lambda)$
are simply the coefficients of the expansion of $[\omega]$ in
terms of the $[C_a]$.  To find the residues we need not actually
find $\lambda$; it suffices to understand the cohomology class of
$\omega$ by any method that may be available.

For instance, if $X$ were compact, we could proceed as follows.
First compute the intersection matrix
\eqn\intma{M_{ab}=\#(C_a\cdot C_b)}
(that is, the number of intersection points of $C_a$ and $C_b$,
after perhaps perturbing the $C_a$ so that they intersect generically).
This is an invertible matrix.  Second, calculate the periods
\eqn\mingtam{c_a=\int_{C_a}\omega.}
Then
\eqn\untam{[\omega]= \sum_{a,b}c_aM^{-1}{}_{ab}[C_b].}
Comparing to \ginno, we get
\eqn\pintma{{\rm Res}_{C_a}(\lambda)=-{1\over 2\pi i}\sum_b
M^{-1}_{ab}c_b.}

We will eventually follow that strategy after compactifying $X$ and
modifying $\omega$ so as to have no pole at infinity.  With this in
view, we will somewhat loosely and prematurely call the $c_b$ the
``periods'' of $\omega$.

\subsec{The Cohomology Of $X$}

It will be useful to know something
about the cohomology of the complex manifold $X$ described
by the equation
\eqn\umoc{y^2=F(x,u)=(x-e_1u)(x-e_2u)(x-e_3u)+{\rm lower\,\,
order\,\,terms}.}
It is helpful to compactify $X$, which will be needed anyway to do the
calculation just mentioned.  We do this by introducing another variable
$v$ and making the equation homogeneous.  First we extend $F$ to a
polynomial $F(x,u,v)$ homogeneous of degree 3 (such that $F(x,u,1)$ is
the original $F$).  We could now consider the homogeneous equation
$vy^2=F(x,u,v)$, with $x,u,v,y$ all of degree 1.  However, things work
out more easily if we instead take $x,u,v,y$ to be of degree $1,1,1,2$;
so we study the homogenous equation
\eqn\yumoc{y^2=v \,F(x,u,v).}

It is helpful to first look at a more general equation
\eqn\xumoc{y^2=G(x,u,v)}
with a generic, irreducible $G$ homogeneous of degree 4.  The variety
$Z$ defined by this equation has the following properties.  There is a
${\bf Z}_2$ symmetry $\alpha:y\to -y$.  The differential form of
interest, $\omega= dx\,du/y$, is odd under the symmetry, so we are
mainly interested in the part of the cohomology of $Z$ that is odd.
Using methods familiar to physicists {}from the study of Calabi-Yau
manifolds,
\foot{See \ref\hubsch{T. Hubsch, {\it Calabi-Yau Manifolds:
A Bestiary For Physicists} (World-Scientific, 1992)} for an introduction
to the requisite methods.  $Z$ has $b_1=b_3=0$ by the Lefschetz
hyperplane theorem.  Its Euler characteristic is 10 so $b_2=8$.  The
$\alpha$-invariant part of the cohomology of $Z$ can be computed {}from
$\alpha$-invariant differential forms, so coincides with the cohomology
of ${\bf CP}^2$.  Hence, the part of $H^2(Z)$ that is even under
$\alpha$ is one dimensional, and the odd part is seven dimensional.} one
can show that the odd part of $H^2(Z)$ is seven dimensional.

The situation is somewhat different for the special quartic polynomial
$G(x,u,v)=v\,F(x,u,v)$, since the manifold $\bar X$ defined by \umoc\
has (for generic $F$) three singularities, at $y=v=x-e_iu=0$, for
$i=1,2,$ or $3$.  To understand the structure of the singularities, set
$u=1$ by scaling, and set $w=(x-e_iu)(e_1-e_2)(e_1-e_3)$.  The behavior
near the singularity is then
\eqn\nearsing{y^2= vw+{\rm higher\,\,order\,\,terms}.}
This is known as an ${\bf A}_1$ singularity.  It is actually
a ${\bf Z}_2$ orbifold singularity,\foot{This was noted in
section 3 where we encountered the same type of singularity in
a different way.}  which is a harmless kind of singularity for our
purposes; for instance, there is no special subtlety in describing
$H^2(\bar X,{\bf C})$ by differential forms.

However, when a complex surface  develops an ${\bf A}_1$ singularity,
the second Betti number drops by 1.  Since the odd part of $H^2(Z)$ is
seven dimensional, and $\bar X$ is a specialization of $Z$ to a case
with three ${\bf A}_1$ singularities, the odd part of $H^2(\bar X)$ is
four dimensional.\foot{It is the odd part whose dimension drops when the
singularity develops, since the even part, which is generated by the
Kahler class, certainly survives.} As we will see, this occurrence of
the number four is no coincidence: it is related to the fact that
conformal invariance with matter hyperdoublets requires $N_f=4$.

\bigskip
\noindent{\it Extending $\omega$}

Let us discuss the behavior of $\omega$ under compactification.
The homogeneous version of $\omega$ is
\eqn\impodo{\omega={v \,dx\,du+x\,du\,dv+u\,dv\,dx\over vy}.}
The point of this formula is that (i) it reduces to the old one
if we set $v=1$; (ii) it is invariant under scaling of the homogeneous
coordinates
\eqn\scallo{\eqalign{\delta x&= \epsilon x\cr
\delta u&= \epsilon u\cr
\delta v&= \epsilon v\cr
\delta y&= 2\epsilon y;\cr}}
(iii) it vanishes if contracted with the vector field in \scallo\ so it
can be interpreted as a pull-back {}from the weighted projective
space of $(x,u,v,y)$.

``Infinity'' in $\bar X$ is just the region with $v=0$ (which one misses
if one sets $v=1$).  It is evident in \impodo\ that $\omega$ has a pole
at $v=0$.  The equation $y^2=v\,F(x,u,v)$ shows that near $v=0$ on the
double cover, $v\sim y^2$, so $y$ is the good coordinate near $v=0$.  As
$dv/v\sim 2\,dy/y$, $\omega$ looks near $y=0$, in, say, a coordinate
system with $u=1$, like
\eqn\underro{\omega\sim {dy\over y^2} \,dx.}
Thus, there is a pole at $y=0$, but the residue vanishes.  Because the
residue vanishes, $\omega$ can be interpreted as a cohomology class not
just on $X$ but on $\bar X$.

One could modify $\omega$ near infinity, preserving the fact that it is
closed, but losing the fact that it is holomorphic and of type $(2,0)$,
so that $\omega$ extends over infinity as a closed two-form and so
defines a cohomology class of $\bar X$.  The ability to so interpret
$\omega$ makes it possible to calculate using intersections and periods,
as we will.

\subsec{Finding the Curves}

Now we would like to find a suitable set of curves $C_a$ on which
$\lambda$ will have poles.  In this section we set $v=1$ and work on the
uncompactified manifold $X$ given by $y^2=F(x,u)$.

To guess what the $C_a$  may be, let us look back to some of the models
treated earlier, for instance the massive $N_f=2$ theory.  In equation
\lambdo, we found an explicit formula for $\lambda$ in this theory;
it had poles at
\eqn\refpoles{\eqalign{x& = \pm {1\over 8}\Lambda_2{}^2 \cr
                       y& = \pm {i\over 8}\Lambda_2{}^2(m_1\pm
		       m_2).\cr}}

This can be interpreted as follows.  The equation $y^2=F(x,u)$ describes
a double cover of the $x-u$ plane.  The equation
\eqn\welwo{x=\pm \Lambda_2{}^2/8} describes
(for a given choice of the sign) a line in that plane.  A generic line
$L$ in the $x-u$ plane would be described by
\eqn\reslin{x=\beta u +\theta,}
for some $\beta$, $\theta$.  The double cover $y^2=F(x,u)$, restricted
to the line $L$, would be given by an equation
\eqn\redlun{y^2=g(u)}
where generically $g(u)$ is not the square of a polynomial.  That being
so, the double cover of $L$ is generically an irreducible complex curve,
which is obviously invariant under $\alpha:y\to -y$.  That is not what
we want, because $\omega$ is {\it odd} under $\alpha$; we need cycles on
$X$ that are odd.  For the special case that the parameters in \reslin\
are as in \welwo, the double cover \redlun\ reduces to $y^2=-(m_1\pm
m_2)^2 \Lambda_2{}^4/64$. Here the right hand side is the square of a
polynomial (in fact a constant), and so the double cover consists of two
branches, with $y=i(m_1\pm m_2)\Lambda_2{}^2/8$ or $y=-i(m_1\pm m_2)
\Lambda_2{}^2/8$.  If we call the two branches $D_{+,\pm}$ and
$D{-,\pm}$ (the second subscript denotes the sign for $x$ in \welwo, and
the first is the sign of $y$), then the differences
$C_\pm=D_{+,\pm}-D_{-,\pm}$ are divisors that are odd under $\alpha$.
The explicit determination of the residue of $\lambda $ in section 15
amounted to expressing the cohomology class $[\omega]$ as a linear
combination of $C_+$ and $C_-$.

We can imitate this in the case at hand
\eqn\casea{y^2=(x-e_1u)(x-e_2u)(x-e_3u) +F_1(x,u) }
(where $F_1(x,u)$ is of degree $\leq 2$ in $x$ and $u$).  We introduce
the line $L$ of \reslin, and consider again the double cover of $L$
deduced {}from \casea.  This is described by an equation $y^2=g(u)$
where for generic $u$, $g(u)$ is cubic.  If $g$ is cubic, it cannot be a
square, so $y^2=g(u)$ describes an irreducible cover of $L$, necessarily
invariant under $y\leftrightarrow -y$.

Therefore, we must adjust the coefficients in \reslin\ to kill the cubic
term in $g(u)$.  This requires that we set $\beta=e_i$ for some $i$.
Then $g$ is quadratic in $u$ and the equation for the double cover is of
the general form
\eqn\pilo{y^2=Au^2+Bu+C.}
The polynomial on the right hand side of \pilo\ is a square when
and only when the discriminant vanishes,
\eqn\nilo{0=\Delta = B^2-4AC,}
When $\Delta=0$, \pilo\ can be written in the form $y^2=A(u-t)^2$ for
some $t$, and its solutions consist of two branches $y=\pm \sqrt
A(u-t)$.  If we call these branches $D_+$ and $D_-$, we get a divisor
$C=D_+-D_-$ that is odd under $\alpha$.

Now actually, $A$ is linear in $\theta$, $B$ is quadratic, and $C$ is
cubic.  So the discriminant is quartic in $\theta$ and has four zeroes
$\theta_a$.  For each we get a line $L_a$ whose double cover has two
components $D_{\pm, a}$. So we get four odd divisors
$C_a=D_{+,a}-D_{-,a}$.  As we discussed in the last subsection, four is
the dimension of the odd part of the cohomology of $X$, so there are as
many $C_a$ as we would need for a basis of that odd part.  We will see
that they are a basis.

But actually, the above construction began by setting $\beta=e_i$ for
some fixed $i$.  We could have carried out the above steps for any
$i=1,2$, or 3.  So making the dependence on $i$ explicit, we have
divisors $D^{(i)}{}_{\pm, a}$, and $\alpha$-odd divisors
$C^{(i)}{}_{a}$.  Of course, ${\rm Spin}(8)$ triality permutes the $(i)$
superscript.

\bigskip
\noindent{\it Intersection Pairings}

We would like to prove that the $C^{(i)}{}_a$ of fixed $i$ do furnish a
basis of the odd part of the cohomology.  Since there are four of them,
it suffices to prove that they are linearly independent; for this
purpose, it is enough to work on $\bar X$ (where the intersection
pairings are topological invariants) and prove that the matrix of
intersection pairings is non-degenerate.

So we have to calculate $C^{(i)}{}_a\cap C^{(i)}{}_b$.  For $a\not = b$,
this vanishes, for the following reason.  The divisors $D^{(i)}{}_{\pm ,
a}$ are given by
\eqn\twoeq{\eqalign{x & = e_i u+\theta_a v\cr
                    y & = \pm \sqrt A (u-t_av)v.\cr}}
(Since we are working on $\bar X$, we have restored $v$.  In doing this,
we made the equations homogeneous, remembering that $x,u,v,y$
have degree $1,1,1,2$.) Any two
of these curves (for distinct $a$ and $b$, but regardless of the
independent choices of $\pm$ signs) meet precisely at the ${\bf Z}_2$
orbifold point $x-e_1u=y=v=0$, so the pairings of these curves with
distinct $a$ and $b$ are
\eqn\somno{D^{(i)}{}_{\pm ,a}\cap D^{(i)}{}_{\pm ,b}= {1\over 2}}
where the two $\pm $ signs are chosen independently.
(The $1/2$ comes because an intersection at the orbifold point
is counted with weight $1/2$.)
Recalling that $C^{(i)}{}_a=D^{(i)}{}_{+,a}
-D^{(i)}{}_{-,a}$
we get  $C^{(i)}{}_a\cap C^{(i)}{}_b=0$ for $a\not= b$.
However, $C^{(i)}{}_a\cap C^{(i)}{}_a=-4$.\foot{The value is obviously
independent of $i$ and $a$.   That it is $-4$ requires a slightly
more detailed analysis that we omit.  The value $-4$ in fact
follows {}from our determination below that
$C^{(i)}{}_a\cap C^{(j)}{}_b=\pm 2$
for $i\not=j$.}  Putting these results together,
\eqn\basiso{C^{(i)}{}_a\cap C^{(i)}{}_b=-4\delta_{ab}.}
This matrix is in particular nondegenerate, showing
that for fixed $i$, the $C^{(i)}{}_a$ give a basis of the relevant
piece of the cohomology.

\bigskip
\noindent{\it Triality}

This result may seem esoteric, but by extending it a bit, we
will see ${\rm Spin}(8)$ triality at work in the classical geometry.
To this end, we want to calculate the intersections $C^{(i)}{}_a\cap
C^{(j)}{}_b$
for $i\not= j$.  We claim that
\eqn\refo{C^{(i)}{}_a\cap C^{(j)}{}_b=\pm 2,~~{\rm for}~i\not= j.}
Granted this, let us see how the statement is connected with
${\rm Spin}(8)$ triality.

Since the $C^{(i)}{}_a$ of fixed $i$ are a basis of the odd part of the
cohomology, the $C^{(j)}{}_a$ can be expanded as linear combinations of
the $C^{(i)}{}_a$.  In fact, we will expand everything in terms of the
$C^{(1)}{}_a$.  Comparing \refo\ to \basiso, we see that the expansion
coefficients are all $\pm 1/2$.  In the above we adopted no particular
convention as to what was $D^{(i)}{}_{+,a}$ and what was
$D^{(i)}{}_{-,a}$, so we have not fixed the signs of the $C^{(i)}{}_a$.
Picking an arbitrary sign for $C^{(2)}{}_1$, we can fix the signs of all
the $C^{(1)}{}_a$ by requiring that the expansion of $C^{(2)}{}_1$ has
all plus signs:
\eqn\nonon{C^{(2)}{}_1={1\over 2}\left(C^{(1)}{}_1+C^{(1)}{}_2
+C^{(1)}{}_3+C^{(1)}{}_4\right).}
Now, we consider $C^{(2)}{}_a$ with $a>1$.
We can carry out a relabeling of the $a$ index, since its meaning has
not been fixed in any special way.
The $C^{(2)}{}_a, \,\,a>1$ are orthogonal
to $C^{(2)}{}_1$, and have expansion coefficients $\pm 1/2$ in terms
of the $C^{(1)}{}_a$.  These conditions are enough to ensure that,
up to permutations on the $a$ index and changes in sign of
$C^{(2)}{}_a$,
\eqn\ilmo{\eqalign{C^{(2)}{}_2 & = {1\over 2}\left(C^{(1)}{}_1
+C^{(1)}{}_2-C^{(1)}{}_3-C^{(1)}{}_4\right)\cr
C^{(2)}{}_3 & = {1\over 2}\left(C^{(1)}{}_1
-C^{(1)}{}_2+C^{(1)}{}_3-C^{(1)}{}_4\right)\cr
C^{(2)}{}_4 & = {1\over 2}\left(C^{(1)}{}_1
-C^{(1)}{}_2-C^{(1)}{}_3+C^{(1)}{}_4\right).\cr}}

It remains to consider the $C^{(3)}{}_a$.  They must have expansion
coefficients of $\pm 1/2$ in terms of either the $C^{(1)}{}_a$
or the $C^{(2)}{}_a$.  This ensures that, up to permutations and
sign changes of the $C^{(3)}{}_a$, they are given by
\eqn\milmo{\eqalign{C^{(3)}{}_1 & = {1\over 2}\left(C^{(1)}{}_1
+C^{(1)}{}_2+C^{(1)}{}_3-C^{(1)}{}_4\right)\cr
C^{(3)}{}_2 & = {1\over 2}\left(C^{(1)}{}_1
+C^{(1)}{}_2-C^{(1)}{}_3+C^{(1)}{}_4\right)\cr
C^{(3)}{}_3 & = {1\over 2}\left(C^{(1)}{}_1
-C^{(1)}{}_2+C^{(1)}{}_3+C^{(1)}{}_4\right)\cr
C^{(3)}{}_4 & = {1\over 2}\left(-C^{(1)}{}_1
+C^{(1)}{}_2+C^{(1)}{}_3+C^{(1)}{}_4\right).\cr}}

Let us compare this to ${\rm Spin}(8)$ triality.  The $N_f=4$ theory has
four masses $m_1,m_2,m_3,m_4$ which are the ``eigenvalues'' of a mass
matrix that is in the adjoint representation of ${\rm Spin}(8)$.  So
they transform under triality like the weights of ${\rm Spin}(8)$.
Under the exchange of the vector with the positive chirality spinor
(leaving the negative chirality spinor fixed) the masses transform by a
formula that was already presented in \triagent:
\eqn\helpme{\eqalign{
m'{}_1 & = {1\over 2}\left(m_1+m_2+m_3+m_4\right) \cr
m'{}_2 & = {1\over 2}\left(m_1+m_2-m_3-m_4\right) \cr
m'{}_3 & = {1\over 2}\left(m_1-m_2+m_3-m_4\right) \cr
m'{}_4 & = {1\over 2}\left(m_1-m_2-m_3+m_4\right) .\cr }}
Under the exchange of the vector with the other spinor, the masses
transform to
\eqn\nelpme{\eqalign{
m''{}_1 & = {1\over 2}\left(m_1+m_2+m_3-m_4\right) \cr
m''{}_2 & = {1\over 2}\left(m_1+m_2-m_3+m_4\right) \cr
m''{}_3 & = {1\over 2}\left(m_1-m_2+m_3+m_4\right) \cr
m''{}_4 & = {1\over 2}\left(-m_1+m_2+m_3+m_4\right). \cr }}
These formulas have precisely the same structure as \nonon, \ilmo,
and \milmo!

This means the following.  Suppose that we introduce
masses $m_i,\,i=1\dots 4$ in the $N_f=4$ theory and determine the
equation $y^2=F(x,u,m_i,\tau)$ by requiring that the ``periods'' of
$\omega=  (\sqrt 2/8\pi)\,dx\,\,du/y$
are proportional to the $m_i$ in the sense that
\eqn\pilfo{[\omega]=-{1\over \sqrt 2}\sum_a m_a[C^{(1)}{}_a].}
The above equations will then ensure that it is also true that
\eqn\upilfo{[\omega]=-{1\over \sqrt 2}\sum_a m'_a[C^{(2)}{}_a]
=-{1\over \sqrt 2}\sum_a m''_a[C^{(3)}{}_a].}
In brief, $\omega$ is triality invariant; this is achieved
because triality permutes $C^{(1)}{}_a, C^{(2)}{}_a$, and $C^{(3)}{}_a$
and also $m_a,m'{}_a , $ and $m''{}_a$ in the same way.
In this sense, the
structure of the complex manifold $X$ makes triality invariance
of the physics possible.

Actually, the formulas \helpme\ and \nelpme\ were not quite uniquely
determined, since triality is only uniquely defined up to a Weyl
transformation.  The Weyl group acts on the $m$'s by permutations and
sign changes, so the arbitrariness that was fixed to write the triality
transformation as in \helpme\ and \nelpme\ has the same structure as the
arbitrariness that was fixed in writing the transformations of the
$C$'s.

\upilfo\ could similarly be modified by a Weyl transformation
(permutations and pairwise sign changes of the $m$'s) without spoiling
triality.  This would not be an essential change.  It would also be
possible to multiply the right hand side by a constant.  Since we have
not proved that the $C^{(i)}{}_a$ are an {\it integral} basis for the
cohomology, and also because we have not analyzed the monodromies of $a$
and $a_D$ precisely enough, we cannot assert now that \upilfo\ is
correctly normalized.  (This may be the reason for a factor of two that
will appear later.)

It remains to justify \refo.  Two lines $L:x=e_iu+\theta_{i\,a}$ and
$L':x=e_ju+\theta_{j\,b}$ with $i\not= j$ are not parallel and intersect
at a point $P$ on the $x-u$ plane.  On the double cover $y^2=F(x,u)$
there are two points $P_\pm$ lying above $P$.  Each
double cover $D^{(i)}{}_{\pm, a}$ of $L$ and each double cover
$D^{(j)}{}_{\pm,b}$ of $L'$ contains either $P_+$ or $P_-$.
If for instance, we fix conventions so that $P_+$ is contained
in $D^{(i)}{}_{+,a}$ and $D^{(j)}{}_{+,b}$ and $P_-$ in the others,
then the $D^{(i)}{}_{+,a}\cap D^{(j)}{}_{+,b}=D^{(i)}
{}_{-,a}\cap D^{(j)}{}_{-,b}=1$,
while the other intersections are zero.  So $C^{(i)}{}_a\cap
C^{(j)}{}_b=2$.
With other conventions, we could get $C^{(i)}{}_a\cap C^{(j)}{}_b=-2$.
So the intersections are $\pm 2$, as claimed in \refo.
(There is no way to pick conventions so that the intersection is $+2$
for all $i,j,a,b$.)

\subsec{Determination Of The Equation For $N_f=4$}

We have finally assembled the tools to determine the precise function
$F$ in our equations $y^2=F(x,u)$.  First we do this for the theory with
$N_f=4$.  We will determine the expansion of $\omega$ in terms of the
$C^{(i)}{}_a$ for some given $i$; we may as well pick $i=1$.  (The
expansions in terms of the $C^{(j)}{}_a$ for $j\not= 1$ are then
determined by the above triality formulas.)  By requiring that $\omega
=-{1\over \sqrt 2}\sum_a m_a[C^{(1)}{}_a]$, $F$ will be determined.

In coordinates with $v=1$, the cubic part of $F$ is
$(x-e_1u)(x-e_2u)(x-e_3u)$.  Since $e_1$ is in any case singled
out by the decision to expand in the $C^{(1)}{}_a$, it is convenient
to make the change of variables $x-e_1u\to x$.  The lines $x=e_1u+\theta$
are now described simply by
\eqn\innok{x=\theta.}
Also, if we set
\eqn\kinno{\alpha=e_2-e_1,\,\,\,\,\,\,\,\,\beta=e_3-e_1,}
the cubic part of $F$ is $x(x-\alpha u)(x-\beta u)$.  The quadratic
part of $F$ is a linear combination of $x^2,$ $xu$, and $u^2$.
We can eliminate any two of the three by shifting $x$ and $u$ by
constants.  We choose to set the coefficients of $xu$ and $u^2$ to zero.
The general structure is then
\eqn\fis{{F}=x(x-\alpha u)(x-\beta u) +ax^2+bx+cu +d.}

Restricted to the line  \innok, one has
\eqn\gis{{F}=\theta\alpha\beta u^2+(c-\theta^2(\alpha+\beta))u
+\theta^3+a\theta^2+b\theta+d.}
If we write the right hand side as $Au^2+Bu+C$, then the discriminant
$\Delta =B^2-4AC$ is
\eqn\hinnok{\Delta=\theta^4(\alpha-\beta)^2-4a\alpha\beta\theta^3
+(-2c(\alpha+\beta)-4b\alpha\beta)\theta^2-4\alpha\beta d\theta+c^2.}
$\Delta$ has four roots $\theta_a$, $a=1\dots 4$.
For $\theta=\theta_a$, the
equation $y^2=F$ takes the form
\eqn\binnok{y^2=\theta_a\alpha\beta(u-u_0)^2}
for some $u_0$.  Restoring the $v$ dependence, in homogenous
coordinates, the equations for the divisors $D^{(1)}{}_{\pm,a}$
take the form
\eqn\jinno{\eqalign{x& =\theta v\cr
                    y&=\pm (\theta_a\alpha\beta)^{1/2}v(u-u_0v).\cr}}

\bigskip
\noindent{\it Computation Of Periods}

Now we want to expand $[\omega]$ in terms of these divisors.
The main point is to study the behavior of $\omega$ near $\infty$
(that is, $v=0$), where we can set $u=1$.
The equation $y^2=v F(x,u,v)$ becomes in this coordinate system
\eqn\yunny{{y^2}= vx(x-\alpha)(x-\beta)+av^2x^2+bv^3x+cv^3 +dv^4.}
So near  the singularity at $x=v=0$, we get
\eqn\unny{2{y\cdot dy}=v\alpha\beta\,dx+\dots}
where the $\dots$ are terms proportional to $dv$ (which will drop out
when we compute $\omega$) or $v^2,vx$ (which are negligible near
$v=x=0$).  Inserting \unny\ in $\omega=(\sqrt 2/8\pi)\,\,dv\,dx/vy$, we
get
\eqn\punny{\omega\sim {\sqrt 2\over 4\pi}{dv\,dy\over v^2\alpha\beta}.}
This means that near $v=0$ we can write $\omega = d\lambda$
with
\eqn\kuku{\lambda=-{\sqrt 2\over 4\pi}{y\,\,dv\over v^2\alpha\beta}.}

Let $X_\epsilon$ be the region with $|v|<\epsilon$.  We want to modify
$\omega$ inside $X_\epsilon$ to eliminate the pole at $v=0$, while
preserving the fact that $\omega$ is closed.  The $\omega$ so modified
extends over a neighborhood of the singularity at $x=v=0$ in the
compactification of $X$.  ($\omega$ could be extended over all of $\bar
X$, but that will not be necessary.)

To make the modification, let $\lambda'=f(v)\lambda$, with $f$ a
smooth function such that $f(v)=1$ for $|v|>\epsilon$ and $f\sim
|v|^2$ for $v\to 0$.  Now, leave $\omega$ unchanged outside of
$X_\epsilon$, but inside $X_\epsilon$ take
\eqn\nuku{\omega=d\lambda'.}

We now have all the information required to compute the desired
residues, which, according to \mingtam\ and \pintma,
are equivalent to the integrals of $\omega$ over $C^{(1)}{}_a
=D^{(1)}{}_{+,a}-D^{(1)}{}_{-,a}$.  The integral over $C^{(1)}{}_a$
is twice the integral over $D=D^{(1)}{}_{+,a}$, using the
symmetry under $y\leftrightarrow -y$:
\eqn\umingtam{c_a=\int_{C^{(1)}{}_a}\omega=2\int_D\omega.}
Let $D_\epsilon$ be the part of $D$ with $|v|<\epsilon$.
Then
\eqn\omingtam{\int_D\omega=\int_{D_\epsilon}\omega,}
as outside $D_\epsilon$, $\omega$ is of type $(2,0)$.
Inside $D_\epsilon$, $\omega=d\lambda'$, so
\eqn\ymingtam{\eqalign{\int_{D_\epsilon}\omega&=\int_{D_\epsilon}d\lambda'
=\oint_{|v|=\epsilon}\lambda'=\oint_{|v|=\epsilon}\lambda
=- {\sqrt 2\over 4\pi}\oint_{|v|=\epsilon}{y dv\over v^2\alpha\beta}\cr &
=-{\sqrt 2\over 4\pi}
\oint_{|v|=\epsilon}{\sqrt{\theta_a\alpha\beta} \,\,dv\over v\alpha\beta}
=-{ i \over \sqrt 2}\sqrt{\theta_a\over\alpha\beta}.\cr}}
The next to last step uses the
fact that near $v=0$ on $D$,
\eqn\reflo{ y\sim  (\theta_a\alpha\beta)^{1/2}v}
according to \jinno.
So
\eqn\caeq{c_a=-\sqrt 2 i\sqrt{\theta_a\over\alpha\beta}.}

Now we can determine the desired residues {}from \pintma\ -- using the
fact that the matrix $M$ is $M=-4$ according to \basiso.  We get
\eqn\nores{{\rm Res}_{C^{(1)}{}_a}\lambda=-{1\over 4\pi\sqrt 2}
       \sqrt{\theta_a\over\alpha\beta}.}
Here of course $\theta_a$ is any of the roots of the discriminant
\hinnok.

\bigskip
\noindent{\it Final Steps}

On the other hand, we want the residues to be $m_a/2\pi i\sqrt 2$, with
$m_a$ the masses.  So the four zeroes $\theta_a$ of the discriminant
should be $\theta_a=-4\alpha\beta m_a{}^2$.
For the discriminant to have these roots means that \hinnok\ can
be rewritten as follows:
\eqn\umping{\theta^4(\alpha-\beta)^2-4a\alpha\beta\theta^3
+(-2c(\alpha+\beta)-4b\alpha\beta)\theta^2-4\alpha\beta d\theta+c^2
=(\alpha-\beta)^2\prod_{a=1}^4\left(\theta+4\alpha\beta m_a{}^2\right).}
Simply by equating the coefficients of different powers of $\theta$,
we now determine all the unknown quantities:\foot{In solving the
equations, one has to take a square root; a sign change in the square
root is equivalent to a change in sign of one of the four masses.  A
similar and related choice was needed in section 16 when we broke the
symmetry between $e_2$ and $e_3$.}
\eqn\tired{\eqalign{a& =-(\alpha-\beta)^2\sum_am_a{}^2\cr
                    b& = -4(\alpha-\beta)^2\alpha\beta\sum_{a<b}
                          m_a{}^2m_b{}^2 +8\alpha\beta
 (\alpha^2-\beta^2)\prod_{a=1}^4m_a\cr
                    c& = -16
           (\alpha-\beta)\alpha^2\beta^2\prod_{a=1}^4m_a\cr
                    d& = -16(\alpha-\beta)^2\alpha^2\beta^2
                    \sum_{a<b<c}m_a{}^2m_b{}^2m_c{}^2.\cr}}

So we determine finally the equation governing the low energy
behavior of the $N_f=4$ theory:
\eqn\longeq{\eqalign{
{y^2}=&x(x-\alpha u)(x-\beta u)-{(\alpha-\beta)^2}
x^2\sum_am_a{}^2\cr &+x\left( -4(\alpha-\beta)^2\alpha\beta\sum_{a<b}
                          m_a{}^2m_b{}^2+8\alpha\beta
        (\alpha^2-\beta^2)\prod_{a=1}^4m_a\right)\cr &
-16u(\alpha-\beta)\alpha^2\beta^2\prod_{a=1}^4m_a
-16(\alpha-\beta)^2\alpha^2\beta^2
                    \sum_{a<b<c}m_a{}^2m_b{}^2m_c{}^2.\cr}}

In particular, the right hand side is a polynomial in $x,$ $u$, and the
$m_i$, as we assumed in all of our analyses of the various models.  In
fact, \longeq\ can be seen to be equivalent to the result obtained in
section 16 (the manifestly triality invariant expression \mostgeneralf\
and its shifted form \nonsyma), verifying that the result of section 16
is compatible with the residue condition.  To see that, substitute in
\nonsyma\ $x\rightarrow x+c_1u$ and use $c_1={3\over 2}
e_1=-\half(\alpha+\beta)$ and $c_2=\half(e_3-e_2) =\half(\beta-\alpha)$
to derive \longeq.  Actually, to agree with section 16, it is also
necessary to divide $m$ by 2; we do not understand the origin of this
discrepancy, but it may have to do with subtleties in normalizing the
topological computation that were mentioned in the next to last
paragraph of section 17.3.

Most of $SL(2,{\bf Z})$ invariance is obvious in \longeq\ since
$\alpha=e_2-e_1$ and $\beta=e_3-e_1$ are modular forms of weight two for
$\Gamma(2)$.  Actually, the full $SL(2,{\bf Z})$ and triality are
guaranteed, for the following reason.  The condition on the residues of
$\omega$ is consistent with $S$-duality and triality, as we saw in
equation \upilfo. Furthermore, in using that condition to find $F$, the
``boundary'' condition, given by the cubic function
$(x-e_1u)(x-e_2u)(x-e_3u)$, is also $S$-dual and triality-invariant.
Moreover, the condition on $\omega$ led to a unique determination of
$F$, up to the possibility of redefining $x$ and adding a constant to
$u$.  Therefore, up to such transformations, the solution must have all
the symmetries of the boundary conditions and, in particular, triality
and $SL(2,{\bf Z})$.

\subsec{The $N=4$ Theory}

In equation \eqofour, we determined the general structure of the curve
for $N=4$:
\eqn\genstro{y^2=(x-e_1 \tilde u-e_1{}^2f)(x-e_2 \tilde u-e_2{}^2f)
(x-e_3 \tilde u-e_3{}^2f).}
We then showed $f$ to be $\tau$-independent by an indirect argument
involving the residues and showed it to be $m^2/4$ by comparing to the
behavior at infinity.  Our purpose here is to analyze the residues
for this curve more directly and completely.

First we discuss the classical geometry of the situation.  The complex
manifold $X$ given by the above equation has three singularities where
\eqn\sinof{0=y=x-e_i \tilde u-e_i{}^2f=x-e_j \tilde u-e_j{}^2f}
for any distinct $i,j$.  These are ${\bf A}_1$ singularities.  We
already know that for generic cubic $F$, the part of the cohomology of
the complex manifold $y^2=F(x, \tilde u)$ that is odd under
$y\leftrightarrow -y$ is four dimensional.  Each time an ${\bf A}_1$
singularity appears, the dimension of that part of the cohomology
decreases by one.  For the special case of the equation in \eqofour,
there are three singularities, so the odd part of the cohomology is one
dimensional.  The number one is no coincidence; the mass-deformed $N=4$
theory has one mass parameter $m$.

Since the relevant part of the cohomology is one dimensional,
there is one period or residue, and we will use it to determine $f$.

As in the discussion of $N_f=4$, we consider a line in the $x- \tilde u$
plane of the form
\eqn\knin{x=e_1 \tilde u+e_1{}^2f+\theta.}
Restricted to that line, the equation
\genstro\ becomes
\eqn\pnin{y^2=\theta\left(\theta+(e_1-e_2) \tilde
u+(e_1{}^2-e_2{}^2)f\right) \left(\theta+(e_1-e_3) \tilde
u+(e_1{}^2-e_3{}^2)f\right).}
The condition that the right hand side is a perfect square gives
\eqn\unin{\theta=-(e_1-e_2)(e_1-e_3)f.}
For that value of $\theta$, we get two curves $D_+$ and $D_-$ given
by \knin\ together with
\eqn\ynin{y=\pm \left(\theta(e_1-e_2)(e_1-e_3)\right)^{1/2}( \tilde u-
\tilde u_0).}
Taking $C=D_+-D_-$ gives one divisor odd under $y\leftrightarrow -y$.
One is enough as the odd part of the cohomology is one dimensional.
One can show that
\eqn\znin{C\cap C=-2.}

Now we can repeat the derivation of \nores.  The only real difference in
determining $c=\int_C\omega$ is that now $\omega=(\sqrt 2/4\pi)
dv\,dx/vy$.  So we get
\eqn\wesdo{c=-2i\sqrt 2\sqrt{\theta\over (e_1-e_2)(e_1-e_3)}.}
In determining the residue $\lambda$, we must also remember that
according to \znin, the intersection matrix is now $M=-2$ instead of
$-4$. So we get
\eqn\resdo{{\rm Res}_C(\lambda)=-{i\sqrt 2\over 2\pi i}
\left({\theta\over (e_1-e_2)
(e_1-e_3)}\right)^{1/2}.}
Setting this residue equal to $m/2\pi i\sqrt 2$, we get
$\theta^{1/2}=im\left((e_1-e_2)(e_1-e_3)\right)^{1/2}/ 2$.
Comparing to \unin\ gives $f=m^2/4$, so finally the curve of the
mass-deformed $N=4$ theory is
\eqn\hesdo{y^2=(x-e_1 \tilde u-{1\over 4}e_1{}^2m^2)
(x-e_2 \tilde u-{1\over 4}e_2{}^2m^2)
(x-e_3 \tilde u -{1\over 4}e_3{}^2m^2),}
in agreement with the result obtained in section 16.2.

\vskip36pt
\centerline{{\bf Acknowledgements}}

We would like to thank T. Banks, K. Intriligator, R. Leigh, G. Moore,
R. Plesser and S. Shenker for useful discussions and P. Deligne for
explanations that made possible the analysis in section 17.  This work
was supported in part by DOE grant \#DE-FG05-90ER40559 and in part by
NSF grant \#PHY92-45317.

\listrefs

\end